\documentclass[aip,graphicx]{revtex4-1}
\usepackage{amsfonts,amssymb,amsmath,amsthm,graphicx}
\usepackage{url}
\usepackage{comment}
\usepackage[separate-uncertainty = true,multi-part-units = brackets]{siunitx}
\usepackage[colorlinks=true, citecolor=blue,urlcolor=blue,linkcolor=blue,filecolor=black]{hyperref}
\usepackage[T1]{fontenc}
\draft 
\usepackage{nowidow}
\usepackage{ulem}

\begin{document}

\title{A Quantum Key Distribution Testbed using a\\ Plug\&Play Telecom-wavelength Single-Photon Source} 

\author{Timm\,Gao}
\affiliation{Institute of Solid State Physics, Technische Universit\"at Berlin, 10623 Berlin, Germany}

\author{Lucas\,Rickert}
\affiliation{Institute of Solid State Physics, Technische Universit\"at Berlin, 10623 Berlin, Germany}

\author{Felix\,Urban}
\affiliation{Institute of Solid State Physics, Technische Universit\"at Berlin, 10623 Berlin, Germany}

\author{Jan\,Gro{\ss}e}
\affiliation{Institute of Solid State Physics, Technische Universit\"at Berlin, 10623 Berlin, Germany}

\author{Nicole\,Srocka}
\affiliation{Institute of Solid State Physics, Technische Universit\"at Berlin, 10623 Berlin, Germany}

\author{Sven\,Rodt}
\affiliation{Institute of Solid State Physics, Technische Universit\"at Berlin, 10623 Berlin, Germany}

\author{Anna\,Musia{\l}}
\affiliation{Department of Experimental Physics, Wroclaw University of Science and Technology, 50-370 Wroc{\l}aw, Poland}

\author{Kinga\,{\.{Z}}o{\l}nacz}
\affiliation{Department of Optics and Photonics, Wroclaw University of Science and Technology, 50-370 Wroc{\l}aw, Poland}

\author{Pawe{\l}\,Mergo}
\affiliation{Institute of Chemical Sciences, Maria Curie Sklodowska University, 20-031 Lublin, Poland}

\author{Kamil\,Dybka}
\affiliation{Fibrain Sp. z o.o., 36-062 Zaczernie, Poland}

\author{Wac{\l}aw\,Urba\'{n}czyk}
\affiliation{Department of Optics and Photonics, Wroclaw University of Science and Technology, 50-370 Wroc{\l}aw, Poland}

\author{Grzegorz\,S{\k{e}}k}
\affiliation{Department of Experimental Physics, Wroclaw University of Science and Technology, 50-370 Wroc{\l}aw, Poland}

\author{Sven\,Burger}
\affiliation{Zuse Institute Berlin, 14195 Berlin, Germany}

\author{Stephan\,Reitzenstein}
\affiliation{Institute of Solid State Physics, Technische Universit\"at Berlin, 10623 Berlin, Germany}

\author{Tobias\,Heindel}
\email[Corresponding author: ]{tobias.heindel@tu-berlin.de}
\affiliation{Institute of Solid State Physics, Technische Universit\"at Berlin, 10623 Berlin, Germany}

\date{\today}

\begin{abstract}
	Deterministic solid-state quantum light sources are considered key building blocks for future communication networks. While several proof-of-principle experiments of quantum communication using such sources have been realized, most of them required large setups often involving liquid helium infrastructure or bulky closed-cycle cryotechnology. In this work, we report on the first quantum key distribution (QKD) testbed using a compact benchtop quantum dot single-photon source operating at telecom wavelengths. The plug\&play device emits single-photon pulses at O-band wavelengths (\SI{1321}{\nano\meter}) and is based on a directly fiber-pigtailed deterministically-fabricated quantum dot device integrated into a compact Stirling cryocooler. The Stirling is housed in a 19-inch rack module including all accessories required for stand-alone operation. Implemented in a simple QKD testbed emulating the BB84 protocol with polarization coding, we achieve an antibunching of $g^{(2)}(0) = 0.10\pm0.01$ and a raw key rate of up to \SI{4.72(13)}{\kilo\hertz} using an external pump laser. In this setting, we further evaluate the performance of our source in terms of the quantum bit error ratios, secure key rates, and tolerable losses expected in full implementations of QKD also accounting for finite key size effects. Furthermore, we investigate optimal settings for a two-dimensional temporal acceptance window applied on receiver side, resulting in predicted tolerable losses up to \SI{23.19}{\decibel}. Not least, we compare our results with previous proof-of-concept QKD experiments using quantum dot single-photon sources. Our study represents an important step forward in the development of fiber-based quantum-secured communication networks exploiting sub-Poissonian quantum light sources.
\end{abstract}


\maketitle 

\section{Introduction}
Quantum light sources emitting single- and entangled photon states on the push of a button are building blocks for advanced photonic quantum technologies and quantum communication in particular. They carry the potential to boost the performance of quantum key distribution (QKD) systems \cite{Waks2002a} and to enable secure communication at global scales by exploiting quantum repeater architectures \cite{Briegel1998}. Among the several promising species of quantum emitters in the solid-state (see Refs. \cite{Aharonovich2016, Arakawa2020,Zhang2020} for detatiled reviews), epitaxial semiconductor quantum dots (QD) stand out due to the highest reported single photon purity in terms of $g^{(2)}(0)$ \cite{Miyazawa2016, Schweickert2018}, high photon indistinguishability and brightness of single- and entangled-photon states \cite{Liu2019, Wang2019, Tomm2021}, and their speed enabling GHz clock rates \cite{Schlehahn2016, Shooter2020}. Due to a lack of efficient and practical quantum light sources, most implementations of QKD are still carried out with faint lasers \cite{Bennett1992}, referred to as weak coherent pulses (WCPs). To compensate for the resulting multi-photon contributions, so-called decoy-state protocols have been invented \cite{Wang2005, Lo2005}, which define the current state-of-the-art in direct point-to-point QKD \cite{Boaron2018}. Still, due to their Poissonian light-states, WCP-based implementations are fundamentally limited in their efficiency. From the tremendous progress seen in the field of QD-based quantum light sources, however, renewed prospects arise for real-world implementations of quantum communication benefitting from sub-Poissonian light states.

While previously the performance and yield of QD-based nanophotonic devices was limited by the random spatial distribution of the quantum emitters and the inhomogeneous spectral broadening of the emitter ensemble (both resulting from the stochastic nature of the self-assembly process during epitaxial growth), nowadays deterministic fabrication technologies enable much higher degrees of control. Exploiting marker-based \cite{Badolato2005} or in-situ \cite{Dousse2008,Gschrey2013} lithography techniques, photonic microstructures can be fabricated deterministically around single pre-selected emitters resulting in high performance and high yield of devices (e.g. Refs. \cite{Gschrey2015,Somaschi2016,Heindel2017}). The advancements of deterministic fabrication technologies, as summarized in a recent review article \cite{Rodt2020}, are expected to have large impact on the application of deterministic quantum light sources in photonic quantum technologies and are also a key to the results reported in our work. 

Moreover, QD-based sources demonstrated their potential for quantum communication in several proof-of-principle experiments using single-photon \cite{Waks2002,Intallura2009,Collins2010,Heindel2012,Rau2014,Takemoto2015} and entangled-photon \cite{Dzurnak2015,Basset2021,Schimpf2021} states. We refer the interested reader to Ref. \cite{Vajner2021} for an extensive recent review article on implementations of quantum communication using QDs. For the remainder of this article, we consider the first QKD protocol ever invented, known as BB84 \cite{Bennett1984}.
Here, one party, Alice, sends single photons randomly prepared in four different polarization states (spanning two conjugate bases) via the quantum channel to the second party, Bob, who detects the photons randomly and independent of Alice in the two different polarization bases. Assuming a single-photon source (SPS) with ideal photon statistics ($g^{(2)}(0)=0$), the key parameters to consider in this prepare-and-measure configuration are the sifted key rate, the quantum bit error rate (QBER), and the secure key rate. The sifted key refers to the fraction of the transmitted (and detected) raw key after excluding (i.e. sifting) the bits prepared and measured in different polarization basis (resulting in a probabilistic projection). The QBER quantifies the amount of erroneous bits in the sifted key and determines how much the sifted key needs to be shrunk in the classical post-processing (error correction and privacy amplification), which results finally in the secure key rate usable for data encryption. If the SPS has a finite $g^{(2)}(0)$, which is always the case in practical implementations, a finite multi-photon emission probability needs to be further taken in to account in the security analysis \cite{Waks2002a}. In addition, one distinguishes scenarios assuming that an infinitely long key can be generated, corresponding to the so-called asymptotic limit, and more realistic scenarios taking finite-key size effects into account \cite{Scarani2009}.

Despite the immense progress in the field, the need for large optical setups and bulky cryotechnology, involving liquid helium infrastructure or complex closed-cycle refrigerators, is often considered a major drawback of QDs or other types of "cryogenic" emitters if it comes to applications. On the other hand, the advances in the performance of QD-based quantum light sources and the prospects for device integration triggered the interest in developing more practical user-friendly sources for applications outside shielded laboratories. First efforts in this direction utilized fiber-coupled QD samples employing fiber-bundles containing about 600 individual fiber cores to spatially post-select a single emitter \cite{Xu2007}. More recently, the direct fiber-coupling of photonic nanostructures with embedded QDs, such as photonic wires \cite{Cadeddu2016} and optically \cite{Snijders2018} as well as electrically \cite{Rickert2021} pumped micropillar cavities, has been reported. Still, these efforts relied on conventional bulky cryotechnology. In previous work, our groups pioneered and advanced the integration of QD-devices in Stirling cryocoolers for stand-alone operation \cite{Schlehahn2015, Schlehahn2018, Musial2020}, enabling much more compact setups. Applications of these user-friendly sources in implementations of quantum information, however, have not been demonstrated so far.

In this article, we report on the application of a benchtop plug\&play QD-SPS operating at telecom wavelengths in experimental tests for quantum communication. To this end, the outline of our article is as follows: In Section~\ref{sec:Testbed} we first present the overall QKD-testbed used to emulate the BB84 protocol for the studies in this work before providing details on the transmitter- and receiver-module in distinct subsections. Subsection~\ref{sec:Alice} presents the developed benchtop QD-SPS providing push-button single-photon pulses in a standard Telecom fiber and provide a basic characterization of its spectral and quantum-optical properties. The receiver-module is discussed in Subsection~\ref{sec:Bob}. Section~\ref{sec:Results} discusses the experimental results achieved with our QKD testbed. Here, we start with a basic performance evaluation in Subsection~\ref{sec:Basics} quantifying the achievable QBER, the suppression of multi-photon emission events via $g^{(2)}(0)$, and the mean photon number per pulse $\mu$. In Subsection~\ref{sec:TempFiltering}, we analyze the achievable secure key rates, both in the asymptotic and finite-key size regime. We apply temporal filtering to optimize the achievable performance and to maximize the tolerable transmission loss, i.e. transmission distance. The results obtained are discussed and compared with previous implementations of QKD in Subsection~\ref{sec:Compare}. Our article closes with a summary and an outlook in Section~\ref{sec:Summary}.

\section{Quantum Key Distribution Testbed}\label{sec:Testbed}

\begin{figure*}
	\includegraphics{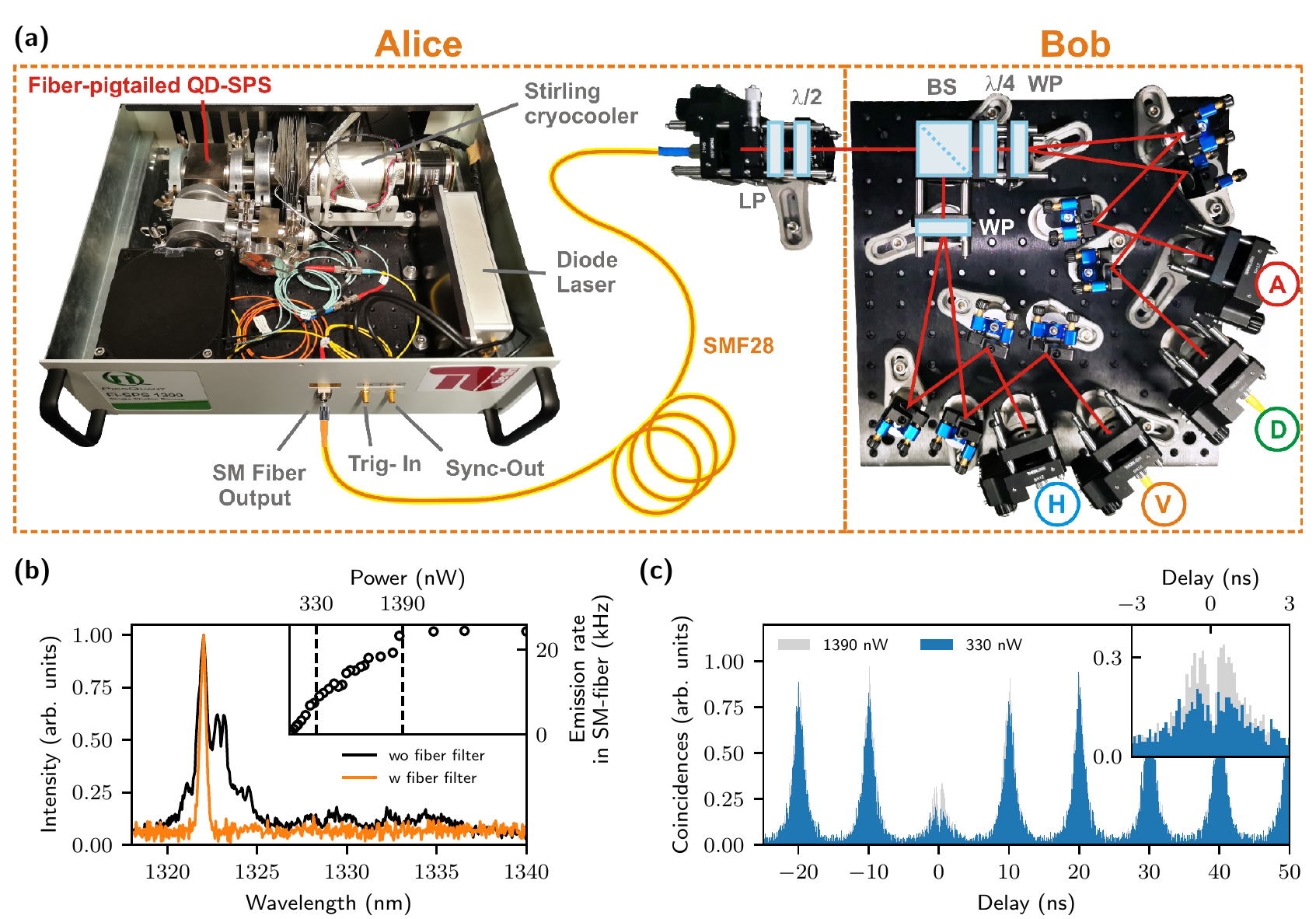}
	\caption{BB84-QKD testbed using a compact stand-alone quantum dot (QD) single-photon source (SPS) integrated in a \SI{19}{inch} rack-module and polarization encoding. (a)~The transmitter Alice comprises a deterministically fiber-coupled QD-SPS integrated in a Stirling cryocooler, a pulsed diode-laser for excitation, and a fiber-based spectral filter (not shown in the image). Alice sends pulses with fixed polarization (H, V, D, A) set by a linear polarizer (LP) and a  lambda-half ($\lambda$/2) waveplate to the receiver Bob. The latter consists of a 4-state polarization analyzer realized with a 50:50 beamsplitter (BS), a compensating $\lambda$-quarter ($\lambda/4$) waveplate, and two Wollaston-prisms (WP) for spatial separation of orthogonal polarizations. (b)~Output spectra of the fiber-coupled SPS with (orange) and without (black) spectral filtering using a fiber-based bandpass filter set to a center wavelength of \SI{1321}{\nano\meter}. Spectra were recorded at a temperature set-point of the Stiriling cryocooler of \SI{40}{\kelvin}. (c)~Second-order autocorrelation $g^{(2)}(\tau)$ histogram under pulsed excitation at \SI{785}{\nano\meter} with a repetition rate of \SI{100}{\mega\hertz} using the internal laser (time-bin width: \SI{100}{\pico\second}). Depending on the excitation power (see inset), more or less pronounced recapture processes are observed near $\tau = 0$ due to the above-band excitation scheme.}
	\label{fig:figure1}
\end{figure*}
The QKD testbed developed for the studies in this work is illustrated in Fig.~\ref{fig:figure1}\,(a). On the transmitter side, Alice comprises a benchtop optically triggered source of flying qubits operating in the telecom O-band. Here, single-photon pulses are provided via a standard Telecom fiber and generated inside a \SI{19}{inch} rack module housing all components required for this complex task, including a cryocooler, a pump laser, and a spectral filter. This benchtop source module enables cryogenic operation, by integrating a deterministically-fabricated fiber-pigtailed QD-device into a compact Stirling cryocooler. Details on the transmitter module are provided in Section \ref{sec:Alice}. To evaluate the performance of our quantum-light source for applications in quantum communication, we prepare single-photon pulses in four polarization states, required for implementations of the BB84 protocol, using a linear polarizer and a $\lambda/2$ waveplate. Note, while full implementations of QKD require a dynamic modulation of the polarization states (e.g. via electro-optical modulators), we statically prepare the polarization states in sequential measurement runs in our testbed, which allows us to determine all key-parameters for QKD and their limits. The polarization qubits are then sent to the receiver module in a back-to-back configuration, i.e. without additional loss in the quantum channel. On the receiver side, Bob comprises a four- (4-) state polarization decoder designed for operation at Telecom wavelengths. Here, photons in the four BB84-states are decoded in four different output fibers connected to a 4-channel single-photon detection system. For details on the receiver module, we refer to Section \ref{sec:Bob}.

\subsection{Transmitter-Module (Alice)}\label{sec:Alice}
At the heart of our transmitter module, the SPS comprises a single InGaAs/GaAs QD which was preselected and deterministically integrated into a photonic microstructure using 3D in-situ electron-beam lithography \cite{Gschrey2015}. The quantum emitter is embedded in a monolithic micromesa structure (height: \SI{705}{\nano\meter}, diameter: \SI{2}{\micro\meter}) integrated on top of a distributed Bragg-reflector serving as bottom mirror ($15\times$ \SI{118}{\nano\meter} Al$_{0.92}$Ga$_{0.08}$As and \SI{102}{\nano\meter} GaAs) for increasing the photon collection efficiency. To shift the emission of the QDs in this sample to the telecom O-band, a strain-reducing InGaAs-layer was applied during the epitaxial growth via metal-organic vapor deposition (MOCVD) \cite{Guffarth2001}. For integration in the Stirling cryocooler (CryoTel GT, Sunpower Inc.), the QD-device is permanently coupled to a custom-made Ge-doped optical single-mode~(SM) fiber with high numerical aperture (NA = 0.42), whose core is thermally expanded onto a standard SMF-28e fiber \cite{Zolnacz2019}. The design parameters for the combined micromesa-fiber assembly follow results of numerical simulations based on finite-element methods \cite{Schneider2018} using an Bayesian optimization algorithm \cite{Schneider2019}. The air-cooled linear-piston Stirling cryocooler with fiber-optical feedthroughs allows for cryogen-free operation of the SPS down to \SI{40}{\kelvin} (at \SI{100}{\watt} cooling power) while fitting into a 19-inch rack box. For optically pumping the quantum emitter, two different lasers can be coupled to the sample via a 1x2 SM fiber-coupler with a split ration of 90:10. While the laser emission reaches the sample via the \SI{10}{\percent}-port, the sample emission is collected via the \SI{90}{\percent}-port for further use. Two options are available for the laser source: (1) a compact diode laser with fixed wavelength for above-bandgap excitation (wavelength: \SI{785}{\nano\meter}, pulsewidth: \SI{70}{\pico\second}, repetition rate: variable, PicoQuant GmbH) which is integrated in the 19-inch rack module, or (2) an external spectrally-tunable laser system (pulsewidth: \SI{2}{\pico\second}, repetition rate: \SI{80}{\mega\hertz}, APE GmbH) allowing for quasi-resonant excitation of the single quantum emitter (emission wavelength: \SI{1321}{\nano\meter}) in an excited state (p-shell) at a wavelength of \SI{1247.9}{\nano\meter}. Both lasers are coupled to SM fibers allowing for a simple connection to the 1x2 fiber-coupler inside the rack module. While option (1) is used for the pre-characterization of the SPS presented in this section (see further below), option (2) is used for the actual QKD tests in Section~\ref{sec:Results}, as quasi-resonant excitation enables an improved single-photon purity being beneficial in QKD. For high-transmission and compact spectral filtering, a fiber-based bandpass filter with adjustable center wavelength is used (full-width at half-maximum \SI{0.5}{\nano\meter}, WL Photonics Inc.) is integrated in the 19-inch rack module, which removes spectral contributions of background emitters and residual scattered light of the excitation laser. Additionally, we applied polarization suppression to reduce the laser background.

Figure~\ref{fig:figure1}\,(b) shows the spectral characteristic of our transmitter module at the SM-fiber output with and without fiber-based bandpass filter, respectively, using the integrated diode laser triggering at a clock rate of \SI{100}{\mega\hertz}. The corresponding second-order autocorrelation histogram $g^{(2)}(\tau)$ recorded at the spectrally filtered output signal, is shown in Fig.~\ref{fig:figure1}\,(c) for excitation powers of the quantum emitter at (\SI{1390}{\nano\watt}) and well below (\SI{330}{\nano\watt}) saturation, respectively. For this pre-characterization measurement, a standard fiber-coupled Hanbury-Brown and Twiss setup was used in combination with superconducting nanowire single-photon detectors (SNSPDs) (Single Quantum Eos, Single Quantum BV) with a temporal resolution between \SI{35}{\pico\second} and \SI{100}{\pico\second}. Analyzing the raw measured coincidences within the full repetition period, we extract antibunching values of $g^{(2)}(0) = 0.49\pm0.01$ at saturation and $g^{(2)}(0) = 0.37\pm0.02$ at \SI{330}{\nano\watt}, clearly indicating single-photon emission with sub-Poissonian photon statistics ($g^{(2)}(0)<0.5$). In this operation mode, however, the $g^{(2)}(0)$ value of our source, i.e. the suppression of multi-photon emission events, is strongly limited by recapture processes \cite{Aichele_2004, Dalgarno2008}. This can be seen from the significant contribution of coincidences close to $\tau=0$ (cf. inset in Fig.~\ref{fig:figure1}\,(c)), resulting from the applied above-band excitation scheme. Compared to Ref.~\cite{Musial2020} the $g^{(2)}(0)$ value stated here is somewhat higher, which has two possible reasons. Firstly, as we used different fiber-pigtailed QD-devices in both studies, the spectral background contributions to the selected emission line can be different (cf. Fig.~\ref{fig:figure1}\,(b)). Secondly, and in contrast to Ref.~\cite{Musial2020}, we did neither apply background subtraction nor fitting, as in the context of quantum communication all detection events within a given temporal acceptance time window contribute to the key material, independent of whether they origin from a signal photon or an erroneous/background event. Note, that we also have to make this worst-case assumption concerning the photon statistics of our source, when applying temporal filtering to optimize the performance of our testbed (cf. Subsection~\ref{sec:TempFiltering}). To improve the single-photon purity for the QKD tests in Section~\ref{sec:Results}, we apply resonant p-shell excitation of the QD using the optional external laser resulting in $g^{(2)}(0) = 0.10$~$\pm$~$0.01$. Importantly, the function of this external laser can be integrated in the transmitter module by either using more compact lasers with the required performance or even lasers integrated on-chip with the QD-devices. Prospects for higher degrees of laser integration will be discussed in Section~\ref{sec:Summary}.

Noteworthy, the dimensions of our benchtop transmitter module ($(40\times45\times13)\,\textrm{cm}^{3}$), providing triggered single-photon pulses in a SMF28 fiber, are for the first time competitive with commercial laser-based QKD systems \cite{Stucki_2002}$^,$\footnote{For companies see e.g., ID Quantique SA, Toshiba Europe Limited, MagiQ Technologies.}. This is a relative reduction to only \SI{14}{\percent} of the O-band QD-SPS's volume reported in Ref.~\cite{Musial2020} ($(45\times60\times60)\,\textrm{cm}^{3}$).

\subsection{Receiver-Module (Bob)}\label{sec:Bob}
The receiver module comprises a 4-state polarization decoder with the four output ports connected to the SNSPD-based detection system. Time-tagging electronics (quTAG, qutools GmbH) are used for recording the detection events which are transferred to a personal computer for further processing. Inside the polarization decoder, we realize a passive basis choice using a non-polarizing beamsplitter (BS) with a nominal splitting ratio of 50:50. The discrimination of the polarization after the BS is realized with two Wollaston prisms (WPs) for the detection in the H-V and D-A basis, respectively. The WPs spatially separate two orthogonal polarizations by an angle of \SI{20}{\degree} with an extinction ratio of up to 1:100,000 and \SI{90}{\percent} transmission. For detection in the D-A basis, the respective WP is rotated by an angle of \SI{45}{\degree} to match the prism's axes with the incoming diagonally or anti-diagonally polarized light. This configuration allows for a low-profile of the overall optical setup. To minimize residual depolarization effects, the rotated WP is implemented in the transmission path of the BS in combination with a quarter wave plate compensating for slight ellipticities introduced by the BS. Note, that although our source is developed for QKD via telecom fibers, we use free-space optics inside the receiver module, due to the achievable lower intrinsic QBER and larger spectral bandwidth, both at comparable transmission losses. The polarization decoder is connected to the SNSPDs via four single-mode fibers (type SMF-28). For high coupling efficiencies to these SM fibers, each output port of the decoder features two mirrors for aligning the four beams after the WPs. 
The overall transmission loss of the polarization decoder $\eta_\mathrm{Bob}$ was experimentally determined to \SI{-5.2}{\decibel}, including the optical losses of \SI{-3.0}{\decibel} and \SI{-2.2}{\decibel} due to the SNSPD's finite detection efficiency.
The time-tags recorded at the four ports of the receiver module are evaluated using a homemade software package (see Ref. \cite{Kupko2020} for details).\\

\section{Results}\label{sec:Results}
\subsection{Basic QKD Performance Evaluation}\label{sec:Basics}

\begin{figure}
	\includegraphics{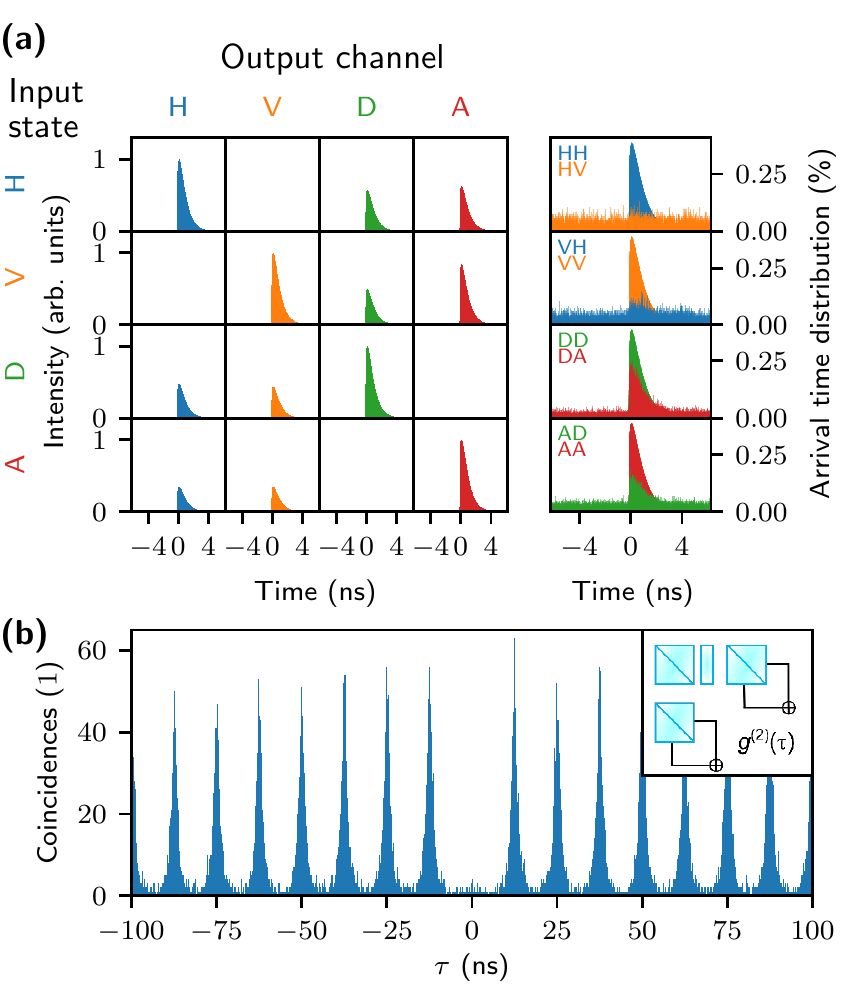}
	\caption{Characterization of the receiver module Bob using triggered single-photon pulses from Alice as input. (a)~Measured photon arrival time distributions at the four detection channels of Bob for single-photon input-polarizations of H, V, D, and A (time-bin width: \SI{1}{\pico\second}). Measurement data in the left 4x4 matrix are normalized to the maximum of the time-trace of the respective diagonal element. The right panel shows the two data sets for the given input polarization basis for the two respective detection channel of the two bases, and each distribution being normalized to the sum of all events in the respective detection channel (e.g., 'HV' corresponds to the detection channel for V-Polarization measured with input H-polarization.) This photon arrival probability distribution reveals erroneous detection events due to optical imperfections in Bob. (b)~Second-order autocorrelation $g^{(2)}(\tau)$ histogram under pulsed resonant p-shell excitation at \SI{1247.9}{\nano\meter} $g^{(2)}(0) = 0.10\pm0.01$ (time-bin width: \SI{50}{\pico\second}, acquisition time: \SI{5}{hours}). The autocorrelation was obtained by correlating the time-tags recorded at all four detection channels of Bob as shown in the inset.}
	\label{fig:figure2}
\end{figure}
The most important performance parameter of a QKD system is the secret key rate. In this work, we are investigating a practical SPS with a finite multi-photon emission probability $p_{\mathrm{m}}$. The secret key rate for the asymptotic case (i.e. infinite key length) is hence given by \cite{10.5555/2011586.2011587, Chaiwongkhot_2017}
\begin{equation}
	S_{\infty} = S_{\mathrm{sift}}\left[A\left(1-h(e/A)\right)-f_{\mathrm{EC}}h(e)\right].
	\label{eq:S_infty} 
\end{equation}
Here, $S_{\mathrm{sift}}$ is the sifted key rate, $A = (p_{\mathrm{click}}-p_{\mathrm{m}})/p_{\mathrm{click}}$ the single-photon detection probability calculated from multi-photon emission probability $p_{\mathrm{m}}$ and overall click probability $p_{\mathrm{click}}~\approx~\mu\cdot~T~\cdot~\eta_{\mathrm{Bob}}~+~p_{\mathrm{dc}}$. Here, $\mu$ denotes the mean photon number per pulse which Alice couples to the quantum channel, $\eta_{\mathrm{Bob}}$ the efficiency of the receiver module including the detector losses, $T$ is the link transmission, and $p_{\mathrm{dc}}$ the dark count probability. The quantum bit error ratio (QBER) \footnote{Note, that a more widely used term in the literature is quantum bit error rate (QBER) in units of $s^{-1}$. As the QBER entering the binary Shannon entropy in Eq.~1 (denoted as $e$) must be a probability (cf. Subsection~\ref{sec:Basics}), we consistently use the quantum bit error ratio, defined as the ratio of erroneous bits to all detected bits in our work.} is denoted as $e$, $h(e)$ the binary Shannon-entropy and $f_{\mathrm{EC}}$ is the error correction efficiency. For an ideal SPS (i.e., $p_{\mathrm{m}} = 0$) the factor $A$ equals unity, so $A$ is a correction term for realistic SPSs.

First, we investigate the performance of our Bob module which is characterized by its transmission loss and its contribution to the overall QBER. For this purpose, we sequentially prepared single-photon pulses in horizontal~(H), vertical~(V), diagonal~(D), and antidiagonal~(A) polarization and recorded events at all four detection channels of our receiver module. The resulting time-resolved experimental data are summarized in Fig.~\ref{fig:figure2}\,(a) in a 4x4 matrix, where each row corresponds to the input-polarization and columns to the respective detection channel. Histograms within one row are normalized by the maximum of the time-trace of the respective diagonal element (e.g., all first-row distributions are normalized by the maximum of the HH distribution (blue), while all second-row distributions are normalized by the maximum of the VV distribution (orange), etc.). The expected behavior of the polarization decoder is clearly observed. For a given input-state, almost all photons are detected in the correct channel of the corresponding basis, i.e., diagonal elements of the matrix, whereas a probabilistic projection is observed in the conjugate basis. Erroneous detection events at the wrong channel are only visible, if the time-resolved measurements are renormalized to the arrival time probability distribution for both output channels in each basis, as depicted in the right panel of Fig.~\ref{fig:figure2}\,(a). Here, each distribution is normalized to the sum of all events registered in the respective channel (e.g., the HV distribution (orange) is normalized by all clicks registered in the V-channel). In this representation, one can qualitatively distinguish different error contributions. Noise stemming from detector dark counts or stray light results in uncorrelated background noise (cf. HV arrival time distribution (orange)). In our case, the errors due to noise accumulate to a total of \SI{42(1)}{\hertz} at the chosen working point of the SNSPDs (mean detection efficiency \SI{-2.2}{\decibel}). In contrast, optical imperfections inside Bob like finite extinction ratios of polarization optics, result in erroneous events correlated in time with the signal events. These errors are almost negligible in the rectilinear bases but become noticeable in the diagonal basis, e.g., red arrival time distribution in the third row (D-input and A-output). Here, temporarily correlated events are superimposed to the uncorrelated background noise. The discrimination between the two error contributions mentioned above will be exploited in section III.B, to effectively enhance the signal-to-noise-ratio via temporal filtering (cf. Fig.~\ref{fig:figure3}). From the experimental data in the right panel of Fig.~\ref{fig:figure2}\,(a), the contribution to the QBER for a given input polarization can be calculated from the number of erroneous events in the wrong channel divided by the total number of events in the sifted basis, e.g., $\mathrm{QBER}_{\mathrm{H}} = N_\mathrm{V}/(N_\mathrm{H}+N_\mathrm{V})$. The corresponding QBERs of the respective polarization-channels range between $\mathrm{QBER}_{\mathrm{H}} = \SI{0.35}{\percent}$ and $\mathrm{QBER}_{\mathrm{D}} = \SI{0.82}{\percent}$. Additional deviations from an ideal setup are due to detection efficiency mismatches between channels (c.f. imbalance between the matrix elements VD and VA in Fig.~\ref{fig:figure2}\,(a)), which need to be taken into account in full implementation of QKD~\cite{Lydersen2010}.

Next, we consider the multi-photon emission probability~$p_{\mathrm{m}}$ of our sub-Poissonian quantum light source. This is not as straightforward as in implementations using weak coherent pulses obeying Poissonian statistics. We estimate the upper bound $p_{\mathrm{m}} = \mu^{2}g^{(2)}(0)/2$ according to Ref.~\cite{Waks2002a}. Note, that finding tighter bounds to $p_{\mathrm{m}}$ could further boost the performance of single-photon QKD in future \cite{Gr_nwald_2019, Chavez-Mackay2020}. In a similar way, recent records in the field of decoy-state QKD \cite{Boaron2018} benefited from lower bounds in the finite-key analysis \cite{Rusca2018}. To determine the photon statistics inside the quantum channel of our QKD testbed, we calculated $g^{(2)}(\tau)$ by correlating all time-tags registered at Bobs' four output channels as displayed in Fig.~\ref{fig:figure2}\,(b) \cite{Kupko2020}. Here, recapture processes are strongly suppressed compared to the above-band excitation shown in Fig.~\ref{fig:figure1}\,(c), due to the resonant p-shell excitation of the quantum emitter. We extract an antibunching value of $g^{(2)}(0) = 0.10\pm0.01$ for this measurement by integrating the coincidences over the full repetition period of \SI{12.49}{\nano\second}. This represents a significant improvement compared to Ref.~\cite{Musial2020}, where a background-corrected value of $g^{(2)}(0) = 0.15$ was reported for a Stirling-based QD-SPS emitting at O-band wavelengths under above-bandgap excitation. Other fiber-coupled sources achieved similar values ($g^{(2)}(0) = 0.09$) \cite{Lee2019}, but at lower temperatures using liquid helium infrastructure. Limiting factors to the $g^{(2)}(0)$ observed in our work are related to a finite suppression and internal reflections of the laser emission as well as residual background contributions from other QD emission lines (cf. Fig.~\ref{fig:figure1}\,(c)). Applying temporal filtering as a post-process to the recorded timestamps, we estimate the contribution of scattered laser photons and events due to QD re-excitation to the overall signal to \SI{2.5}{\percent}. Removing these events results in a reduction of $g^{(2)}(0)$ from 0.10 to 0.06, with the remaining imperfection attributed to the residual background emission of nearby QD emission lines. The finite laser suppression in our experiment is mainly caused by reflections and scattering events associated with the FC/PC connectors in combination with a finite transmission of the bandpassfilter at the laser wavelength in our all-fiber-coupled measurement configuration. Straightforward improvements include the use of FC/APC connectors, a bandpassfilter with better side-band suppression ratios, and the use of coherent excitation schemes and/or QD-devices with improved spectral purity. With these improvements, our all-fiber benchtop device is anticipated to become competitive or even surpassing state-of-the-art values of $g^{(2)}(0) = 0.027$\cite{Srocka2020} (at \SI{10}{\kelvin}) and $g^{(2)}(0) = 0.026$\cite{Kolatschek2021} (at \SI{40}{\kelvin}) for deterministically and non-deterministically fabricated O-band QD-devices, respectively, obtained in experiments using standard bulky cryotechnology, high-NA microscope objectives, and using grating spectrometers for spectral filtering.
Not least, the $g^{(2)}(\tau)$ in the present work indicates a blinking behavior \cite{Jahn2015} which can be described by the ``on-off'' model from Ref.~\cite{Santori2001}. The respective time constants are $\tau_{\mathrm{on}} = \SI{482.3(25)}{\nano\second}$ and $\tau_{\mathrm{off}} = \SI{275.1(10)}{\nano\second}$ yielding a QD ``on''-state efficiency of \SI{57}{\percent} reducing the efficiency of our source (see also Supplementary Material, Figure~\ref{fig:blinking}). This blinking effect is not a fundamental limit and can possibly be overcome by optimizing the QD growth or embedding QDs in charge tunable devices \cite{Zhai2020}. In addition, we observe a higher photon detection rate, i.e., raw key rate, of \SI{4.72(13)}{\kilo\hertz} compared to \SI{1.15}{\kilo\hertz} in Ref.~\cite{Musial2020} due to reduced losses in the detection apparatus. From the raw key rate of $S_{\mathrm{raw}} = \SI{4.72(13)}{\kilo\hertz}$, the transmission of the Bob module $\eta_{\mathrm{Bob}} = 0.3$, and the clock rate $f = \SI{80}{\mega\hertz}$, we determine the mean photon number per pulse into the quantum channel via $\mu = S_{\mathrm{raw}}\eta_{\mathrm{Bob}}^{-1}f^{-1} = 0.0002$. We note here that employing our source in a full implementation would yield a slightly lower $\mu$ due to the use of an electro-optical modulator with typical internal losses of about \SI{-2}{\decibel}.

\subsection{Optimization via temporal filtering}\label{sec:TempFiltering}

\begin{figure*}[htp]
	\includegraphics{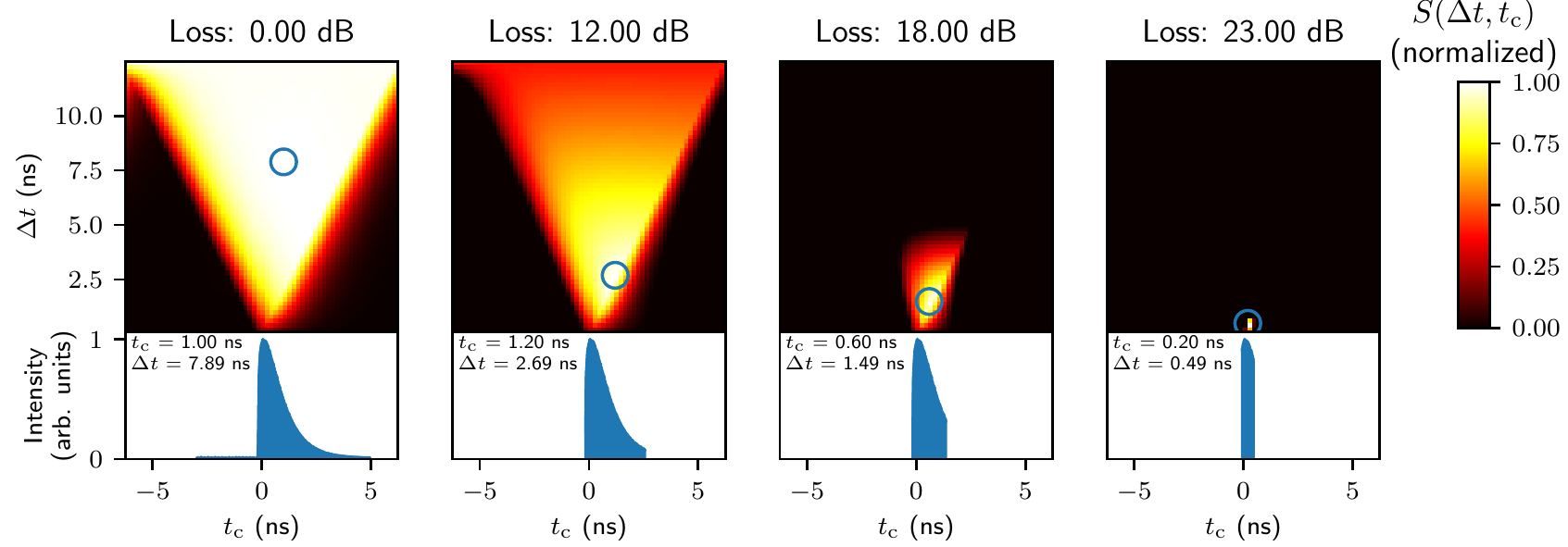}
	\caption{2D temporal filtering for optimization of the secret key rate. For the optimization, a subset of the measurement in Figure~\ref{fig:figure2}\,(b) was used for reasonable computing times. Expected secret key rate fraction $S(\Delta t,t_{\mathrm{c}})$ from Eq.~(\ref{eq:S_infty}) as a function of the temporal width $\Delta t$ and the center $t_{\mathrm{c}}$ of the acceptance time window. $S(\Delta t,t_{\mathrm{c}})$ is normalized to the maximum achievable secret key rate for each map individually. For the analysis the QBER and the sifted key fraction were considered in a two-dimensional parameter space ($\Delta t$, $t_{\mathrm{c}}$), while $g^{(2)}(0)$ was fixed to its unfiltered value (cf. discussion in main text). From the parameters at \SI{0}{\decibel} transmission losses, three different loss scenarios (\SI{12}{\decibel}, \SI{18}{\decibel}, \SI{23}{\decibel}) are simulated. A blue circle marks the optimal parameter sets resulting in a maximal secure key rate. Additionally, the evaluated optimal acceptance time window is depicted in the time-resolved measurements in the lower panels.}
	\label{fig:figure3}
\end{figure*}
Based on the parameter analysis from our QKD testbed, the secret key rate expected in a full implementation of QKD can be estimated. In this section, we evaluate optimal performance regimes by applying a routine for 2D temporal filtering recently reported by our group in Ref.~\cite{Kupko2020}. As the quantum emitter has a relatively short lifetime (\SI{0.9}{\nano\second}) compared to the pulse repetition period (\SI{12.49}{\nano\second}), we can efficiently reduce the contribution of clicks due to detector noise by reducing the temporal acceptance window. This results in a reduction of the QBER - ideally down to the limit of below \SI{0.4}{\percent} caused by optical imperfections in the worst channel. On the other hand, the applied temporal filter will also reduce the number of bits available for key distillation. Choosing the optimal trade-off between QBER and sifted key material, the secure key can be maximized for a given channel loss (see Supplementary Material, Figure~\ref{fig:QBER_Frac} for details). Due to the asymmetric shape of the photon arrival time distribution, it is beneficial to vary both the center $t_{\mathrm{c}}$ and the width $\Delta t$ of the acceptance time window. Important with respect to the protocol security is, that although we use temporal filtering in the following to maximize the secret key rate, we always make the worst-case assumption concerning the photon statistics of our source by using the full unfiltered $g^{(2)}(0)$ value. While in principle an additional performance gain is achievable by exploiting the temporally filtered $g^{(2)}(0)$, this would open security loopholes without additional extensions to the protocol implementation (see Ref.~\cite{Kupko2020} for details). Figure~\ref{fig:figure3} depicts the estimated asymptotic secret key rate $S(\Delta t, t_{\mathrm{c}})$ calculated according to Eq.~(\ref*{eq:S_infty}) in heat-maps for different loss regimes together with the resulting optimal temporal acceptance window (cf. lower panels and circle in heat-maps) resulting in the largest possible secret key rate. We used our experimental parameters $g^{(2)}(0)$ and $\mu$ as stated above. The dark count probability $p_{\mathrm{dc}}$ scales linearly with the acceptance time window. In contrast, the reduced click rate cannot be modelled so easily due to the distribution of the arrival time probabilities. The fraction of the signal resulting in different values of $p_{\mathrm{click}}$ was therefore evaluated for each acceptance time window individually (c.f. Ref.~\cite{Kupko2020}, Supplementary Note 3) by rejecting detection events outside the temporal filter. As Eq.~(\ref{eq:S_infty}) is a function of the link loss $T$, we can simulate the expected key rate for different loss regimes and still apply the temporal filtering. Accounting for the temporal resolution of our experimental setup (convolution of the detector response and the time-tagging device) we used a lower limit for the acceptance time window of \SI{90}{\pico\second}. Already for a back-to-back transmission (i.e. zero channel loss) discarding some of the sifted key can lead to higher $S(\Delta t, t_{\mathrm{c}})$ due to the improved signal-to-noise ratio. But not only does the 2D optimization increase $S(\Delta t, t_{\mathrm{c}})$, it allows also to extend the achievable transmission distance. This is exemplarily depicted in the \SI{18}{\decibel} and \SI{23}{\decibel} case in Fig.~(\ref{fig:figure3}). Black regions indicate acceptance time windows for which no secret key can be exchanged. In those higher loss regimes, without applying any temporal filtering, no secret key could be exchanged. Thus, this optimization results in an effective range extension.
\begin{figure*}[htp]
	\includegraphics{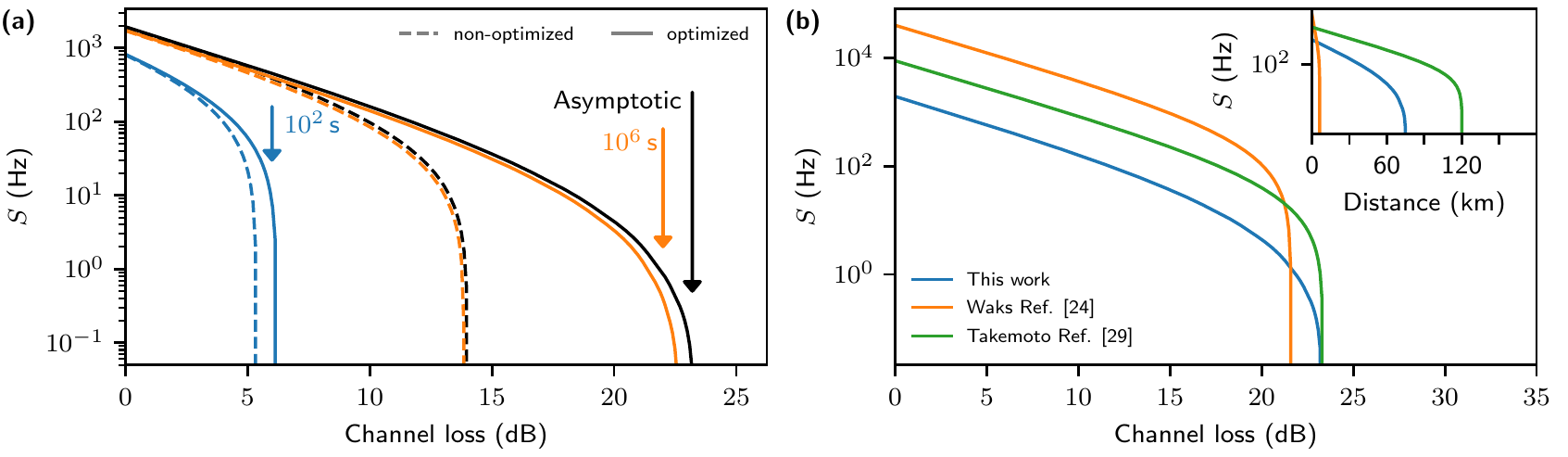}
	\caption{Rate-loss diagrams considering our experimental data from Fig.~\ref{fig:figure3}. (a) Comparison between asymptotic (Eq.~(\ref*{eq:S_infty})) and finite (Eq.~(\ref{eq:S_finite})) key rate. The secret key rate $S$ in full QKD-implementations is calculated from our experimental parameters. The values without optimizations are depicted with dashed lines, the optimized values with solid lines, respectively. (b) Comparison of our optimized parameter set with previous QD-based QKD implementations for fiber channel. We assume \SI{0.31}{\decibel/\kilo\meter} for our source at \SI{1321}{\nano\meter}, for Ref.~\cite{Takemoto2015} \SI{0.194}{\decibel/\kilo\meter} at \SI{1580.5}{\nano\meter}, and for Ref.~\cite{Waks2002} \SI{3.5}{\decibel/\kilo\meter} at \SI{877}{\nano\meter}.}
	\label{fig:figure4}
\end{figure*}
\begin{table}
	\caption{\label{tab:Parameters} Relevant parameters to calculate the secret key rates in Fig.\ref{fig:figure4}~(b). The parameters for Ref.~\cite{Waks2002} and Ref.~\cite{Takemoto2015} are based on the respective published experimental data. The efficiency $\eta_{\mathrm{Bob}}$ and $p_{\mathrm{dc}}$ are lower when applying individual filtering.}
	\begin{tabular}{|c|c|c|c|}
		\hline
		& Ref.~\cite{Waks2002} & Ref.~\cite{Takemoto2015} & This Work\\
		\hline
		$\lambda$ (nm) & 877 & 1580.5 & 1321 \\
		\hline
		$\mu$ & 0.007 & 0.009 & 0.0002 \\
		\hline
		$p_{\mathrm{dc}}$ & $10.5\cdot10^{-7}$ & $3\cdot10^{-7}$ & $5.25\cdot10^{-7}$ \\
		\hline
		$e_{\mathrm{detector}}$ & 0.025 & 0.023 & 0.010 \\
		\hline
		$\eta_{\mathrm{Bob}}$ & 0.24 & 0.048 & 0.3 \\
		\hline
		$g^{(2)}(0)$ & 0.14 & 0.0051 & 0.10 \\
		\hline
	\end{tabular}
\end{table}

In practical scenarios it is not sufficient to consider only the asymptotic case, as the limited transmission time or transferred data size limits not only the achievable rate at a given distance but also the achievable distance itself. This becomes particular relevant for QKD in free-space optical systems with a relative movement between transmitters and receivers (e.g. airplanes \cite{Nauerth2013}, satellites \cite{Liao2017, Yin2017, Yin2020}), at which stable link-conditions cannot be maintained for a long time. The loss budget of fiber-based connections needs to include finite key size effects especially in regimes with high losses and therefore long transmission times. Adapting the finite size key corrections of Ref.~\cite{Cai_2009} the key rate in Eq.~(\ref{eq:S_infty}) is modified:
\begin{equation}
	S_{\mathrm{finite}}(N) = nA\left(1-h(\tilde{e}/A)\right)-nf_{\mathrm{EC}}h(e) - \Delta(n).
	\label{eq:S_finite} 
\end{equation}
First, one has to take into account that not all of the sifted key $S_{\mathrm{sift}}$ can be used for the secret key but a fraction of it has to be discarded for parameter estimation. Out of the sifted events a number of $m$ is taken for parameter estimation and the remaining $n$ bits are used for the key generation. Taking only a finite sample $m$ might lead to deviations between the estimated and the actual error rate. This is taken into consideration by
\begin{equation}
	\tilde{e} = e+\sqrt{(\ln{(1/\epsilon_{\mathrm{PE}})}+2\ln{(m+1)})/2m}
\end{equation}
where $\epsilon_{\mathrm{PE}}$ is the probability for such deviations to occur. Additional finite corrections are subsumed as $\Delta(n)$
\begin{equation}
	\Delta(n) = 7n\sqrt{\frac{1}{n}\log_2{\frac{2}{\tilde{\epsilon}}}}+2\log_2{\frac{1}{\epsilon_{\mathrm{PA}}}}+\log_2{\frac{2}{\epsilon_{\mathrm{EC}}}},
\end{equation}
with the overall security parameter $\epsilon = \epsilon_{\mathrm{EC}}+\tilde{\epsilon}+\epsilon_{\mathrm{PE}}+\epsilon_{\mathrm{PA}}$ including the failure probabilities for the error correction $\epsilon_{\mathrm{EC}}$, the entropy estimation $\tilde{\epsilon}$, the parameter estimation $\epsilon_{\mathrm{PE}}$, and the privacy amplification $\epsilon_{\mathrm{PA}}$ \cite{Cai_2009}.

The rate-loss diagram in Fig.~\ref{fig:figure4}~(a) comparatively illustrates the asymptotic and finite-size secret key rates $S$ calculated via Eq.~(\ref{eq:S_infty}) and Eq.~(\ref{eq:S_finite}), respectively, considering our experimentally determined parameters. Additionally, we consider each case with and without applying optimized 2D temporal filters. To this end we maximize $S$ for each loss individually, by choosing optimal parameter sets as exemplary depicted in Fig.~\ref{fig:figure3} (see Supplementary Material, Fig.~\ref{fig:parameter_opt} for the optimized parameter sets). This contrasts with Ref.~\cite{Kupko2020}, where we evaluated rate-loss dependencies for exemplary settings of the acceptance window. For a qualitative understanding, each curve can be divided into two regimes: the absorption limited regime at small channel loss and the noise limited regime at large channel loss. At small channel loss, $S$ follows a linear trend (in logarithmic scaling) due to the absorption in the quantum channel. In the regime of larger losses, this trend changes to a multi-exponential drop of $S$ once the probability to register an event due to detector dark counts becomes comparable to the probability to detect a signal photon ($p_\text{dc}\approx p_\text{signal}$). Quantitatively, we extract a maximal tolerable loss for our test-system of \SI{13.96}{\decibel} in the non optimized asymptotic case. However, with 2D temporal filtering individually optimized for each loss, one does not only get a slightly higher secret key rate back-to-back but can additionally extend the tolerable losses up to \SI{23.19}{\decibel}. In a state-of-the-art single-mode fiber with losses of \SI{0.31}{\decibel}/km at \SI{1300}{\nano\meter} this corresponds to a range extension of almost \SI{30}{\kilo\meter} from \SI{45.0}{\kilo\meter} to \SI{74.8}{\kilo\meter}. Taking finite key size effects into account Alice has to send for a minimum transmission time of \SI{18}{seconds} to yield a positive $S$ in the back-to-back case. Increasing the transmission time by one order of magnitude to \SI{100}{seconds} would already allow for \SI{6.8}{\decibel} of loss in the optimized case. With unpractical transmission times of \SI{1e+6}{seconds} one could almost reach the asymptotic case. By comparing the two cases of rather short and long transmission times, one can see that the range extension gained by applying 2D temporal filtering increases with the transmission time. The implementation and discussion of finite-size effects presented above is an important step towards practical implementations of single-photon QKD, also compared to our previous work in Ref.~\cite{Kupko2020} discussing key rates only in the asymptotic limit.

\subsection{Comparison and Discussion}\label{sec:Compare}

Finally, we compare the extrapolated performance of our testbed with selected previous implementations of QKD. As mentioned already in the introduction, several QKD-experiments using QD-sources have been reported (as summarized in Ref.\cite{Vajner2021}). For our comparison, we select the QKD experiments reported in Ref.~\cite{Waks2002} and Ref.~\cite{Takemoto2015} employing QD-based SPSs operating at \SI{877}{nm} and \SI{1580.5}{nm}, respectively. This choice was made to cover, together with our source, the technically most relevant wavelength ranges ($1^\text{st}$, $2^\text{nd}$, and $3^\text{rd}$ telecom window) for fiber-based applications and due to the high secure key rates achieved in these experiments. We assume implementations in SM fiber with losses of \SI{0.31}{\decibel/\kilo\meter} in case of our work, \SI{0.194}{\decibel/\kilo\meter} for Ref.~\cite{Takemoto2015}, and \SI{3.5}{\decibel/\kilo\meter} for Ref.~\cite{Waks2002}, corresponding to the typical losses in optical fibers at the implementation's specified wavelengths. For the sake of comparison we calculated the key rates for the asymptotic case, representing the ultimate limit for each system. The parameters used for the comparison are summarized in Tab.~\ref{tab:Parameters}. Figure~\ref{fig:figure4}~(b) displays the resulting curves in a rate-loss diagram and the corresponding rate-distance diagram as inset. While the first proof-of-principle implementation by Waks~et~al. in 2002 achieved a higher $\mu = 0.007$ compared to our present stand-alone device, their wavelength is not ideal for fiber based implementations. Therefore, the achievable distance in optical fiber is strongly limited (cf. Figure~\ref{fig:figure4}~(b), inset), although their implementation reaches a maximal tolerable loss of \SI{21.6}{\decibel}. Among the three QKD experiments, our optimized case achieves the highest tolerable loss of \SI{23.19}{\decibel}, due to low dark count levels and low intrinsic QBERs caused by our receiver module. One should note, that the latter will increase slightly for a full implementations using our SPS, due to additional contributions to the QBER caused by dynamic state preparation via EOMs. To date, the QKD system reported by Takemoto et al. in Ref.~\cite{Takemoto2015} achieves the highest possible transmission distance exceeding \SI{100}{\kilo\meter}, due to the low fiber-loss at C-band wavelengths. Please note, that deviations in the achievable transmission distance compared to the original references are related to the use of slightly different rate equations or $p_{\mathrm{m}}$ and a different $f_{\mathrm{EC}}$. Not least, also the previous implementations cited here might benefit from 2D temporal filtering, which could allow for performance improvements. While the efficiency of our all-fiber-coupled benchtop SPS is so far not compatible with the state-of-the-art in terms of $\mu$ inside the quantum channel, we anticipate substantially enhanced performance in the future employing the improvements discussed in the conclusion below.

In terms of practicability and compactness, our QKD testbed, with a size of the transmitter comparable to commercial laser-based QKD systems, represents a milestone compared to previous QKD experiments using sub-Poissonian quantum light sources. In terms of the performance, our source and any sub-Poissonian QKD experiment reported to date is not yet competitive with WCP-based implementations. The current record for direct point-to-point QKD at C-band wavelength reported by Boaron et al. \cite{Boaron2018} achieved a tolerable loss of \SI{71.9}{\decibel} corresponding to a communication distance of \SI{421}{\kilo\meter} in ultra-low loss optical fiber. For single-photon QKD, significant improvements are required in both, the efficiency $\mu$ and $g^{(2)}(0)$ to become competitive or even surpass WCP-based implementations. In WCP-based implementations obeying Poisson statistics, there is a fundamental upper limit of $\mu = 0.37$ (cf. Ref.~\cite{Kupko2020} for details). For deterministic single-photon sources, $\mu$ is in principle only limited by technological challenges, e.g., photon-collection and fiber-coupling efficiencies. Recently a QD-SPS with an SM fiber-coupled efficiency of \SI{57}{\percent} and $g^{(2)}(0) = 0.021$ was reported at a wavelength of \SI{919}{\nano\meter} \cite{Tomm2021} and the authors claimed \SI{80}{\percent} efficiency within reach. Still, overcoming the limit of \SI{37}{\percent} inside the quantum channel, including losses for qubit-state preparation, is a challenge in practical implementation. While the efficiency of a SPS is important in the regime of low channel losses, low multi-photon probabilities~$p_\mathrm{m}$ become the limiting factor at high channel losses \cite{Waks2002}. Therefore, it is important to improve $g^{(2)}(0)$ as far as possible without reducing the efficiency $\mu$. In contrast to sources based on spontaneous parametric down conversion, no fundamental trade-off between low $g^{(2)}(0)$ values and high efficiencies exists for QD sources. Therefore, further improvements in device technology will enable to simultaneously achieve record low $g^{(2)}(0)$ values (c.f. $g^{(2)}(0) = \num{7.5(16)d-5}$ in Ref.~\cite{Schweickert2018}) and record efficiencies in future QD devices. In addition, as mentioned in section III.A, finding tighter bounds for estimating~$p_\mathrm{m}$ can result in a boost of the performance of sub-Poissonian QKD. Important to note, is that the discussion above refers to point-to-point QKD in a prepare-and-measure configuration. In recent entanglement-based QKD experiments using QD-based sources, it was shown that already much lower efficiencies than $0.37$ are sufficient to beat QKD implementations based on spontaneous parametric down conversion (cf. Ref.~\cite{Schimpf2021}, Supplemental Material).

\section{Summary \& Conclusions}\label{sec:Summary}

We demonstrated the first test experiments evaluating a compact and plug\&play benchtop solid state based SPS for applications in QKD. For this purpose, we developed the most compact QD-SPS operating in the telecom O-band reported to date. The source module housed in a \SI{19}{inch} rack unit comprises a fiber-pigtailed QD integrated into a Stirling cryocooler, and a fiber-based band-pass filter. On receiver side of the QKD testbed, we employed a 4-state polarization decoder optimized for O-band wavelenghts allowing for low intrinsic QBER. Using this testbed, we demonstrated that maximal tolerable losses of \SI{23.19}{\decibel} are achievable in full implementations with this system, if optimization routines are applied for a 2D temporal acceptance time window. We further investigated the impact of finite key size effects to the performance of our testbed and compared our results with previous implementations of QKD using QD-SPSs. 

The benchtop transmitter module developed in this work, with a size competitive to commercial laser-based QKD systems, represents a milestone compared to previous QD-based QKD experiments in terms of compactness and practicability, while offering a comparable performance level. In terms of efficiency and multi-photon emission probability, the performance is not yet sufficient to compete with state-of-the-art decoy-state QKD systems using WCPs. To boost the performance of sub-Poissonian QKD in the future, advanced fiber-coupled SPSs based on circular Bragg grating (CBG) cavities can be employed, which show prospects for photon collection efficiencies exceeding \SI{80}{\percent} in SM fiber-pigtailed configuration \cite{Rickert2019}. Moreover, exploiting microcavities with slight ellipticities, highly polarized SPSs can be realized \cite{Wang2019} to further reduce the losses inside the sender module. Another route for improvements concerns the implementation of GHz clock rates, e.g., by using electrically driven cavity-enhanced QD-SPSs \cite{Schlehahn2016}, which can be directly fiber-pigtailed using a technique recently developed in our group \cite{Rickert2021}. Possible future improvements concern the integration of lasers in the transmitter module, which are suitable for the coherent excitation of quantum emitters. Two routes appear promising in this context. The first option is the use of more compact pulsed lasers suitable for the coherent excitation of quantum emitters, which can be achieved by combining spectrally tunable continuous-wave lasers with intensity modulators \cite{Dada2016}. Here, also spectrally tunable fiber-coupled QD-devices \cite{Snijders2018} are interesting and provide an additional tuning knob. The second option, offering the highest degree of device integration, is the use of lasers integrated on-chip nearby the quantum-light emitting device itself \cite{Munnelly2017, Lee2017} - a concept that was previously tested in installed fiber networks \cite{Xiang2020}.
Not least, lower operating temperatures could be achieved by advances in the development of compact cryocoolers, enabling the generation of photons with high indistinguishability \cite{Thoma2016}, paving the way for implementations of quantum communication beyond direct point-to-point links \cite{Braunstein2012, Lo2012} using benchtop deterministic quantum light sources. Although this work focused on direct point-to-point QKD, the progress achieved, and prospects discussed in this work are important for practical implementations of advanced schemes of quantum information, e.g., quantum repeaters \cite{Briegel1998}, where quantum advantages against classical light states get even more apparent.

\section*{Supplementary Material}\label{sec:SM}
See the supplementary material in section for additional experimental data on the blinking behavior (Fig.~\ref{fig:blinking}) and the effect of temporal filtering on the signal fraction and QBER as recorded for all four detection channels (Fig.~\ref{fig:QBER_Frac}). In addition, we provide the optimal parameter sets for each individual temporal acceptance window applied for optimizing the rate-loss dependencies in Fig.~\ref{fig:figure4}~(a) (Fig.~\ref{fig:parameter_opt}).

\begin{acknowledgments}
We greatfully acknowledge financial support by the German Federal Ministry of Education and Research (BMBF) via project “QuSecure” (Grant No. 13N14876) within the funding program Photonic Research Germany and by the Einstein Foundation via the Einstein Research Unit “Quantum Devices”. The development of the QD-SPS has further been supported by the project “FI‐SEQUR” jointly financed by the European Regional Development Fund (EFRE) of the European Union in the framework of the programme to promote research, innovation, and technologies (Pro FIT) in Germany, and the National Centre for Research and Development in Poland within the 2nd Poland‐Berlin Photonics Programme (Grant No. 2/POLBER‐2/2016)). S. Rodt and S. Reitzenstein further acknowledge support by the German Science Foundation (DFG) via CRC 787 and S. Burger support by the BMBF via project “MODAL/NanoLab” (Grant No.: 05M20ZBM). Not least, support by the Polish National Agency for Academic Exchange is greatfully acknowledged.
\end{acknowledgments}

\section*{Author Declarations}

\subsection*{Author contributions}
TK designed and set up the QKD testbed with support by FU, ran the experiments together with LR, and evaluated the experimental data. LR assembled parts of the stand-alone single-photon source, integrated the fiber-pigtailed QD, and operated the source. JG, NS, and S Rodt grew and processed the QD sample under supervision of S Reitzenstein who conceived the QD-SPS development and fabrication. AM and KZ fiber-pigtailed the QD under supervision of WU and GS using the high-NA specialty fiber developed by PM and KD. KD connected the specialty and the standard SM-fiber. SB contributed numerically optimized devices designs for processing the QD sample. TK, LR, and TH wrote the manuscript with input from all authors. TK and LR contributed equally to this work. TH conceived the experiments presented in this work and supervised all efforts on the QKD testbed.

\subsection*{Conflict of interest}
The authors have no conflicts to disclose.


\begin{thebibliography}{77}%
\makeatletter
\providecommand \@ifxundefined [1]{%
 \@ifx{#1\undefined}
}%
\providecommand \@ifnum [1]{%
 \ifnum #1\expandafter \@firstoftwo
 \else \expandafter \@secondoftwo
 \fi
}%
\providecommand \@ifx [1]{%
 \ifx #1\expandafter \@firstoftwo
 \else \expandafter \@secondoftwo
 \fi
}%
\providecommand \natexlab [1]{#1}%
\providecommand \enquote  [1]{``#1''}%
\providecommand \bibnamefont  [1]{#1}%
\providecommand \bibfnamefont [1]{#1}%
\providecommand \citenamefont [1]{#1}%
\providecommand \href@noop [0]{\@secondoftwo}%
\providecommand \href [0]{\begingroup \@sanitize@url \@href}%
\providecommand \@href[1]{\@@startlink{#1}\@@href}%
\providecommand \@@href[1]{\endgroup#1\@@endlink}%
\providecommand \@sanitize@url [0]{\catcode `\\12\catcode `\$12\catcode
  `\&12\catcode `\#12\catcode `\^12\catcode `\_12\catcode `\%12\relax}%
\providecommand \@@startlink[1]{}%
\providecommand \@@endlink[0]{}%
\providecommand \url  [0]{\begingroup\@sanitize@url \@url }%
\providecommand \@url [1]{\endgroup\@href {#1}{\urlprefix }}%
\providecommand \urlprefix  [0]{URL }%
\providecommand \Eprint [0]{\href }%
\providecommand \doibase [0]{http://dx.doi.org/}%
\providecommand \selectlanguage [0]{\@gobble}%
\providecommand \bibinfo  [0]{\@secondoftwo}%
\providecommand \bibfield  [0]{\@secondoftwo}%
\providecommand \translation [1]{[#1]}%
\providecommand \BibitemOpen [0]{}%
\providecommand \bibitemStop [0]{}%
\providecommand \bibitemNoStop [0]{.\EOS\space}%
\providecommand \EOS [0]{\spacefactor3000\relax}%
\providecommand \BibitemShut  [1]{\csname bibitem#1\endcsname}%
\let\auto@bib@innerbib\@empty
\bibitem [{\citenamefont {Waks}, \citenamefont {Santori},\ and\ \citenamefont
  {Yamamoto}(2002)}]{Waks2002a}%
  \BibitemOpen
  \bibfield  {author} {\bibinfo {author} {\bibfnamefont {E.}~\bibnamefont
  {Waks}}, \bibinfo {author} {\bibfnamefont {C.}~\bibnamefont {Santori}}, \
  and\ \bibinfo {author} {\bibfnamefont {Y.}~\bibnamefont {Yamamoto}},\
  }\bibfield  {title} {\enquote {\bibinfo {title} {Security aspects of quantum
  key distribution with sub-poisson light},}\ }\href {\doibase
  10.1103/PhysRevA.66.042315} {\bibfield  {journal} {\bibinfo  {journal}
  {Physical Review A}\ }\textbf {\bibinfo {volume} {66}},\ \bibinfo {pages}
  {042315} (\bibinfo {year} {2002})}\BibitemShut {NoStop}%
\bibitem [{\citenamefont {Briegel}\ \emph {et~al.}(1998)\citenamefont
  {Briegel}, \citenamefont {Dür}, \citenamefont {Cirac},\ and\ \citenamefont
  {Zoller}}]{Briegel1998}%
  \BibitemOpen
  \bibfield  {author} {\bibinfo {author} {\bibfnamefont {H.-J.}\ \bibnamefont
  {Briegel}}, \bibinfo {author} {\bibfnamefont {W.}~\bibnamefont {Dür}},
  \bibinfo {author} {\bibfnamefont {J.~I.}\ \bibnamefont {Cirac}}, \ and\
  \bibinfo {author} {\bibfnamefont {P.}~\bibnamefont {Zoller}},\ }\bibfield
  {title} {\enquote {\bibinfo {title} {Quantum repeaters: The role of imperfect
  local operations in quantum communication},}\ }\href {\doibase
  10.1103/physrevlett.81.5932} {\bibfield  {journal} {\bibinfo  {journal}
  {Physical Review Letters}\ }\textbf {\bibinfo {volume} {81}},\ \bibinfo
  {pages} {5932--5935} (\bibinfo {year} {1998})}\BibitemShut {NoStop}%
\bibitem [{\citenamefont {Aharonovich}, \citenamefont {Englund},\ and\
  \citenamefont {Toth}(2016)}]{Aharonovich2016}%
  \BibitemOpen
  \bibfield  {author} {\bibinfo {author} {\bibfnamefont {I.}~\bibnamefont
  {Aharonovich}}, \bibinfo {author} {\bibfnamefont {D.}~\bibnamefont
  {Englund}}, \ and\ \bibinfo {author} {\bibfnamefont {M.}~\bibnamefont
  {Toth}},\ }\bibfield  {title} {\enquote {\bibinfo {title} {Solid-state
  single-photon emitters},}\ }\href {\doibase 10.1038/nphoton.2016.186}
  {\bibfield  {journal} {\bibinfo  {journal} {Nature Photonics}\ }\textbf
  {\bibinfo {volume} {10}},\ \bibinfo {pages} {631--641} (\bibinfo {year}
  {2016})}\BibitemShut {NoStop}%
\bibitem [{\citenamefont {Arakawa}\ and\ \citenamefont
  {Holmes}(2020)}]{Arakawa2020}%
  \BibitemOpen
  \bibfield  {author} {\bibinfo {author} {\bibfnamefont {Y.}~\bibnamefont
  {Arakawa}}\ and\ \bibinfo {author} {\bibfnamefont {M.~J.}\ \bibnamefont
  {Holmes}},\ }\bibfield  {title} {\enquote {\bibinfo {title} {Progress in
  quantum-dot single photon sources for quantum information technologies: A
  broad spectrum overview},}\ }\href {\doibase 10.1063/5.0010193} {\bibfield
  {journal} {\bibinfo  {journal} {Applied Physics Reviews}\ }\textbf {\bibinfo
  {volume} {7}},\ \bibinfo {pages} {021309} (\bibinfo {year}
  {2020})}\BibitemShut {NoStop}%
\bibitem [{\citenamefont {Zhang}\ \emph {et~al.}(2020)\citenamefont {Zhang},
  \citenamefont {Cheng}, \citenamefont {Chou},\ and\ \citenamefont
  {Gali}}]{Zhang2020}%
  \BibitemOpen
  \bibfield  {author} {\bibinfo {author} {\bibfnamefont {G.}~\bibnamefont
  {Zhang}}, \bibinfo {author} {\bibfnamefont {Y.}~\bibnamefont {Cheng}},
  \bibinfo {author} {\bibfnamefont {J.-P.}\ \bibnamefont {Chou}}, \ and\
  \bibinfo {author} {\bibfnamefont {A.}~\bibnamefont {Gali}},\ }\bibfield
  {title} {\enquote {\bibinfo {title} {Material platforms for defect qubits and
  single-photon emitters},}\ }\href {https://doi.org/10.1063/5.0006075}
  {\bibfield  {journal} {\bibinfo  {journal} {Applied Physics Reviews}\
  }\textbf {\bibinfo {volume} {7}},\ \bibinfo {pages} {031308} (\bibinfo {year}
  {2020})}\BibitemShut {NoStop}%
\bibitem [{\citenamefont {Miyazawa}\ \emph {et~al.}(2016)\citenamefont
  {Miyazawa}, \citenamefont {Takemoto}, \citenamefont {Nambu}, \citenamefont
  {Miki}, \citenamefont {Yamashita}, \citenamefont {Terai}, \citenamefont
  {Fujiwara}, \citenamefont {Sasaki}, \citenamefont {Sakuma}, \citenamefont
  {Takatsu}, \citenamefont {Yamamoto},\ and\ \citenamefont
  {Arakawa}}]{Miyazawa2016}%
  \BibitemOpen
  \bibfield  {author} {\bibinfo {author} {\bibfnamefont {T.}~\bibnamefont
  {Miyazawa}}, \bibinfo {author} {\bibfnamefont {K.}~\bibnamefont {Takemoto}},
  \bibinfo {author} {\bibfnamefont {Y.}~\bibnamefont {Nambu}}, \bibinfo
  {author} {\bibfnamefont {S.}~\bibnamefont {Miki}}, \bibinfo {author}
  {\bibfnamefont {T.}~\bibnamefont {Yamashita}}, \bibinfo {author}
  {\bibfnamefont {H.}~\bibnamefont {Terai}}, \bibinfo {author} {\bibfnamefont
  {M.}~\bibnamefont {Fujiwara}}, \bibinfo {author} {\bibfnamefont
  {M.}~\bibnamefont {Sasaki}}, \bibinfo {author} {\bibfnamefont
  {Y.}~\bibnamefont {Sakuma}}, \bibinfo {author} {\bibfnamefont
  {M.}~\bibnamefont {Takatsu}}, \bibinfo {author} {\bibfnamefont
  {T.}~\bibnamefont {Yamamoto}}, \ and\ \bibinfo {author} {\bibfnamefont
  {Y.}~\bibnamefont {Arakawa}},\ }\bibfield  {title} {\enquote {\bibinfo
  {title} {Single-photon emission at 1.5 micrometer from an {InAs}/{InP}
  quantum dot with highly suppressed multi-photon emission probabilities},}\
  }\href {\doibase 10.1063/1.4961888} {\bibfield  {journal} {\bibinfo
  {journal} {Applied Physics Letters}\ }\textbf {\bibinfo {volume} {109}},\
  \bibinfo {pages} {132106} (\bibinfo {year} {2016})}\BibitemShut {NoStop}%
\bibitem [{\citenamefont {Schweickert}\ \emph {et~al.}(2018)\citenamefont
  {Schweickert}, \citenamefont {Jöns}, \citenamefont {Zeuner}, \citenamefont
  {da~Silva}, \citenamefont {Huang}, \citenamefont {Lettner}, \citenamefont
  {Reindl}, \citenamefont {Zichi}, \citenamefont {Trotta}, \citenamefont
  {Rastelli},\ and\ \citenamefont {Zwiller}}]{Schweickert2018}%
  \BibitemOpen
  \bibfield  {author} {\bibinfo {author} {\bibfnamefont {L.}~\bibnamefont
  {Schweickert}}, \bibinfo {author} {\bibfnamefont {K.~D.}\ \bibnamefont
  {Jöns}}, \bibinfo {author} {\bibfnamefont {K.~D.}\ \bibnamefont {Zeuner}},
  \bibinfo {author} {\bibfnamefont {S.~F.~C.}\ \bibnamefont {da~Silva}},
  \bibinfo {author} {\bibfnamefont {H.}~\bibnamefont {Huang}}, \bibinfo
  {author} {\bibfnamefont {T.}~\bibnamefont {Lettner}}, \bibinfo {author}
  {\bibfnamefont {M.}~\bibnamefont {Reindl}}, \bibinfo {author} {\bibfnamefont
  {J.}~\bibnamefont {Zichi}}, \bibinfo {author} {\bibfnamefont
  {R.}~\bibnamefont {Trotta}}, \bibinfo {author} {\bibfnamefont
  {A.}~\bibnamefont {Rastelli}}, \ and\ \bibinfo {author} {\bibfnamefont
  {V.}~\bibnamefont {Zwiller}},\ }\bibfield  {title} {\enquote {\bibinfo
  {title} {On-demand generation of background-free single photons from a
  solid-state source},}\ }\href {\doibase 10.1063/1.5020038} {\bibfield
  {journal} {\bibinfo  {journal} {Applied Physics Letters}\ }\textbf {\bibinfo
  {volume} {112}},\ \bibinfo {pages} {093106} (\bibinfo {year}
  {2018})}\BibitemShut {NoStop}%
\bibitem [{\citenamefont {Liu}\ \emph {et~al.}(2019)\citenamefont {Liu},
  \citenamefont {Su}, \citenamefont {Wei}, \citenamefont {Yao}, \citenamefont
  {da~Silva}, \citenamefont {Yu}, \citenamefont {Iles-Smith}, \citenamefont
  {Srinivasan}, \citenamefont {Rastelli}, \citenamefont {Li},\ and\
  \citenamefont {Wang}}]{Liu2019}%
  \BibitemOpen
  \bibfield  {author} {\bibinfo {author} {\bibfnamefont {J.}~\bibnamefont
  {Liu}}, \bibinfo {author} {\bibfnamefont {R.}~\bibnamefont {Su}}, \bibinfo
  {author} {\bibfnamefont {Y.}~\bibnamefont {Wei}}, \bibinfo {author}
  {\bibfnamefont {B.}~\bibnamefont {Yao}}, \bibinfo {author} {\bibfnamefont
  {S.~F.~C.}\ \bibnamefont {da~Silva}}, \bibinfo {author} {\bibfnamefont
  {Y.}~\bibnamefont {Yu}}, \bibinfo {author} {\bibfnamefont {J.}~\bibnamefont
  {Iles-Smith}}, \bibinfo {author} {\bibfnamefont {K.}~\bibnamefont
  {Srinivasan}}, \bibinfo {author} {\bibfnamefont {A.}~\bibnamefont
  {Rastelli}}, \bibinfo {author} {\bibfnamefont {J.}~\bibnamefont {Li}}, \ and\
  \bibinfo {author} {\bibfnamefont {X.}~\bibnamefont {Wang}},\ }\bibfield
  {title} {\enquote {\bibinfo {title} {A solid-state source of strongly
  entangled photon pairs with high brightness and indistinguishability},}\
  }\href {https://doi.org/10.1038/s41565-019-0435-9} {\bibfield  {journal}
  {\bibinfo  {journal} {Nature Nanotechnology}\ }\textbf {\bibinfo {volume}
  {14}},\ \bibinfo {pages} {586--593} (\bibinfo {year} {2019})}\BibitemShut
  {NoStop}%
\bibitem [{\citenamefont {Wang}\ \emph {et~al.}(2019)\citenamefont {Wang},
  \citenamefont {Hu}, \citenamefont {Chung}, \citenamefont {Qin}, \citenamefont
  {Yang}, \citenamefont {Li}, \citenamefont {Liu}, \citenamefont {Zhong},
  \citenamefont {He}, \citenamefont {Ding}, \citenamefont {Deng}, \citenamefont
  {Dai}, \citenamefont {Huo}, \citenamefont {H\"{o}fling}, \citenamefont {Lu},\
  and\ \citenamefont {Pan}}]{Wang2019}%
  \BibitemOpen
  \bibfield  {author} {\bibinfo {author} {\bibfnamefont {H.}~\bibnamefont
  {Wang}}, \bibinfo {author} {\bibfnamefont {H.}~\bibnamefont {Hu}}, \bibinfo
  {author} {\bibfnamefont {T.-H.}\ \bibnamefont {Chung}}, \bibinfo {author}
  {\bibfnamefont {J.}~\bibnamefont {Qin}}, \bibinfo {author} {\bibfnamefont
  {X.}~\bibnamefont {Yang}}, \bibinfo {author} {\bibfnamefont {J.-P.}\
  \bibnamefont {Li}}, \bibinfo {author} {\bibfnamefont {R.-Z.}\ \bibnamefont
  {Liu}}, \bibinfo {author} {\bibfnamefont {H.-S.}\ \bibnamefont {Zhong}},
  \bibinfo {author} {\bibfnamefont {Y.-M.}\ \bibnamefont {He}}, \bibinfo
  {author} {\bibfnamefont {X.}~\bibnamefont {Ding}}, \bibinfo {author}
  {\bibfnamefont {Y.-H.}\ \bibnamefont {Deng}}, \bibinfo {author}
  {\bibfnamefont {Q.}~\bibnamefont {Dai}}, \bibinfo {author} {\bibfnamefont
  {Y.-H.}\ \bibnamefont {Huo}}, \bibinfo {author} {\bibfnamefont
  {S.}~\bibnamefont {H\"{o}fling}}, \bibinfo {author} {\bibfnamefont {C.-Y.}\
  \bibnamefont {Lu}}, \ and\ \bibinfo {author} {\bibfnamefont {J.-W.}\
  \bibnamefont {Pan}},\ }\bibfield  {title} {\enquote {\bibinfo {title}
  {On-demand semiconductor source of entangled photons which simultaneously has
  high fidelity, efficiency, and indistinguishability},}\ }\href
  {https://link.aps.org/doi/10.1103/PhysRevLett.122.113602} {\bibfield
  {journal} {\bibinfo  {journal} {Physical Review Letters}\ }\textbf {\bibinfo
  {volume} {122}} (\bibinfo {year} {2019})}\BibitemShut {NoStop}%
\bibitem [{\citenamefont {Tomm}\ \emph {et~al.}(2021)\citenamefont {Tomm},
  \citenamefont {Javadi}, \citenamefont {Antoniadis}, \citenamefont {Najer},
  \citenamefont {Löbl}, \citenamefont {Korsch}, \citenamefont {Schott},
  \citenamefont {Valentin}, \citenamefont {Wieck}, \citenamefont {Ludwig},\
  and\ \citenamefont {Warburton}}]{Tomm2021}%
  \BibitemOpen
  \bibfield  {author} {\bibinfo {author} {\bibfnamefont {N.}~\bibnamefont
  {Tomm}}, \bibinfo {author} {\bibfnamefont {A.}~\bibnamefont {Javadi}},
  \bibinfo {author} {\bibfnamefont {N.~O.}\ \bibnamefont {Antoniadis}},
  \bibinfo {author} {\bibfnamefont {D.}~\bibnamefont {Najer}}, \bibinfo
  {author} {\bibfnamefont {M.~C.}\ \bibnamefont {Löbl}}, \bibinfo {author}
  {\bibfnamefont {A.~R.}\ \bibnamefont {Korsch}}, \bibinfo {author}
  {\bibfnamefont {R.}~\bibnamefont {Schott}}, \bibinfo {author} {\bibfnamefont
  {S.~R.}\ \bibnamefont {Valentin}}, \bibinfo {author} {\bibfnamefont {A.~D.}\
  \bibnamefont {Wieck}}, \bibinfo {author} {\bibfnamefont {A.}~\bibnamefont
  {Ludwig}}, \ and\ \bibinfo {author} {\bibfnamefont {R.~J.}\ \bibnamefont
  {Warburton}},\ }\bibfield  {title} {\enquote {\bibinfo {title} {A bright and
  fast source of coherent single photons},}\ }\href {\doibase
  10.1038/s41565-020-00831-x} {\bibfield  {journal} {\bibinfo  {journal}
  {Nature Nanotechnology}\ }\textbf {\bibinfo {volume} {16}},\ \bibinfo {pages}
  {399--403} (\bibinfo {year} {2021})}\BibitemShut {NoStop}%
\bibitem [{\citenamefont {Schlehahn}\ \emph {et~al.}(2016)\citenamefont
  {Schlehahn}, \citenamefont {Thoma}, \citenamefont {Munnelly}, \citenamefont
  {Kamp}, \citenamefont {H\"{o}fling}, \citenamefont {Heindel}, \citenamefont
  {Schneider},\ and\ \citenamefont {Reitzenstein}}]{Schlehahn2016}%
  \BibitemOpen
  \bibfield  {author} {\bibinfo {author} {\bibfnamefont {A.}~\bibnamefont
  {Schlehahn}}, \bibinfo {author} {\bibfnamefont {A.}~\bibnamefont {Thoma}},
  \bibinfo {author} {\bibfnamefont {P.}~\bibnamefont {Munnelly}}, \bibinfo
  {author} {\bibfnamefont {M.}~\bibnamefont {Kamp}}, \bibinfo {author}
  {\bibfnamefont {S.}~\bibnamefont {H\"{o}fling}}, \bibinfo {author}
  {\bibfnamefont {T.}~\bibnamefont {Heindel}}, \bibinfo {author} {\bibfnamefont
  {C.}~\bibnamefont {Schneider}}, \ and\ \bibinfo {author} {\bibfnamefont
  {S.}~\bibnamefont {Reitzenstein}},\ }\bibfield  {title} {\enquote {\bibinfo
  {title} {An electrically driven cavity-enhanced source of indistinguishable
  photons with 61\% overall efficiency},}\ }\href
  {https://doi.org/10.1063/1.4939831} {\bibfield  {journal} {\bibinfo
  {journal} {APL Photonics}\ }\textbf {\bibinfo {volume} {1}} (\bibinfo {year}
  {2016})}\BibitemShut {NoStop}%
\bibitem [{\citenamefont {Shooter}\ \emph {et~al.}(2020)\citenamefont
  {Shooter}, \citenamefont {Xiang}, \citenamefont {Müller}, \citenamefont
  {Skiba-Szymanska}, \citenamefont {Huwer}, \citenamefont {Griffiths},
  \citenamefont {Mitchell}, \citenamefont {Anderson}, \citenamefont {Müller},
  \citenamefont {Krysa}, \citenamefont {Stevenson}, \citenamefont {Heffernan},
  \citenamefont {Ritchie},\ and\ \citenamefont {Shields}}]{Shooter2020}%
  \BibitemOpen
  \bibfield  {author} {\bibinfo {author} {\bibfnamefont {G.}~\bibnamefont
  {Shooter}}, \bibinfo {author} {\bibfnamefont {Z.-H.}\ \bibnamefont {Xiang}},
  \bibinfo {author} {\bibfnamefont {J.~R.~A.}\ \bibnamefont {Müller}},
  \bibinfo {author} {\bibfnamefont {J.}~\bibnamefont {Skiba-Szymanska}},
  \bibinfo {author} {\bibfnamefont {J.}~\bibnamefont {Huwer}}, \bibinfo
  {author} {\bibfnamefont {J.}~\bibnamefont {Griffiths}}, \bibinfo {author}
  {\bibfnamefont {T.}~\bibnamefont {Mitchell}}, \bibinfo {author}
  {\bibfnamefont {M.}~\bibnamefont {Anderson}}, \bibinfo {author}
  {\bibfnamefont {T.}~\bibnamefont {Müller}}, \bibinfo {author} {\bibfnamefont
  {A.~B.}\ \bibnamefont {Krysa}}, \bibinfo {author} {\bibfnamefont {R.~M.}\
  \bibnamefont {Stevenson}}, \bibinfo {author} {\bibfnamefont {J.}~\bibnamefont
  {Heffernan}}, \bibinfo {author} {\bibfnamefont {D.~A.}\ \bibnamefont
  {Ritchie}}, \ and\ \bibinfo {author} {\bibfnamefont {A.~J.}\ \bibnamefont
  {Shields}},\ }\bibfield  {title} {\enquote {\bibinfo {title} {1\text{GHz}
  clocked distribution of electrically generated entangled photon pairs},}\
  }\href {\doibase 10.1364/oe.405466} {\bibfield  {journal} {\bibinfo
  {journal} {Optics Express}\ }\textbf {\bibinfo {volume} {28}},\ \bibinfo
  {pages} {36838} (\bibinfo {year} {2020})}\BibitemShut {NoStop}%
\bibitem [{\citenamefont {Bennett}\ \emph {et~al.}(1992)\citenamefont
  {Bennett}, \citenamefont {Bessette}, \citenamefont {Brassard}, \citenamefont
  {Salvail},\ and\ \citenamefont {Smolin}}]{Bennett1992}%
  \BibitemOpen
  \bibfield  {author} {\bibinfo {author} {\bibfnamefont {C.~H.}\ \bibnamefont
  {Bennett}}, \bibinfo {author} {\bibfnamefont {F.}~\bibnamefont {Bessette}},
  \bibinfo {author} {\bibfnamefont {G.}~\bibnamefont {Brassard}}, \bibinfo
  {author} {\bibfnamefont {L.}~\bibnamefont {Salvail}}, \ and\ \bibinfo
  {author} {\bibfnamefont {J.}~\bibnamefont {Smolin}},\ }\bibfield  {title}
  {\enquote {\bibinfo {title} {Experimental quantum cryptography},}\ }\href
  {\doibase 10.1007/BF00191318} {\bibfield  {journal} {\bibinfo  {journal}
  {Journal of Cryptology}\ }\textbf {\bibinfo {volume} {5}},\ \bibinfo {pages}
  {3--28} (\bibinfo {year} {1992})}\BibitemShut {NoStop}%
\bibitem [{\citenamefont {Wang}(2005)}]{Wang2005}%
  \BibitemOpen
  \bibfield  {author} {\bibinfo {author} {\bibfnamefont {X.-B.}\ \bibnamefont
  {Wang}},\ }\bibfield  {title} {\enquote {\bibinfo {title} {Beating the
  photon-number-splitting attack in practical quantum cryptography},}\ }\href
  {\doibase 10.1103/PhysRevLett.94.230503} {\bibfield  {journal} {\bibinfo
  {journal} {Physical Review Letters}\ }\textbf {\bibinfo {volume} {94}},\
  \bibinfo {pages} {230503} (\bibinfo {year} {2005})}\BibitemShut {NoStop}%
\bibitem [{\citenamefont {Lo}, \citenamefont {Ma},\ and\ \citenamefont
  {Chen}(2005)}]{Lo2005}%
  \BibitemOpen
  \bibfield  {author} {\bibinfo {author} {\bibfnamefont {H.-K.}\ \bibnamefont
  {Lo}}, \bibinfo {author} {\bibfnamefont {X.}~\bibnamefont {Ma}}, \ and\
  \bibinfo {author} {\bibfnamefont {K.}~\bibnamefont {Chen}},\ }\bibfield
  {title} {\enquote {\bibinfo {title} {Decoy state quantum key distribution},}\
  }\href {\doibase 10.1103/PhysRevLett.94.230504} {\bibfield  {journal}
  {\bibinfo  {journal} {Physical Review Letters}\ }\textbf {\bibinfo {volume}
  {94}},\ \bibinfo {pages} {230504} (\bibinfo {year} {2005})}\BibitemShut
  {NoStop}%
\bibitem [{\citenamefont {Boaron}\ \emph {et~al.}(2018)\citenamefont {Boaron},
  \citenamefont {Boso}, \citenamefont {Rusca}, \citenamefont {Vulliez},
  \citenamefont {Autebert}, \citenamefont {Caloz}, \citenamefont {Perrenoud},
  \citenamefont {Gras}, \citenamefont {Bussi\`eres}, \citenamefont {Li},
  \citenamefont {Nolan}, \citenamefont {Martin},\ and\ \citenamefont
  {Zbinden}}]{Boaron2018}%
  \BibitemOpen
  \bibfield  {author} {\bibinfo {author} {\bibfnamefont {A.}~\bibnamefont
  {Boaron}}, \bibinfo {author} {\bibfnamefont {G.}~\bibnamefont {Boso}},
  \bibinfo {author} {\bibfnamefont {D.}~\bibnamefont {Rusca}}, \bibinfo
  {author} {\bibfnamefont {C.}~\bibnamefont {Vulliez}}, \bibinfo {author}
  {\bibfnamefont {C.}~\bibnamefont {Autebert}}, \bibinfo {author}
  {\bibfnamefont {M.}~\bibnamefont {Caloz}}, \bibinfo {author} {\bibfnamefont
  {M.}~\bibnamefont {Perrenoud}}, \bibinfo {author} {\bibfnamefont
  {G.}~\bibnamefont {Gras}}, \bibinfo {author} {\bibfnamefont {F.}~\bibnamefont
  {Bussi\`eres}}, \bibinfo {author} {\bibfnamefont {M.-J.}\ \bibnamefont {Li}},
  \bibinfo {author} {\bibfnamefont {D.}~\bibnamefont {Nolan}}, \bibinfo
  {author} {\bibfnamefont {A.}~\bibnamefont {Martin}}, \ and\ \bibinfo {author}
  {\bibfnamefont {H.}~\bibnamefont {Zbinden}},\ }\bibfield  {title} {\enquote
  {\bibinfo {title} {Secure quantum key distribution over 421 km of optical
  fiber},}\ }\href {\doibase 10.1103/PhysRevLett.121.190502} {\bibfield
  {journal} {\bibinfo  {journal} {Phys. Rev. Lett.}\ }\textbf {\bibinfo
  {volume} {121}},\ \bibinfo {pages} {190502} (\bibinfo {year}
  {2018})}\BibitemShut {NoStop}%
\bibitem [{\citenamefont {Badolato}\ \emph {et~al.}(2005)\citenamefont
  {Badolato}, \citenamefont {Hennessy}, \citenamefont {Atat\"ure},
  \citenamefont {Dreiser}, \citenamefont {Hu}, \citenamefont {Petroff},\ and\
  \citenamefont {Imamo\u{g}lu}}]{Badolato2005}%
  \BibitemOpen
  \bibfield  {author} {\bibinfo {author} {\bibfnamefont {A.}~\bibnamefont
  {Badolato}}, \bibinfo {author} {\bibfnamefont {K.}~\bibnamefont {Hennessy}},
  \bibinfo {author} {\bibfnamefont {M.}~\bibnamefont {Atat\"ure}}, \bibinfo
  {author} {\bibfnamefont {J.}~\bibnamefont {Dreiser}}, \bibinfo {author}
  {\bibfnamefont {E.}~\bibnamefont {Hu}}, \bibinfo {author} {\bibfnamefont
  {P.~M.}\ \bibnamefont {Petroff}}, \ and\ \bibinfo {author} {\bibfnamefont
  {A.}~\bibnamefont {Imamo\u{g}lu}},\ }\bibfield  {title} {\enquote {\bibinfo
  {title} {Deterministic coupling of single quantum dots to single nanocavity
  modes},}\ }\href {https://www.science.org/doi/abs/10.1126/science.1109815}
  {\bibfield  {journal} {\bibinfo  {journal} {Science}\ }\textbf {\bibinfo
  {volume} {308}},\ \bibinfo {pages} {1158--1161} (\bibinfo {year}
  {2005})}\BibitemShut {NoStop}%
\bibitem [{\citenamefont {Dousse}\ \emph {et~al.}(2008)\citenamefont {Dousse},
  \citenamefont {Lanco}, \citenamefont {Suffczy\ifmmode~\acute{n}\else
  \'{n}\fi{}ski}, \citenamefont {Semenova}, \citenamefont {Miard},
  \citenamefont {Lema\^{\i}tre}, \citenamefont {Sagnes}, \citenamefont
  {Roblin}, \citenamefont {Bloch},\ and\ \citenamefont
  {Senellart}}]{Dousse2008}%
  \BibitemOpen
  \bibfield  {author} {\bibinfo {author} {\bibfnamefont {A.}~\bibnamefont
  {Dousse}}, \bibinfo {author} {\bibfnamefont {L.}~\bibnamefont {Lanco}},
  \bibinfo {author} {\bibfnamefont {J.}~\bibnamefont
  {Suffczy\ifmmode~\acute{n}\else \'{n}\fi{}ski}}, \bibinfo {author}
  {\bibfnamefont {E.}~\bibnamefont {Semenova}}, \bibinfo {author}
  {\bibfnamefont {A.}~\bibnamefont {Miard}}, \bibinfo {author} {\bibfnamefont
  {A.}~\bibnamefont {Lema\^{\i}tre}}, \bibinfo {author} {\bibfnamefont
  {I.}~\bibnamefont {Sagnes}}, \bibinfo {author} {\bibfnamefont
  {C.}~\bibnamefont {Roblin}}, \bibinfo {author} {\bibfnamefont
  {J.}~\bibnamefont {Bloch}}, \ and\ \bibinfo {author} {\bibfnamefont
  {P.}~\bibnamefont {Senellart}},\ }\bibfield  {title} {\enquote {\bibinfo
  {title} {Controlled light-matter coupling for a single quantum dot embedded
  in a pillar microcavity using far-field optical lithography},}\ }\href
  {\doibase 10.1103/PhysRevLett.101.267404} {\bibfield  {journal} {\bibinfo
  {journal} {Physical Review Letters.}\ }\textbf {\bibinfo {volume} {101}},\
  \bibinfo {pages} {267404} (\bibinfo {year} {2008})}\BibitemShut {NoStop}%
\bibitem [{\citenamefont {Gschrey}\ \emph {et~al.}(2013)\citenamefont
  {Gschrey}, \citenamefont {Gericke}, \citenamefont {Schüßler}, \citenamefont
  {Schmidt}, \citenamefont {Schulze}, \citenamefont {Heindel}, \citenamefont
  {Rodt}, \citenamefont {Strittmatter},\ and\ \citenamefont
  {Reitzenstein}}]{Gschrey2013}%
  \BibitemOpen
  \bibfield  {author} {\bibinfo {author} {\bibfnamefont {M.}~\bibnamefont
  {Gschrey}}, \bibinfo {author} {\bibfnamefont {F.}~\bibnamefont {Gericke}},
  \bibinfo {author} {\bibfnamefont {A.}~\bibnamefont {Schüßler}}, \bibinfo
  {author} {\bibfnamefont {R.}~\bibnamefont {Schmidt}}, \bibinfo {author}
  {\bibfnamefont {J.-H.}\ \bibnamefont {Schulze}}, \bibinfo {author}
  {\bibfnamefont {T.}~\bibnamefont {Heindel}}, \bibinfo {author} {\bibfnamefont
  {S.}~\bibnamefont {Rodt}}, \bibinfo {author} {\bibfnamefont {A.}~\bibnamefont
  {Strittmatter}}, \ and\ \bibinfo {author} {\bibfnamefont {S.}~\bibnamefont
  {Reitzenstein}},\ }\bibfield  {title} {\enquote {\bibinfo {title} {In situ
  electron-beam lithography of deterministic single-quantum-dot mesa-structures
  using low-temperature cathodoluminescence spectroscopy},}\ }\href
  {https://doi.org/10.1063/1.4812343} {\bibfield  {journal} {\bibinfo
  {journal} {Applied Physics Letters}\ }\textbf {\bibinfo {volume} {102}},\
  \bibinfo {pages} {251113} (\bibinfo {year} {2013})}\BibitemShut {NoStop}%
\bibitem [{\citenamefont {Gschrey}\ \emph {et~al.}(2015)\citenamefont
  {Gschrey}, \citenamefont {Thoma}, \citenamefont {Schnauber}, \citenamefont
  {Seifried}, \citenamefont {Schmidt}, \citenamefont {Wohlfeil}, \citenamefont
  {Kr\"{u}ger}, \citenamefont {Schulze}, \citenamefont {Heindel}, \citenamefont
  {Burger}, \citenamefont {Schmidt}, \citenamefont {Strittmatter},
  \citenamefont {Rodt},\ and\ \citenamefont {Reitzenstein}}]{Gschrey2015}%
  \BibitemOpen
  \bibfield  {author} {\bibinfo {author} {\bibfnamefont {M.}~\bibnamefont
  {Gschrey}}, \bibinfo {author} {\bibfnamefont {A.}~\bibnamefont {Thoma}},
  \bibinfo {author} {\bibfnamefont {P.}~\bibnamefont {Schnauber}}, \bibinfo
  {author} {\bibfnamefont {M.}~\bibnamefont {Seifried}}, \bibinfo {author}
  {\bibfnamefont {R.}~\bibnamefont {Schmidt}}, \bibinfo {author} {\bibfnamefont
  {B.}~\bibnamefont {Wohlfeil}}, \bibinfo {author} {\bibfnamefont
  {L.}~\bibnamefont {Kr\"{u}ger}}, \bibinfo {author} {\bibfnamefont {J.~H.}\
  \bibnamefont {Schulze}}, \bibinfo {author} {\bibfnamefont {T.}~\bibnamefont
  {Heindel}}, \bibinfo {author} {\bibfnamefont {S.}~\bibnamefont {Burger}},
  \bibinfo {author} {\bibfnamefont {F.}~\bibnamefont {Schmidt}}, \bibinfo
  {author} {\bibfnamefont {A.}~\bibnamefont {Strittmatter}}, \bibinfo {author}
  {\bibfnamefont {S.}~\bibnamefont {Rodt}}, \ and\ \bibinfo {author}
  {\bibfnamefont {S.}~\bibnamefont {Reitzenstein}},\ }\bibfield  {title}
  {\enquote {\bibinfo {title} {Highly indistinguishable photons from
  deterministic quantum-dot microlenses utilizing three-dimensional in situ
  electron-beam lithography},}\ }\href {\doibase 10.1038/ncomms8662} {\bibfield
   {journal} {\bibinfo  {journal} {Nature Communications}\ }\textbf {\bibinfo
  {volume} {6}},\ \bibinfo {pages} {7662} (\bibinfo {year} {2015})}\BibitemShut
  {NoStop}%
\bibitem [{\citenamefont {Somaschi}\ \emph {et~al.}(2016)\citenamefont
  {Somaschi}, \citenamefont {Giesz}, \citenamefont {De~Santis}, \citenamefont
  {Loredo}, \citenamefont {Almeida}, \citenamefont {Hornecker}, \citenamefont
  {Portalupi}, \citenamefont {Grange}, \citenamefont {Ant\'{o}n}, \citenamefont
  {Demory}, \citenamefont {G\'{o}mez}, \citenamefont {Sagnes}, \citenamefont
  {Lanzillotti-Kimura}, \citenamefont {Lema\'{i}tre}, \citenamefont {Auffeves},
  \citenamefont {White}, \citenamefont {Lanco},\ and\ \citenamefont
  {Senellart}}]{Somaschi2016}%
  \BibitemOpen
  \bibfield  {author} {\bibinfo {author} {\bibfnamefont {N.}~\bibnamefont
  {Somaschi}}, \bibinfo {author} {\bibfnamefont {V.}~\bibnamefont {Giesz}},
  \bibinfo {author} {\bibfnamefont {L.}~\bibnamefont {De~Santis}}, \bibinfo
  {author} {\bibfnamefont {J.~C.}\ \bibnamefont {Loredo}}, \bibinfo {author}
  {\bibfnamefont {M.~P.}\ \bibnamefont {Almeida}}, \bibinfo {author}
  {\bibfnamefont {G.}~\bibnamefont {Hornecker}}, \bibinfo {author}
  {\bibfnamefont {S.~L.}\ \bibnamefont {Portalupi}}, \bibinfo {author}
  {\bibfnamefont {T.}~\bibnamefont {Grange}}, \bibinfo {author} {\bibfnamefont
  {C.}~\bibnamefont {Ant\'{o}n}}, \bibinfo {author} {\bibfnamefont
  {J.}~\bibnamefont {Demory}}, \bibinfo {author} {\bibfnamefont
  {C.}~\bibnamefont {G\'{o}mez}}, \bibinfo {author} {\bibfnamefont
  {I.}~\bibnamefont {Sagnes}}, \bibinfo {author} {\bibfnamefont {N.~D.}\
  \bibnamefont {Lanzillotti-Kimura}}, \bibinfo {author} {\bibfnamefont
  {A.}~\bibnamefont {Lema\'{i}tre}}, \bibinfo {author} {\bibfnamefont
  {A.}~\bibnamefont {Auffeves}}, \bibinfo {author} {\bibfnamefont {A.~G.}\
  \bibnamefont {White}}, \bibinfo {author} {\bibfnamefont {L.}~\bibnamefont
  {Lanco}}, \ and\ \bibinfo {author} {\bibfnamefont {P.}~\bibnamefont
  {Senellart}},\ }\bibfield  {title} {\enquote {\bibinfo {title} {Near-optimal
  single-photon sources in the solid state},}\ }\href {\doibase
  10.1038/nphoton.2016.23} {\bibfield  {journal} {\bibinfo  {journal} {Nat.
  Photon.}\ }\textbf {\bibinfo {volume} {10}},\ \bibinfo {pages} {340–345}
  (\bibinfo {year} {2016})}\BibitemShut {NoStop}%
\bibitem [{\citenamefont {Heindel}\ \emph {et~al.}(2017)\citenamefont
  {Heindel}, \citenamefont {Thoma}, \citenamefont {von Helversen},
  \citenamefont {Schmidt}, \citenamefont {Schlehahn}, \citenamefont {Gschrey},
  \citenamefont {Schnauber}, \citenamefont {Schulze}, \citenamefont
  {Strittmatter}, \citenamefont {Beyer}, \citenamefont {Rodt}, \citenamefont
  {Carmele}, \citenamefont {Knorr},\ and\ \citenamefont
  {Reitzenstein}}]{Heindel2017}%
  \BibitemOpen
  \bibfield  {author} {\bibinfo {author} {\bibfnamefont {T.}~\bibnamefont
  {Heindel}}, \bibinfo {author} {\bibfnamefont {A.}~\bibnamefont {Thoma}},
  \bibinfo {author} {\bibfnamefont {M.}~\bibnamefont {von Helversen}}, \bibinfo
  {author} {\bibfnamefont {M.}~\bibnamefont {Schmidt}}, \bibinfo {author}
  {\bibfnamefont {A.}~\bibnamefont {Schlehahn}}, \bibinfo {author}
  {\bibfnamefont {M.}~\bibnamefont {Gschrey}}, \bibinfo {author} {\bibfnamefont
  {P.}~\bibnamefont {Schnauber}}, \bibinfo {author} {\bibfnamefont {J.~H.}\
  \bibnamefont {Schulze}}, \bibinfo {author} {\bibfnamefont {A.}~\bibnamefont
  {Strittmatter}}, \bibinfo {author} {\bibfnamefont {J.}~\bibnamefont {Beyer}},
  \bibinfo {author} {\bibfnamefont {S.}~\bibnamefont {Rodt}}, \bibinfo {author}
  {\bibfnamefont {A.}~\bibnamefont {Carmele}}, \bibinfo {author} {\bibfnamefont
  {A.}~\bibnamefont {Knorr}}, \ and\ \bibinfo {author} {\bibfnamefont
  {S.}~\bibnamefont {Reitzenstein}},\ }\bibfield  {title} {\enquote {\bibinfo
  {title} {A bright triggered twin-photon source in the solid state},}\
  }\href@noop {} {\bibfield  {journal} {\bibinfo  {journal} {Nature
  Communications}\ }\textbf {\bibinfo {volume} {8}},\ \bibinfo {pages} {14870}
  (\bibinfo {year} {2017})}\BibitemShut {NoStop}%
\bibitem [{\citenamefont {Rodt}, \citenamefont {Reitzenstein},\ and\
  \citenamefont {Heindel}(2020)}]{Rodt2020}%
  \BibitemOpen
  \bibfield  {author} {\bibinfo {author} {\bibfnamefont {S.}~\bibnamefont
  {Rodt}}, \bibinfo {author} {\bibfnamefont {S.}~\bibnamefont {Reitzenstein}},
  \ and\ \bibinfo {author} {\bibfnamefont {T.}~\bibnamefont {Heindel}},\
  }\bibfield  {title} {\enquote {\bibinfo {title} {Deterministically fabricated
  solid-state quantum-light sources},}\ }\href {\doibase
  10.1088/1361-648x/ab5e15} {\bibfield  {journal} {\bibinfo  {journal} {Journal
  of Physics: Condensed Matter}\ }\textbf {\bibinfo {volume} {32}},\ \bibinfo
  {pages} {153003} (\bibinfo {year} {2020})}\BibitemShut {NoStop}%
\bibitem [{\citenamefont {Waks}\ \emph {et~al.}(2002)\citenamefont {Waks},
  \citenamefont {Inoue}, \citenamefont {Santori}, \citenamefont {Fattal},
  \citenamefont {Vuckovic}, \citenamefont {Solomon},\ and\ \citenamefont
  {Yamamoto}}]{Waks2002}%
  \BibitemOpen
  \bibfield  {author} {\bibinfo {author} {\bibfnamefont {E.}~\bibnamefont
  {Waks}}, \bibinfo {author} {\bibfnamefont {K.}~\bibnamefont {Inoue}},
  \bibinfo {author} {\bibfnamefont {C.}~\bibnamefont {Santori}}, \bibinfo
  {author} {\bibfnamefont {D.}~\bibnamefont {Fattal}}, \bibinfo {author}
  {\bibfnamefont {J.}~\bibnamefont {Vuckovic}}, \bibinfo {author}
  {\bibfnamefont {G.~S.}\ \bibnamefont {Solomon}}, \ and\ \bibinfo {author}
  {\bibfnamefont {Y.}~\bibnamefont {Yamamoto}},\ }\bibfield  {title} {\enquote
  {\bibinfo {title} {Secure communication: Quantum cryptography with a photon
  turnstile},}\ }\href {\doibase 10.1038/420762a} {\bibfield  {journal}
  {\bibinfo  {journal} {Nature}\ }\textbf {\bibinfo {volume} {420}},\ \bibinfo
  {pages} {762} (\bibinfo {year} {2002})}\BibitemShut {NoStop}%
\bibitem [{\citenamefont {Intallura}\ \emph {et~al.}(2009)\citenamefont
  {Intallura}, \citenamefont {Ward}, \citenamefont {Karimov}, \citenamefont
  {Yuan}, \citenamefont {See}, \citenamefont {Atkinson}, \citenamefont
  {Ritchie},\ and\ \citenamefont {Shields}}]{Intallura2009}%
  \BibitemOpen
  \bibfield  {author} {\bibinfo {author} {\bibfnamefont {P.~M.}\ \bibnamefont
  {Intallura}}, \bibinfo {author} {\bibfnamefont {M.~B.}\ \bibnamefont {Ward}},
  \bibinfo {author} {\bibfnamefont {O.~Z.}\ \bibnamefont {Karimov}}, \bibinfo
  {author} {\bibfnamefont {Z.~L.}\ \bibnamefont {Yuan}}, \bibinfo {author}
  {\bibfnamefont {P.}~\bibnamefont {See}}, \bibinfo {author} {\bibfnamefont
  {P.}~\bibnamefont {Atkinson}}, \bibinfo {author} {\bibfnamefont {D.~A.}\
  \bibnamefont {Ritchie}}, \ and\ \bibinfo {author} {\bibfnamefont {A.~J.}\
  \bibnamefont {Shields}},\ }\bibfield  {title} {\enquote {\bibinfo {title}
  {Quantum communication using single photons from a semiconductor quantum dot
  emitting at a telecommunication wavelength},}\ }\href {\doibase
  10.1088/1464-4258/11/5/054005} {\bibfield  {journal} {\bibinfo  {journal}
  {Journal of Optics A: Pure and Applied Optics}\ }\textbf {\bibinfo {volume}
  {11}},\ \bibinfo {pages} {054005} (\bibinfo {year} {2009})}\BibitemShut
  {NoStop}%
\bibitem [{\citenamefont {Collins}\ \emph {et~al.}(2010)\citenamefont
  {Collins}, \citenamefont {Clarke}, \citenamefont {Fern\'{a}ndez},
  \citenamefont {Gordon}, \citenamefont {Makhonin}, \citenamefont {Timpson},
  \citenamefont {Tahraoui}, \citenamefont {Hopkinson}, \citenamefont {Fox},
  \citenamefont {Skolnick},\ and\ \citenamefont {Buller}}]{Collins2010}%
  \BibitemOpen
  \bibfield  {author} {\bibinfo {author} {\bibfnamefont {R.~J.}\ \bibnamefont
  {Collins}}, \bibinfo {author} {\bibfnamefont {P.~J.}\ \bibnamefont {Clarke}},
  \bibinfo {author} {\bibfnamefont {V.}~\bibnamefont {Fern\'{a}ndez}}, \bibinfo
  {author} {\bibfnamefont {K.~J.}\ \bibnamefont {Gordon}}, \bibinfo {author}
  {\bibfnamefont {M.~N.}\ \bibnamefont {Makhonin}}, \bibinfo {author}
  {\bibfnamefont {J.~A.}\ \bibnamefont {Timpson}}, \bibinfo {author}
  {\bibfnamefont {A.}~\bibnamefont {Tahraoui}}, \bibinfo {author}
  {\bibfnamefont {M.}~\bibnamefont {Hopkinson}}, \bibinfo {author}
  {\bibfnamefont {A.~M.}\ \bibnamefont {Fox}}, \bibinfo {author} {\bibfnamefont
  {M.~S.}\ \bibnamefont {Skolnick}}, \ and\ \bibinfo {author} {\bibfnamefont
  {G.~S.}\ \bibnamefont {Buller}},\ }\bibfield  {title} {\enquote {\bibinfo
  {title} {Quantum key distribution system in standard telecommunications fiber
  using a short wavelength single photon source},}\ }\href {\doibase
  10.1063/1.3327427} {\bibfield  {journal} {\bibinfo  {journal} {Journal of
  Applied Physics}\ }\textbf {\bibinfo {volume} {107}},\ \bibinfo {eid}
  {073102} (\bibinfo {year} {2010})}\BibitemShut {NoStop}%
\bibitem [{\citenamefont {Heindel}\ \emph {et~al.}(2012)\citenamefont
  {Heindel}, \citenamefont {Kessler}, \citenamefont {Rau}, \citenamefont
  {Schneider}, \citenamefont {F\"{u}rst}, \citenamefont {Hargart},
  \citenamefont {Schulz}, \citenamefont {Eichfelder}, \citenamefont {Roßbach},
  \citenamefont {Nauerth}, \citenamefont {Lermer}, \citenamefont {Weier},
  \citenamefont {Jetter}, \citenamefont {Kamp}, \citenamefont {Reitzenstein},
  \citenamefont {H\"{o}fling}, \citenamefont {Michler}, \citenamefont
  {Weinfurter},\ and\ \citenamefont {Forchel}}]{Heindel2012}%
  \BibitemOpen
  \bibfield  {author} {\bibinfo {author} {\bibfnamefont {T.}~\bibnamefont
  {Heindel}}, \bibinfo {author} {\bibfnamefont {C.~A.}\ \bibnamefont
  {Kessler}}, \bibinfo {author} {\bibfnamefont {M.}~\bibnamefont {Rau}},
  \bibinfo {author} {\bibfnamefont {C.}~\bibnamefont {Schneider}}, \bibinfo
  {author} {\bibfnamefont {M.}~\bibnamefont {F\"{u}rst}}, \bibinfo {author}
  {\bibfnamefont {F.}~\bibnamefont {Hargart}}, \bibinfo {author} {\bibfnamefont
  {W.-M.}\ \bibnamefont {Schulz}}, \bibinfo {author} {\bibfnamefont
  {M.}~\bibnamefont {Eichfelder}}, \bibinfo {author} {\bibfnamefont
  {R.}~\bibnamefont {Roßbach}}, \bibinfo {author} {\bibfnamefont
  {S.}~\bibnamefont {Nauerth}}, \bibinfo {author} {\bibfnamefont
  {M.}~\bibnamefont {Lermer}}, \bibinfo {author} {\bibfnamefont
  {H.}~\bibnamefont {Weier}}, \bibinfo {author} {\bibfnamefont
  {M.}~\bibnamefont {Jetter}}, \bibinfo {author} {\bibfnamefont
  {M.}~\bibnamefont {Kamp}}, \bibinfo {author} {\bibfnamefont {S.}~\bibnamefont
  {Reitzenstein}}, \bibinfo {author} {\bibfnamefont {S.}~\bibnamefont
  {H\"{o}fling}}, \bibinfo {author} {\bibfnamefont {P.}~\bibnamefont
  {Michler}}, \bibinfo {author} {\bibfnamefont {H.}~\bibnamefont {Weinfurter}},
  \ and\ \bibinfo {author} {\bibfnamefont {A.}~\bibnamefont {Forchel}},\
  }\bibfield  {title} {\enquote {\bibinfo {title} {Quantum key distribution
  using quantum dot single-photon emitting diodes in the red and near infrared
  spectral range},}\ }\href {http://stacks.iop.org/1367-2630/14/i=8/a=083001}
  {\bibfield  {journal} {\bibinfo  {journal} {New Journal of Physics}\ }\textbf
  {\bibinfo {volume} {14}},\ \bibinfo {pages} {083001} (\bibinfo {year}
  {2012})}\BibitemShut {NoStop}%
\bibitem [{\citenamefont {Rau}\ \emph {et~al.}(2014)\citenamefont {Rau},
  \citenamefont {Heindel}, \citenamefont {Unsleber}, \citenamefont {Braun},
  \citenamefont {Fischer}, \citenamefont {Frick}, \citenamefont {Nauerth},
  \citenamefont {Schneider}, \citenamefont {Vest}, \citenamefont
  {Reitzenstein}, \citenamefont {Kamp}, \citenamefont {Forchel}, \citenamefont
  {Höfling},\ and\ \citenamefont {Weinfurter}}]{Rau2014}%
  \BibitemOpen
  \bibfield  {author} {\bibinfo {author} {\bibfnamefont {M.}~\bibnamefont
  {Rau}}, \bibinfo {author} {\bibfnamefont {T.}~\bibnamefont {Heindel}},
  \bibinfo {author} {\bibfnamefont {S.}~\bibnamefont {Unsleber}}, \bibinfo
  {author} {\bibfnamefont {T.}~\bibnamefont {Braun}}, \bibinfo {author}
  {\bibfnamefont {J.}~\bibnamefont {Fischer}}, \bibinfo {author} {\bibfnamefont
  {S.}~\bibnamefont {Frick}}, \bibinfo {author} {\bibfnamefont
  {S.}~\bibnamefont {Nauerth}}, \bibinfo {author} {\bibfnamefont
  {C.}~\bibnamefont {Schneider}}, \bibinfo {author} {\bibfnamefont
  {G.}~\bibnamefont {Vest}}, \bibinfo {author} {\bibfnamefont {S.}~\bibnamefont
  {Reitzenstein}}, \bibinfo {author} {\bibfnamefont {M.}~\bibnamefont {Kamp}},
  \bibinfo {author} {\bibfnamefont {A.}~\bibnamefont {Forchel}}, \bibinfo
  {author} {\bibfnamefont {S.}~\bibnamefont {Höfling}}, \ and\ \bibinfo
  {author} {\bibfnamefont {H.}~\bibnamefont {Weinfurter}},\ }\bibfield  {title}
  {\enquote {\bibinfo {title} {Free space quantum key distribution over 500
  meters using electrically driven quantum dot single-photon
  sources{\textemdash}a proof of principle experiment},}\ }\href {\doibase
  10.1088/1367-2630/16/4/043003} {\bibfield  {journal} {\bibinfo  {journal}
  {New Journal of Physics}\ }\textbf {\bibinfo {volume} {16}},\ \bibinfo
  {pages} {043003} (\bibinfo {year} {2014})}\BibitemShut {NoStop}%
\bibitem [{\citenamefont {Takemoto}\ \emph {et~al.}(2015)\citenamefont
  {Takemoto}, \citenamefont {Nambu}, \citenamefont {Miyazawa}, \citenamefont
  {Sakuma}, \citenamefont {Yamamoto}, \citenamefont {Yorozu},\ and\
  \citenamefont {Arakawa}}]{Takemoto2015}%
  \BibitemOpen
  \bibfield  {author} {\bibinfo {author} {\bibfnamefont {K.}~\bibnamefont
  {Takemoto}}, \bibinfo {author} {\bibfnamefont {Y.}~\bibnamefont {Nambu}},
  \bibinfo {author} {\bibfnamefont {T.}~\bibnamefont {Miyazawa}}, \bibinfo
  {author} {\bibfnamefont {Y.}~\bibnamefont {Sakuma}}, \bibinfo {author}
  {\bibfnamefont {T.}~\bibnamefont {Yamamoto}}, \bibinfo {author}
  {\bibfnamefont {S.}~\bibnamefont {Yorozu}}, \ and\ \bibinfo {author}
  {\bibfnamefont {Y.}~\bibnamefont {Arakawa}},\ }\bibfield  {title} {\enquote
  {\bibinfo {title} {Quantum key distribution over 120{\hspace{0.167em}}km
  using ultrahigh purity single-photon source and superconducting single-photon
  detectors},}\ }\href {https://doi.org/10.1038/srep14383} {\bibfield
  {journal} {\bibinfo  {journal} {Scientific Reports}\ }\textbf {\bibinfo
  {volume} {5}},\ \bibinfo {pages} {14383} (\bibinfo {year}
  {2015})}\BibitemShut {NoStop}%
\bibitem [{\citenamefont {Dzurnak}\ \emph {et~al.}(2015)\citenamefont
  {Dzurnak}, \citenamefont {Stevenson}, \citenamefont {Nilsson}, \citenamefont
  {Dynes}, \citenamefont {Yuan}, \citenamefont {Skiba-Szymanska}, \citenamefont
  {Farrer}, \citenamefont {Ritchie},\ and\ \citenamefont
  {Shields}}]{Dzurnak2015}%
  \BibitemOpen
  \bibfield  {author} {\bibinfo {author} {\bibfnamefont {B.}~\bibnamefont
  {Dzurnak}}, \bibinfo {author} {\bibfnamefont {R.~M.}\ \bibnamefont
  {Stevenson}}, \bibinfo {author} {\bibfnamefont {J.}~\bibnamefont {Nilsson}},
  \bibinfo {author} {\bibfnamefont {J.~F.}\ \bibnamefont {Dynes}}, \bibinfo
  {author} {\bibfnamefont {Z.~L.}\ \bibnamefont {Yuan}}, \bibinfo {author}
  {\bibfnamefont {J.}~\bibnamefont {Skiba-Szymanska}}, \bibinfo {author}
  {\bibfnamefont {I.}~\bibnamefont {Farrer}}, \bibinfo {author} {\bibfnamefont
  {D.~A.}\ \bibnamefont {Ritchie}}, \ and\ \bibinfo {author} {\bibfnamefont
  {A.~J.}\ \bibnamefont {Shields}},\ }\bibfield  {title} {\enquote {\bibinfo
  {title} {Quantum key distribution with an entangled light emitting diode},}\
  }\href {\doibase 10.1063/1.4938502} {\bibfield  {journal} {\bibinfo
  {journal} {Applied Physics Letters}\ }\textbf {\bibinfo {volume} {107}},\
  \bibinfo {pages} {261101} (\bibinfo {year} {2015})}\BibitemShut {NoStop}%
\bibitem [{\citenamefont {Basset}\ \emph {et~al.}(2021)\citenamefont {Basset},
  \citenamefont {Valeri}, \citenamefont {Roccia}, \citenamefont {Muredda},
  \citenamefont {Poderini}, \citenamefont {Neuwirth}, \citenamefont {Spagnolo},
  \citenamefont {Rota}, \citenamefont {Carvacho}, \citenamefont {Sciarrino},\
  and\ \citenamefont {Trotta}}]{Basset2021}%
  \BibitemOpen
  \bibfield  {author} {\bibinfo {author} {\bibfnamefont {F.~B.}\ \bibnamefont
  {Basset}}, \bibinfo {author} {\bibfnamefont {M.}~\bibnamefont {Valeri}},
  \bibinfo {author} {\bibfnamefont {E.}~\bibnamefont {Roccia}}, \bibinfo
  {author} {\bibfnamefont {V.}~\bibnamefont {Muredda}}, \bibinfo {author}
  {\bibfnamefont {D.}~\bibnamefont {Poderini}}, \bibinfo {author}
  {\bibfnamefont {J.}~\bibnamefont {Neuwirth}}, \bibinfo {author}
  {\bibfnamefont {N.}~\bibnamefont {Spagnolo}}, \bibinfo {author}
  {\bibfnamefont {M.~B.}\ \bibnamefont {Rota}}, \bibinfo {author}
  {\bibfnamefont {G.}~\bibnamefont {Carvacho}}, \bibinfo {author}
  {\bibfnamefont {F.}~\bibnamefont {Sciarrino}}, \ and\ \bibinfo {author}
  {\bibfnamefont {R.}~\bibnamefont {Trotta}},\ }\bibfield  {title} {\enquote
  {\bibinfo {title} {Quantum key distribution with entangled photons generated
  on demand by a quantum dot},}\ }\href {\doibase 10.1126/sciadv.abe6379}
  {\bibfield  {journal} {\bibinfo  {journal} {Science Advances}\ }\textbf
  {\bibinfo {volume} {7}},\ \bibinfo {pages} {eabe6379} (\bibinfo {year}
  {2021})}\BibitemShut {NoStop}%
\bibitem [{\citenamefont {Schimpf}\ \emph {et~al.}(2021)\citenamefont
  {Schimpf}, \citenamefont {Reindl}, \citenamefont {Huber}, \citenamefont
  {Lehner}, \citenamefont {Silva}, \citenamefont {Manna}, \citenamefont
  {Vyvlecka}, \citenamefont {Walther},\ and\ \citenamefont
  {Rastelli}}]{Schimpf2021}%
  \BibitemOpen
  \bibfield  {author} {\bibinfo {author} {\bibfnamefont {C.}~\bibnamefont
  {Schimpf}}, \bibinfo {author} {\bibfnamefont {M.}~\bibnamefont {Reindl}},
  \bibinfo {author} {\bibfnamefont {D.}~\bibnamefont {Huber}}, \bibinfo
  {author} {\bibfnamefont {B.}~\bibnamefont {Lehner}}, \bibinfo {author}
  {\bibfnamefont {S.~F. C.~D.}\ \bibnamefont {Silva}}, \bibinfo {author}
  {\bibfnamefont {S.}~\bibnamefont {Manna}}, \bibinfo {author} {\bibfnamefont
  {M.}~\bibnamefont {Vyvlecka}}, \bibinfo {author} {\bibfnamefont
  {P.}~\bibnamefont {Walther}}, \ and\ \bibinfo {author} {\bibfnamefont
  {A.}~\bibnamefont {Rastelli}},\ }\bibfield  {title} {\enquote {\bibinfo
  {title} {Quantum cryptography with highly entangled photons from
  semiconductor quantum dots},}\ }\href {\doibase 10.1126/sciadv.abe8905}
  {\bibfield  {journal} {\bibinfo  {journal} {Science Advances}\ }\textbf
  {\bibinfo {volume} {7}},\ \bibinfo {pages} {eabe8905} (\bibinfo {year}
  {2021})}\BibitemShut {NoStop}%
\bibitem [{\citenamefont {Vajner}\ \emph {et~al.}(2021)\citenamefont {Vajner},
  \citenamefont {Rickert}, \citenamefont {Gao}, \citenamefont {Kaymazlar},\
  and\ \citenamefont {Heindel}}]{Vajner2021}%
  \BibitemOpen
  \bibfield  {author} {\bibinfo {author} {\bibfnamefont {D.~A.}\ \bibnamefont
  {Vajner}}, \bibinfo {author} {\bibfnamefont {L.}~\bibnamefont {Rickert}},
  \bibinfo {author} {\bibfnamefont {T.}~\bibnamefont {Gao}}, \bibinfo {author}
  {\bibfnamefont {K.}~\bibnamefont {Kaymazlar}}, \ and\ \bibinfo {author}
  {\bibfnamefont {T.}~\bibnamefont {Heindel}},\ }\bibfield  {title} {\enquote
  {\bibinfo {title} {Quantum communication using semiconductor quantum dots},}\
  }\href@noop {} {\bibfield  {journal} {\bibinfo  {journal} {arXiv2108.13877}\
  } (\bibinfo {year} {2021})}\BibitemShut {NoStop}%
\bibitem [{\citenamefont {Bennett}\ and\ \citenamefont
  {Brassard}(1984)}]{Bennett1984}%
  \BibitemOpen
  \bibfield  {author} {\bibinfo {author} {\bibfnamefont {C.~H.}\ \bibnamefont
  {Bennett}}\ and\ \bibinfo {author} {\bibfnamefont {G.}~\bibnamefont
  {Brassard}},\ }\bibfield  {title} {\enquote {\bibinfo {title} {Quantum
  cryptography: Public key distribution and coin tossing},}\ }\href@noop {}
  {\bibfield  {journal} {\bibinfo  {journal} {Proceedings of IEEE International
  Conference on Computers, Systems and Signal Processing, Bangalore, India}\ ,\
  \bibinfo {pages} {175--179}} (\bibinfo {year} {1984})}\BibitemShut {NoStop}%
\bibitem [{\citenamefont {Scarani}\ \emph {et~al.}(2009)\citenamefont
  {Scarani}, \citenamefont {Bechmann-Pasquinucci}, \citenamefont {Cerf},
  \citenamefont {Du\ifmmode~\check{s}\else \v{s}\fi{}ek}, \citenamefont
  {L\"utkenhaus},\ and\ \citenamefont {Peev}}]{Scarani2009}%
  \BibitemOpen
  \bibfield  {author} {\bibinfo {author} {\bibfnamefont {V.}~\bibnamefont
  {Scarani}}, \bibinfo {author} {\bibfnamefont {H.}~\bibnamefont
  {Bechmann-Pasquinucci}}, \bibinfo {author} {\bibfnamefont {N.~J.}\
  \bibnamefont {Cerf}}, \bibinfo {author} {\bibfnamefont {M.}~\bibnamefont
  {Du\ifmmode~\check{s}\else \v{s}\fi{}ek}}, \bibinfo {author} {\bibfnamefont
  {N.}~\bibnamefont {L\"utkenhaus}}, \ and\ \bibinfo {author} {\bibfnamefont
  {M.}~\bibnamefont {Peev}},\ }\bibfield  {title} {\enquote {\bibinfo {title}
  {The security of practical quantum key distribution},}\ }\href {\doibase
  10.1103/RevModPhys.81.1301} {\bibfield  {journal} {\bibinfo  {journal}
  {Reviews of Modern Physics}\ }\textbf {\bibinfo {volume} {81}},\ \bibinfo
  {pages} {1301--1350} (\bibinfo {year} {2009})}\BibitemShut {NoStop}%
\bibitem [{\citenamefont {Xu}\ \emph {et~al.}(2007)\citenamefont {Xu},
  \citenamefont {Toft}, \citenamefont {Phillips}, \citenamefont {Mar},
  \citenamefont {Hammura},\ and\ \citenamefont {Williams}}]{Xu2007}%
  \BibitemOpen
  \bibfield  {author} {\bibinfo {author} {\bibfnamefont {X.}~\bibnamefont
  {Xu}}, \bibinfo {author} {\bibfnamefont {I.}~\bibnamefont {Toft}}, \bibinfo
  {author} {\bibfnamefont {R.~T.}\ \bibnamefont {Phillips}}, \bibinfo {author}
  {\bibfnamefont {J.}~\bibnamefont {Mar}}, \bibinfo {author} {\bibfnamefont
  {K.}~\bibnamefont {Hammura}}, \ and\ \bibinfo {author} {\bibfnamefont
  {D.~A.}\ \bibnamefont {Williams}},\ }\bibfield  {title} {\enquote {\bibinfo
  {title} {'plug and play' single-photon sources},}\ }\href {\doibase
  10.1063/1.2437727} {\bibfield  {journal} {\bibinfo  {journal} {Applied
  Physics Letters}\ }\textbf {\bibinfo {volume} {90}},\ \bibinfo {pages}
  {061103} (\bibinfo {year} {2007})}\BibitemShut {NoStop}%
\bibitem [{\citenamefont {Cadeddu}\ \emph {et~al.}(2016)\citenamefont
  {Cadeddu}, \citenamefont {Teissier}, \citenamefont {Braakman}, \citenamefont
  {Gregersen}, \citenamefont {Stepanov}, \citenamefont {G{\'{e}}rard},
  \citenamefont {Claudon}, \citenamefont {Warburton}, \citenamefont {Poggio},\
  and\ \citenamefont {Munsch}}]{Cadeddu2016}%
  \BibitemOpen
  \bibfield  {author} {\bibinfo {author} {\bibfnamefont {D.}~\bibnamefont
  {Cadeddu}}, \bibinfo {author} {\bibfnamefont {J.}~\bibnamefont {Teissier}},
  \bibinfo {author} {\bibfnamefont {F.~R.}\ \bibnamefont {Braakman}}, \bibinfo
  {author} {\bibfnamefont {N.}~\bibnamefont {Gregersen}}, \bibinfo {author}
  {\bibfnamefont {P.}~\bibnamefont {Stepanov}}, \bibinfo {author}
  {\bibfnamefont {J.-M.}\ \bibnamefont {G{\'{e}}rard}}, \bibinfo {author}
  {\bibfnamefont {J.}~\bibnamefont {Claudon}}, \bibinfo {author} {\bibfnamefont
  {R.~J.}\ \bibnamefont {Warburton}}, \bibinfo {author} {\bibfnamefont
  {M.}~\bibnamefont {Poggio}}, \ and\ \bibinfo {author} {\bibfnamefont
  {M.}~\bibnamefont {Munsch}},\ }\bibfield  {title} {\enquote {\bibinfo {title}
  {A fiber-coupled quantum-dot on a photonic tip},}\ }\href {\doibase
  10.1063/1.4939264} {\bibfield  {journal} {\bibinfo  {journal} {Applied
  Physics Letters}\ }\textbf {\bibinfo {volume} {108}},\ \bibinfo {pages}
  {011112} (\bibinfo {year} {2016})}\BibitemShut {NoStop}%
\bibitem [{\citenamefont {Snijders}\ \emph {et~al.}(2018)\citenamefont
  {Snijders}, \citenamefont {Frey}, \citenamefont {Norman}, \citenamefont
  {Post}, \citenamefont {Gossard}, \citenamefont {Bowers}, \citenamefont {van
  Exter}, \citenamefont {Löffler},\ and\ \citenamefont
  {Bouwmeester}}]{Snijders2018}%
  \BibitemOpen
  \bibfield  {author} {\bibinfo {author} {\bibfnamefont {H.}~\bibnamefont
  {Snijders}}, \bibinfo {author} {\bibfnamefont {J.}~\bibnamefont {Frey}},
  \bibinfo {author} {\bibfnamefont {J.}~\bibnamefont {Norman}}, \bibinfo
  {author} {\bibfnamefont {V.}~\bibnamefont {Post}}, \bibinfo {author}
  {\bibfnamefont {A.}~\bibnamefont {Gossard}}, \bibinfo {author} {\bibfnamefont
  {J.}~\bibnamefont {Bowers}}, \bibinfo {author} {\bibfnamefont
  {M.}~\bibnamefont {van Exter}}, \bibinfo {author} {\bibfnamefont
  {W.}~\bibnamefont {Löffler}}, \ and\ \bibinfo {author} {\bibfnamefont
  {D.}~\bibnamefont {Bouwmeester}},\ }\bibfield  {title} {\enquote {\bibinfo
  {title} {Fiber-coupled cavity-{QED} source of identical single photons},}\
  }\href {https://doi.org/10.1103/PhysRevApplied.9.031002} {\bibfield
  {journal} {\bibinfo  {journal} {Physical Review Applied}\ }\textbf {\bibinfo
  {volume} {9}},\ \bibinfo {pages} {031002} (\bibinfo {year}
  {2018})}\BibitemShut {NoStop}%
\bibitem [{\citenamefont {Rickert}\ \emph {et~al.}(2021)\citenamefont
  {Rickert}, \citenamefont {Schröder}, \citenamefont {Gao}, \citenamefont
  {Schneider}, \citenamefont {Höfling},\ and\ \citenamefont
  {Heindel}}]{Rickert2021}%
  \BibitemOpen
  \bibfield  {author} {\bibinfo {author} {\bibfnamefont {L.}~\bibnamefont
  {Rickert}}, \bibinfo {author} {\bibfnamefont {F.}~\bibnamefont {Schröder}},
  \bibinfo {author} {\bibfnamefont {T.}~\bibnamefont {Gao}}, \bibinfo {author}
  {\bibfnamefont {C.}~\bibnamefont {Schneider}}, \bibinfo {author}
  {\bibfnamefont {S.}~\bibnamefont {Höfling}}, \ and\ \bibinfo {author}
  {\bibfnamefont {T.}~\bibnamefont {Heindel}},\ }\bibfield  {title} {\enquote
  {\bibinfo {title} {Fiber-pigtailing quantum-dot cavity-enhanced light
  emitting diodes},}\ }\href {\doibase 10.1063/5.0063697} {\bibfield  {journal}
  {\bibinfo  {journal} {Applied Physics Letters}\ }\textbf {\bibinfo {volume}
  {119}},\ \bibinfo {pages} {131104} (\bibinfo {year} {2021})}\BibitemShut
  {NoStop}%
\bibitem [{\citenamefont {Schlehahn}\ \emph {et~al.}(2015)\citenamefont
  {Schlehahn}, \citenamefont {Krüger}, \citenamefont {Gschrey}, \citenamefont
  {Schulze}, \citenamefont {Rodt}, \citenamefont {Strittmatter}, \citenamefont
  {Heindel},\ and\ \citenamefont {Reitzenstein}}]{Schlehahn2015}%
  \BibitemOpen
  \bibfield  {author} {\bibinfo {author} {\bibfnamefont {A.}~\bibnamefont
  {Schlehahn}}, \bibinfo {author} {\bibfnamefont {L.}~\bibnamefont {Krüger}},
  \bibinfo {author} {\bibfnamefont {M.}~\bibnamefont {Gschrey}}, \bibinfo
  {author} {\bibfnamefont {J.-H.}\ \bibnamefont {Schulze}}, \bibinfo {author}
  {\bibfnamefont {S.}~\bibnamefont {Rodt}}, \bibinfo {author} {\bibfnamefont
  {A.}~\bibnamefont {Strittmatter}}, \bibinfo {author} {\bibfnamefont
  {T.}~\bibnamefont {Heindel}}, \ and\ \bibinfo {author} {\bibfnamefont
  {S.}~\bibnamefont {Reitzenstein}},\ }\bibfield  {title} {\enquote {\bibinfo
  {title} {Operating single quantum emitters with a compact stirling
  cryocooler},}\ }\href {\doibase 10.1063/1.4906548} {\bibfield  {journal}
  {\bibinfo  {journal} {Review of Scientific Instruments}\ }\textbf {\bibinfo
  {volume} {86}},\ \bibinfo {pages} {013113} (\bibinfo {year}
  {2015})}\BibitemShut {NoStop}%
\bibitem [{\citenamefont {Schlehahn}\ \emph {et~al.}(2018)\citenamefont
  {Schlehahn}, \citenamefont {Fischbach}, \citenamefont {Schmidt},
  \citenamefont {Kaganskiy}, \citenamefont {Strittmatter}, \citenamefont
  {Rodt}, \citenamefont {Heindel},\ and\ \citenamefont
  {Reitzenstein}}]{Schlehahn2018}%
  \BibitemOpen
  \bibfield  {author} {\bibinfo {author} {\bibfnamefont {A.}~\bibnamefont
  {Schlehahn}}, \bibinfo {author} {\bibfnamefont {S.}~\bibnamefont
  {Fischbach}}, \bibinfo {author} {\bibfnamefont {R.}~\bibnamefont {Schmidt}},
  \bibinfo {author} {\bibfnamefont {A.}~\bibnamefont {Kaganskiy}}, \bibinfo
  {author} {\bibfnamefont {A.}~\bibnamefont {Strittmatter}}, \bibinfo {author}
  {\bibfnamefont {S.}~\bibnamefont {Rodt}}, \bibinfo {author} {\bibfnamefont
  {T.}~\bibnamefont {Heindel}}, \ and\ \bibinfo {author} {\bibfnamefont
  {S.}~\bibnamefont {Reitzenstein}},\ }\bibfield  {title} {\enquote {\bibinfo
  {title} {A stand-alone fiber-coupled single-photon source},}\ }\href
  {https://doi.org/10.1038/s41598-017-19049-4} {\bibfield  {journal} {\bibinfo
  {journal} {Scientific Reports}\ }\textbf {\bibinfo {volume} {8}},\ \bibinfo
  {pages} {1340} (\bibinfo {year} {2018})}\BibitemShut {NoStop}%
\bibitem [{\citenamefont {Musia{\l}}\ \emph {et~al.}(2020)\citenamefont
  {Musia{\l}}, \citenamefont {{\.{Z}}o{\l}nacz}, \citenamefont {Srocka},
  \citenamefont {Kravets}, \citenamefont {Gro{\ss}e}, \citenamefont
  {Olszewski}, \citenamefont {Poturaj}, \citenamefont {W{\'{o}}jcik},
  \citenamefont {Mergo}, \citenamefont {Dybka}, \citenamefont {Dyrkacz},
  \citenamefont {D{\l}ubek}, \citenamefont {Lauritsen}, \citenamefont
  {Bülter}, \citenamefont {Schneider}, \citenamefont {Zschiedrich},
  \citenamefont {Burger}, \citenamefont {Rodt}, \citenamefont
  {Urba{\'{n}}czyk}, \citenamefont {S{\k{e}}k},\ and\ \citenamefont
  {Reitzenstein}}]{Musial2020}%
  \BibitemOpen
  \bibfield  {author} {\bibinfo {author} {\bibfnamefont {A.}~\bibnamefont
  {Musia{\l}}}, \bibinfo {author} {\bibfnamefont {K.}~\bibnamefont
  {{\.{Z}}o{\l}nacz}}, \bibinfo {author} {\bibfnamefont {N.}~\bibnamefont
  {Srocka}}, \bibinfo {author} {\bibfnamefont {O.}~\bibnamefont {Kravets}},
  \bibinfo {author} {\bibfnamefont {J.}~\bibnamefont {Gro{\ss}e}}, \bibinfo
  {author} {\bibfnamefont {J.}~\bibnamefont {Olszewski}}, \bibinfo {author}
  {\bibfnamefont {K.}~\bibnamefont {Poturaj}}, \bibinfo {author} {\bibfnamefont
  {G.}~\bibnamefont {W{\'{o}}jcik}}, \bibinfo {author} {\bibfnamefont
  {P.}~\bibnamefont {Mergo}}, \bibinfo {author} {\bibfnamefont
  {K.}~\bibnamefont {Dybka}}, \bibinfo {author} {\bibfnamefont
  {M.}~\bibnamefont {Dyrkacz}}, \bibinfo {author} {\bibfnamefont
  {M.}~\bibnamefont {D{\l}ubek}}, \bibinfo {author} {\bibfnamefont
  {K.}~\bibnamefont {Lauritsen}}, \bibinfo {author} {\bibfnamefont
  {A.}~\bibnamefont {Bülter}}, \bibinfo {author} {\bibfnamefont {P.-I.}\
  \bibnamefont {Schneider}}, \bibinfo {author} {\bibfnamefont {L.}~\bibnamefont
  {Zschiedrich}}, \bibinfo {author} {\bibfnamefont {S.}~\bibnamefont {Burger}},
  \bibinfo {author} {\bibfnamefont {S.}~\bibnamefont {Rodt}}, \bibinfo {author}
  {\bibfnamefont {W.}~\bibnamefont {Urba{\'{n}}czyk}}, \bibinfo {author}
  {\bibfnamefont {G.}~\bibnamefont {S{\k{e}}k}}, \ and\ \bibinfo {author}
  {\bibfnamefont {S.}~\bibnamefont {Reitzenstein}},\ }\bibfield  {title}
  {\enquote {\bibinfo {title} {Plug{\&}play fiber-coupled 73~{kHz}
  single-photon source operating in the telecom o-band},}\ }\href {\doibase
  10.1002/qute.202000018} {\bibfield  {journal} {\bibinfo  {journal} {Advanced
  Quantum Technologies}\ }\textbf {\bibinfo {volume} {3}},\ \bibinfo {pages}
  {2000018} (\bibinfo {year} {2020})}\BibitemShut {NoStop}%
\bibitem [{\citenamefont {Guffarth}\ \emph {et~al.}(2001)\citenamefont
  {Guffarth}, \citenamefont {Heitz}, \citenamefont {Schliwa}, \citenamefont
  {Stier}, \citenamefont {Ledentsov}, \citenamefont {Kovsh}, \citenamefont
  {Ustinov},\ and\ \citenamefont {Bimberg}}]{Guffarth2001}%
  \BibitemOpen
  \bibfield  {author} {\bibinfo {author} {\bibfnamefont {F.}~\bibnamefont
  {Guffarth}}, \bibinfo {author} {\bibfnamefont {R.}~\bibnamefont {Heitz}},
  \bibinfo {author} {\bibfnamefont {A.}~\bibnamefont {Schliwa}}, \bibinfo
  {author} {\bibfnamefont {O.}~\bibnamefont {Stier}}, \bibinfo {author}
  {\bibfnamefont {N.~N.}\ \bibnamefont {Ledentsov}}, \bibinfo {author}
  {\bibfnamefont {A.~R.}\ \bibnamefont {Kovsh}}, \bibinfo {author}
  {\bibfnamefont {V.~M.}\ \bibnamefont {Ustinov}}, \ and\ \bibinfo {author}
  {\bibfnamefont {D.}~\bibnamefont {Bimberg}},\ }\bibfield  {title} {\enquote
  {\bibinfo {title} {Strain engineering of self-organized inas quantum dots},}\
  }\href {\doibase 10.1103/PhysRevB.64.085305} {\bibfield  {journal} {\bibinfo
  {journal} {Phys. Rev. B}\ }\textbf {\bibinfo {volume} {64}},\ \bibinfo
  {pages} {085305} (\bibinfo {year} {2001})}\BibitemShut {NoStop}%
\bibitem [{\citenamefont {{\.{Z}}o{\l}nacz}\ \emph {et~al.}(2019)\citenamefont
  {{\.{Z}}o{\l}nacz}, \citenamefont {Musia{\l}}, \citenamefont {Srocka},
  \citenamefont {Gro{\ss}e}, \citenamefont {Schlösinger}, \citenamefont
  {Schneider}, \citenamefont {Kravets}, \citenamefont {Mikulicz}, \citenamefont
  {Olszewski}, \citenamefont {Poturaj}, \citenamefont {W{\'{o}}jcik},
  \citenamefont {Mergo}, \citenamefont {Dybka}, \citenamefont {Dyrkacz},
  \citenamefont {D{\l}ubek}, \citenamefont {Rodt}, \citenamefont {Burger},
  \citenamefont {Zschiedrich}, \citenamefont {S{\k{e}}k}, \citenamefont
  {Reitzenstein},\ and\ \citenamefont {Urba{\'{n}}czyk}}]{Zolnacz2019}%
  \BibitemOpen
  \bibfield  {author} {\bibinfo {author} {\bibfnamefont {K.}~\bibnamefont
  {{\.{Z}}o{\l}nacz}}, \bibinfo {author} {\bibfnamefont {A.}~\bibnamefont
  {Musia{\l}}}, \bibinfo {author} {\bibfnamefont {N.}~\bibnamefont {Srocka}},
  \bibinfo {author} {\bibfnamefont {J.}~\bibnamefont {Gro{\ss}e}}, \bibinfo
  {author} {\bibfnamefont {M.~J.}\ \bibnamefont {Schlösinger}}, \bibinfo
  {author} {\bibfnamefont {P.-I.}\ \bibnamefont {Schneider}}, \bibinfo {author}
  {\bibfnamefont {O.}~\bibnamefont {Kravets}}, \bibinfo {author} {\bibfnamefont
  {M.}~\bibnamefont {Mikulicz}}, \bibinfo {author} {\bibfnamefont
  {J.}~\bibnamefont {Olszewski}}, \bibinfo {author} {\bibfnamefont
  {K.}~\bibnamefont {Poturaj}}, \bibinfo {author} {\bibfnamefont
  {G.}~\bibnamefont {W{\'{o}}jcik}}, \bibinfo {author} {\bibfnamefont
  {P.}~\bibnamefont {Mergo}}, \bibinfo {author} {\bibfnamefont
  {K.}~\bibnamefont {Dybka}}, \bibinfo {author} {\bibfnamefont
  {M.}~\bibnamefont {Dyrkacz}}, \bibinfo {author} {\bibfnamefont
  {M.}~\bibnamefont {D{\l}ubek}}, \bibinfo {author} {\bibfnamefont
  {S.}~\bibnamefont {Rodt}}, \bibinfo {author} {\bibfnamefont {S.}~\bibnamefont
  {Burger}}, \bibinfo {author} {\bibfnamefont {L.}~\bibnamefont {Zschiedrich}},
  \bibinfo {author} {\bibfnamefont {G.}~\bibnamefont {S{\k{e}}k}}, \bibinfo
  {author} {\bibfnamefont {S.}~\bibnamefont {Reitzenstein}}, \ and\ \bibinfo
  {author} {\bibfnamefont {W.}~\bibnamefont {Urba{\'{n}}czyk}},\ }\bibfield
  {title} {\enquote {\bibinfo {title} {Method for direct coupling of a
  semiconductor quantum dot to an optical fiber for single-photon source
  applications},}\ }\href {\doibase 10.1364/oe.27.026772} {\bibfield  {journal}
  {\bibinfo  {journal} {Optics Express}\ }\textbf {\bibinfo {volume} {27}},\
  \bibinfo {pages} {26772} (\bibinfo {year} {2019})}\BibitemShut {NoStop}%
\bibitem [{\citenamefont {Schneider}\ \emph {et~al.}(2018)\citenamefont
  {Schneider}, \citenamefont {Srocka}, \citenamefont {Rodt}, \citenamefont
  {Zschiedrich}, \citenamefont {Reitzenstein},\ and\ \citenamefont
  {Burger}}]{Schneider2018}%
  \BibitemOpen
  \bibfield  {author} {\bibinfo {author} {\bibfnamefont {P.-I.}\ \bibnamefont
  {Schneider}}, \bibinfo {author} {\bibfnamefont {N.}~\bibnamefont {Srocka}},
  \bibinfo {author} {\bibfnamefont {S.}~\bibnamefont {Rodt}}, \bibinfo {author}
  {\bibfnamefont {L.}~\bibnamefont {Zschiedrich}}, \bibinfo {author}
  {\bibfnamefont {S.}~\bibnamefont {Reitzenstein}}, \ and\ \bibinfo {author}
  {\bibfnamefont {S.}~\bibnamefont {Burger}},\ }\bibfield  {title} {\enquote
  {\bibinfo {title} {Numerical optimization of the extraction efficiency of a
  quantum-dot based single-photon emitter into a single-mode fiber},}\ }\href
  {\doibase 10.1364/OE.26.008479} {\bibfield  {journal} {\bibinfo  {journal}
  {Optics Express}\ }\textbf {\bibinfo {volume} {26}},\ \bibinfo {pages}
  {8479--8492} (\bibinfo {year} {2018})}\BibitemShut {NoStop}%
\bibitem [{\citenamefont {Schneider}\ \emph {et~al.}(2019)\citenamefont
  {Schneider}, \citenamefont {Garcia~Santiago}, \citenamefont {Soltwisch},
  \citenamefont {Hammerschmidt}, \citenamefont {Burger},\ and\ \citenamefont
  {Rockstuhl}}]{Schneider2019}%
  \BibitemOpen
  \bibfield  {author} {\bibinfo {author} {\bibfnamefont {P.-I.}\ \bibnamefont
  {Schneider}}, \bibinfo {author} {\bibfnamefont {X.}~\bibnamefont
  {Garcia~Santiago}}, \bibinfo {author} {\bibfnamefont {V.}~\bibnamefont
  {Soltwisch}}, \bibinfo {author} {\bibfnamefont {M.}~\bibnamefont
  {Hammerschmidt}}, \bibinfo {author} {\bibfnamefont {S.}~\bibnamefont
  {Burger}}, \ and\ \bibinfo {author} {\bibfnamefont {C.}~\bibnamefont
  {Rockstuhl}},\ }\bibfield  {title} {\enquote {\bibinfo {title} {Benchmarking
  five global optimization approaches for nano-optical shape optimization and
  parameter reconstruction},}\ }\href {\doibase 10.1021/acsphotonics.9b00706}
  {\bibfield  {journal} {\bibinfo  {journal} {ACS Photonics}\ }\textbf
  {\bibinfo {volume} {6}},\ \bibinfo {pages} {2726--2733} (\bibinfo {year}
  {2019})}\BibitemShut {NoStop}%
\bibitem [{\citenamefont {Aichele}, \citenamefont {Zwiller},\ and\
  \citenamefont {Benson}(2004)}]{Aichele_2004}%
  \BibitemOpen
  \bibfield  {author} {\bibinfo {author} {\bibfnamefont {T.}~\bibnamefont
  {Aichele}}, \bibinfo {author} {\bibfnamefont {V.}~\bibnamefont {Zwiller}}, \
  and\ \bibinfo {author} {\bibfnamefont {O.}~\bibnamefont {Benson}},\
  }\bibfield  {title} {\enquote {\bibinfo {title} {Visible single-photon
  generation from semiconductor quantum dots},}\ }\href {\doibase
  10.1088/1367-2630/6/1/090} {\bibfield  {journal} {\bibinfo  {journal} {New
  Journal of Physics}\ }\textbf {\bibinfo {volume} {6}},\ \bibinfo {pages} {90}
  (\bibinfo {year} {2004})}\BibitemShut {NoStop}%
\bibitem [{\citenamefont {Dalgarno}\ \emph {et~al.}(2008)\citenamefont
  {Dalgarno}, \citenamefont {McFarlane}, \citenamefont {Brunner}, \citenamefont
  {Lambert}, \citenamefont {Gerardot}, \citenamefont {Warburton}, \citenamefont
  {Karrai}, \citenamefont {Badolato},\ and\ \citenamefont
  {Petroff}}]{Dalgarno2008}%
  \BibitemOpen
  \bibfield  {author} {\bibinfo {author} {\bibfnamefont {P.~A.}\ \bibnamefont
  {Dalgarno}}, \bibinfo {author} {\bibfnamefont {J.}~\bibnamefont {McFarlane}},
  \bibinfo {author} {\bibfnamefont {D.}~\bibnamefont {Brunner}}, \bibinfo
  {author} {\bibfnamefont {R.~W.}\ \bibnamefont {Lambert}}, \bibinfo {author}
  {\bibfnamefont {B.~D.}\ \bibnamefont {Gerardot}}, \bibinfo {author}
  {\bibfnamefont {R.~J.}\ \bibnamefont {Warburton}}, \bibinfo {author}
  {\bibfnamefont {K.}~\bibnamefont {Karrai}}, \bibinfo {author} {\bibfnamefont
  {A.}~\bibnamefont {Badolato}}, \ and\ \bibinfo {author} {\bibfnamefont
  {P.~M.}\ \bibnamefont {Petroff}},\ }\bibfield  {title} {\enquote {\bibinfo
  {title} {Hole recapture limited single photon generation from a single n-type
  charge-tunable quantum dot},}\ }\href {https://doi.org/10.1063/1.2924315}
  {\bibfield  {journal} {\bibinfo  {journal} {Applied Physics Letters}\
  }\textbf {\bibinfo {volume} {92}},\ \bibinfo {pages} {193103} (\bibinfo
  {year} {2008})}\BibitemShut {NoStop}%
\bibitem [{\citenamefont {Stucki}\ \emph {et~al.}(2002)\citenamefont {Stucki},
  \citenamefont {Gisin}, \citenamefont {Guinnard}, \citenamefont {Ribordy},\
  and\ \citenamefont {Zbinden}}]{Stucki_2002}%
  \BibitemOpen
  \bibfield  {author} {\bibinfo {author} {\bibfnamefont {D.}~\bibnamefont
  {Stucki}}, \bibinfo {author} {\bibfnamefont {N.}~\bibnamefont {Gisin}},
  \bibinfo {author} {\bibfnamefont {O.}~\bibnamefont {Guinnard}}, \bibinfo
  {author} {\bibfnamefont {G.}~\bibnamefont {Ribordy}}, \ and\ \bibinfo
  {author} {\bibfnamefont {H.}~\bibnamefont {Zbinden}},\ }\bibfield  {title}
  {\enquote {\bibinfo {title} {Quantum key distribution over 67 km with a
  plug{\&}play system},}\ }\href {\doibase 10.1088/1367-2630/4/1/341}
  {\bibfield  {journal} {\bibinfo  {journal} {New Journal of Physics}\ }\textbf
  {\bibinfo {volume} {4}},\ \bibinfo {pages} {41--41} (\bibinfo {year}
  {2002})}\BibitemShut {NoStop}%
\bibitem [{Note1()}]{Note1}%
  \BibitemOpen
  \bibinfo {note} {For companies see e.g., ID Quantique SA, Toshiba Europe
  Limited, MagiQ Technologies.}\BibitemShut {Stop}%
\bibitem [{\citenamefont {Kupko}\ \emph {et~al.}(2020)\citenamefont {Kupko},
  \citenamefont {von Helversen}, \citenamefont {Rickert}, \citenamefont
  {Schulze}, \citenamefont {Strittmatter}, \citenamefont {Gschrey},
  \citenamefont {Rodt}, \citenamefont {Reitzenstein},\ and\ \citenamefont
  {Heindel}}]{Kupko2020}%
  \BibitemOpen
  \bibfield  {author} {\bibinfo {author} {\bibfnamefont {T.}~\bibnamefont
  {Kupko}}, \bibinfo {author} {\bibfnamefont {M.}~\bibnamefont {von
  Helversen}}, \bibinfo {author} {\bibfnamefont {L.}~\bibnamefont {Rickert}},
  \bibinfo {author} {\bibfnamefont {J.-H.}\ \bibnamefont {Schulze}}, \bibinfo
  {author} {\bibfnamefont {A.}~\bibnamefont {Strittmatter}}, \bibinfo {author}
  {\bibfnamefont {M.}~\bibnamefont {Gschrey}}, \bibinfo {author} {\bibfnamefont
  {S.}~\bibnamefont {Rodt}}, \bibinfo {author} {\bibfnamefont {S.}~\bibnamefont
  {Reitzenstein}}, \ and\ \bibinfo {author} {\bibfnamefont {T.}~\bibnamefont
  {Heindel}},\ }\bibfield  {title} {\enquote {\bibinfo {title} {Tools for the
  performance optimization of single-photon quantum key distribution},}\ }\href
  {https://doi.org/10.1038/s41534-020-0262-8} {\bibfield  {journal} {\bibinfo
  {journal} {npj Quantum Information}\ }\textbf {\bibinfo {volume} {6}}
  (\bibinfo {year} {2020})}\BibitemShut {NoStop}%
\bibitem [{\citenamefont {Gottesman}\ \emph {et~al.}(2004)\citenamefont
  {Gottesman}, \citenamefont {Lo}, \citenamefont {L\"{u}tkenhaus},\ and\
  \citenamefont {Preskill}}]{10.5555/2011586.2011587}%
  \BibitemOpen
  \bibfield  {author} {\bibinfo {author} {\bibfnamefont {D.}~\bibnamefont
  {Gottesman}}, \bibinfo {author} {\bibfnamefont {H.-K.}\ \bibnamefont {Lo}},
  \bibinfo {author} {\bibfnamefont {N.}~\bibnamefont {L\"{u}tkenhaus}}, \ and\
  \bibinfo {author} {\bibfnamefont {J.}~\bibnamefont {Preskill}},\ }\bibfield
  {title} {\enquote {\bibinfo {title} {Security of quantum key distribution
  with imperfect devices},}\ }\href@noop {} {\bibfield  {journal} {\bibinfo
  {journal} {Quantum Info. Comput.}\ }\textbf {\bibinfo {volume} {4}},\
  \bibinfo {pages} {325–360} (\bibinfo {year} {2004})}\BibitemShut {NoStop}%
\bibitem [{\citenamefont {Chaiwongkhot}\ \emph {et~al.}(2017)\citenamefont
  {Chaiwongkhot}, \citenamefont {Sajeed}, \citenamefont {Lydersen},\ and\
  \citenamefont {Makarov}}]{Chaiwongkhot_2017}%
  \BibitemOpen
  \bibfield  {author} {\bibinfo {author} {\bibfnamefont {P.}~\bibnamefont
  {Chaiwongkhot}}, \bibinfo {author} {\bibfnamefont {S.}~\bibnamefont
  {Sajeed}}, \bibinfo {author} {\bibfnamefont {L.}~\bibnamefont {Lydersen}}, \
  and\ \bibinfo {author} {\bibfnamefont {V.}~\bibnamefont {Makarov}},\
  }\bibfield  {title} {\enquote {\bibinfo {title} {Finite-key-size effect in a
  commercial plug-and-play {QKD} system},}\ }\href {\doibase
  10.1088/2058-9565/aa804b} {\bibfield  {journal} {\bibinfo  {journal} {Quantum
  Science and Technology}\ }\textbf {\bibinfo {volume} {2}},\ \bibinfo {pages}
  {044003} (\bibinfo {year} {2017})}\BibitemShut {NoStop}%
\bibitem [{Note2()}]{Note2}%
  \BibitemOpen
  \bibinfo {note} {Note, that a more widely used term in the literature is
  quantum bit error rate (QBER) in units of $s^{-1}$. As the QBER entering the
  binary Shannon entropy in Eq.~1 (denoted as $e$) must be a probability (cf.
  Subsection~\ref {sec:Basics}), we consistently use the quantum bit error
  ratio, defined as the ratio of erroneous bits to all detected bits in our
  work.}\BibitemShut {Stop}%
\bibitem [{\citenamefont {Lydersen}\ and\ \citenamefont
  {Skaar}(2010)}]{Lydersen2010}%
  \BibitemOpen
  \bibfield  {author} {\bibinfo {author} {\bibfnamefont {L.}~\bibnamefont
  {Lydersen}}\ and\ \bibinfo {author} {\bibfnamefont {J.}~\bibnamefont
  {Skaar}},\ }\bibfield  {title} {\enquote {\bibinfo {title} {Security of
  quantum key distribution with bit and basis dependent detector flaws.}}\
  }\href {https://www.rintonpress.com/journals/doi/QIC10.1-2-5.html} {\bibfield
   {journal} {\bibinfo  {journal} {Quantum Information and Computation}\
  }\textbf {\bibinfo {volume} {10}},\ \bibinfo {pages} {60–76} (\bibinfo
  {year} {2010})}\BibitemShut {NoStop}%
\bibitem [{\citenamefont {Grünwald}(2019)}]{Gr_nwald_2019}%
  \BibitemOpen
  \bibfield  {author} {\bibinfo {author} {\bibfnamefont {P.}~\bibnamefont
  {Grünwald}},\ }\bibfield  {title} {\enquote {\bibinfo {title} {Effective
  second-order correlation function and single-photon detection},}\ }\href
  {\doibase 10.1088/1367-2630/ab3ae0} {\bibfield  {journal} {\bibinfo
  {journal} {New Journal of Physics}\ }\textbf {\bibinfo {volume} {21}},\
  \bibinfo {pages} {093003} (\bibinfo {year} {2019})}\BibitemShut {NoStop}%
\bibitem [{\citenamefont {Chavez-Mackay}, \citenamefont {Gr\"unwald},\ and\
  \citenamefont {Rodr\'{\i}guez-Lara}(2020)}]{Chavez-Mackay2020}%
  \BibitemOpen
  \bibfield  {author} {\bibinfo {author} {\bibfnamefont {J.~R.}\ \bibnamefont
  {Chavez-Mackay}}, \bibinfo {author} {\bibfnamefont {P.}~\bibnamefont
  {Gr\"unwald}}, \ and\ \bibinfo {author} {\bibfnamefont {B.~M.}\ \bibnamefont
  {Rodr\'{\i}guez-Lara}},\ }\bibfield  {title} {\enquote {\bibinfo {title}
  {Estimating the single-photon projection of low-intensity light sources},}\
  }\href {\doibase 10.1103/PhysRevA.101.053815} {\bibfield  {journal} {\bibinfo
   {journal} {Physical Review A}\ }\textbf {\bibinfo {volume} {101}},\ \bibinfo
  {pages} {053815} (\bibinfo {year} {2020})}\BibitemShut {NoStop}%
\bibitem [{\citenamefont {Rusca}\ \emph {et~al.}(2018)\citenamefont {Rusca},
  \citenamefont {Boaron}, \citenamefont {Grünenfelder}, \citenamefont
  {Martin},\ and\ \citenamefont {Zbinden}}]{Rusca2018}%
  \BibitemOpen
  \bibfield  {author} {\bibinfo {author} {\bibfnamefont {D.}~\bibnamefont
  {Rusca}}, \bibinfo {author} {\bibfnamefont {A.}~\bibnamefont {Boaron}},
  \bibinfo {author} {\bibfnamefont {F.}~\bibnamefont {Grünenfelder}}, \bibinfo
  {author} {\bibfnamefont {A.}~\bibnamefont {Martin}}, \ and\ \bibinfo {author}
  {\bibfnamefont {H.}~\bibnamefont {Zbinden}},\ }\bibfield  {title} {\enquote
  {\bibinfo {title} {Finite-key analysis for the 1-decoy state qkd protocol},}\
  }\href {\doibase 10.1063/1.5023340} {\bibfield  {journal} {\bibinfo
  {journal} {Applied Physics Letters}\ }\textbf {\bibinfo {volume} {112}},\
  \bibinfo {pages} {171104} (\bibinfo {year} {2018})}\BibitemShut {NoStop}%
\bibitem [{\citenamefont {Lee}\ \emph {et~al.}(2019)\citenamefont {Lee},
  \citenamefont {Buyukkaya}, \citenamefont {Aghaeimeibodi}, \citenamefont
  {Karasahin}, \citenamefont {Richardson},\ and\ \citenamefont
  {Waks}}]{Lee2019}%
  \BibitemOpen
  \bibfield  {author} {\bibinfo {author} {\bibfnamefont {C.-M.}\ \bibnamefont
  {Lee}}, \bibinfo {author} {\bibfnamefont {M.~A.}\ \bibnamefont {Buyukkaya}},
  \bibinfo {author} {\bibfnamefont {S.}~\bibnamefont {Aghaeimeibodi}}, \bibinfo
  {author} {\bibfnamefont {A.}~\bibnamefont {Karasahin}}, \bibinfo {author}
  {\bibfnamefont {C.~J.~K.}\ \bibnamefont {Richardson}}, \ and\ \bibinfo
  {author} {\bibfnamefont {E.}~\bibnamefont {Waks}},\ }\bibfield  {title}
  {\enquote {\bibinfo {title} {A fiber-integrated nanobeam single photon source
  emitting at telecom wavelengths},}\ }\href {\doibase 10.1063/1.5089907}
  {\bibfield  {journal} {\bibinfo  {journal} {Applied Physics Letters}\
  }\textbf {\bibinfo {volume} {114}},\ \bibinfo {pages} {171101} (\bibinfo
  {year} {2019})}\BibitemShut {NoStop}%
\bibitem [{\citenamefont {Srocka}\ \emph {et~al.}(2020)\citenamefont {Srocka},
  \citenamefont {Mrowi{\'{n}}ski}, \citenamefont {Gro{\ss}e}, \citenamefont
  {von Helversen}, \citenamefont {Heindel}, \citenamefont {Rodt},\ and\
  \citenamefont {Reitzenstein}}]{Srocka2020}%
  \BibitemOpen
  \bibfield  {author} {\bibinfo {author} {\bibfnamefont {N.}~\bibnamefont
  {Srocka}}, \bibinfo {author} {\bibfnamefont {P.}~\bibnamefont
  {Mrowi{\'{n}}ski}}, \bibinfo {author} {\bibfnamefont {J.}~\bibnamefont
  {Gro{\ss}e}}, \bibinfo {author} {\bibfnamefont {M.}~\bibnamefont {von
  Helversen}}, \bibinfo {author} {\bibfnamefont {T.}~\bibnamefont {Heindel}},
  \bibinfo {author} {\bibfnamefont {S.}~\bibnamefont {Rodt}}, \ and\ \bibinfo
  {author} {\bibfnamefont {S.}~\bibnamefont {Reitzenstein}},\ }\bibfield
  {title} {\enquote {\bibinfo {title} {Deterministically fabricated quantum dot
  single-photon source emitting indistinguishable photons in the telecom
  o-band},}\ }\href {\doibase 10.1063/5.0010436} {\bibfield  {journal}
  {\bibinfo  {journal} {Applied Physics Letters}\ }\textbf {\bibinfo {volume}
  {116}},\ \bibinfo {pages} {231104} (\bibinfo {year} {2020})}\BibitemShut
  {NoStop}%
\bibitem [{\citenamefont {Kolatschek}\ \emph {et~al.}(2021)\citenamefont
  {Kolatschek}, \citenamefont {Nawrath}, \citenamefont {Bauer}, \citenamefont
  {Huang}, \citenamefont {Fischer}, \citenamefont {Sittig}, \citenamefont
  {Jetter}, \citenamefont {Portalupi},\ and\ \citenamefont
  {Michler}}]{Kolatschek2021}%
  \BibitemOpen
  \bibfield  {author} {\bibinfo {author} {\bibfnamefont {S.}~\bibnamefont
  {Kolatschek}}, \bibinfo {author} {\bibfnamefont {C.}~\bibnamefont {Nawrath}},
  \bibinfo {author} {\bibfnamefont {S.}~\bibnamefont {Bauer}}, \bibinfo
  {author} {\bibfnamefont {J.}~\bibnamefont {Huang}}, \bibinfo {author}
  {\bibfnamefont {J.}~\bibnamefont {Fischer}}, \bibinfo {author} {\bibfnamefont
  {R.}~\bibnamefont {Sittig}}, \bibinfo {author} {\bibfnamefont
  {M.}~\bibnamefont {Jetter}}, \bibinfo {author} {\bibfnamefont {S.~L.}\
  \bibnamefont {Portalupi}}, \ and\ \bibinfo {author} {\bibfnamefont
  {P.}~\bibnamefont {Michler}},\ }\bibfield  {title} {\enquote {\bibinfo
  {title} {Bright purcell enhanced single-photon source in the telecom o-band
  based on a quantum dot in a circular bragg grating},}\ }\href {\doibase
  10.1021/acs.nanolett.1c02647} {\bibfield  {journal} {\bibinfo  {journal}
  {Nano Letters}\ }\textbf {\bibinfo {volume} {21}},\ \bibinfo {pages}
  {7740--7745} (\bibinfo {year} {2021})}\BibitemShut {NoStop}%
\bibitem [{\citenamefont {Jahn}\ \emph {et~al.}(2015)\citenamefont {Jahn},
  \citenamefont {Munsch}, \citenamefont {B{\'{e}}guin}, \citenamefont
  {Kuhlmann}, \citenamefont {Renggli}, \citenamefont {Huo}, \citenamefont
  {Ding}, \citenamefont {Trotta}, \citenamefont {Reindl}, \citenamefont
  {Schmidt}, \citenamefont {Rastelli}, \citenamefont {Treutlein},\ and\
  \citenamefont {Warburton}}]{Jahn2015}%
  \BibitemOpen
  \bibfield  {author} {\bibinfo {author} {\bibfnamefont {J.-P.}\ \bibnamefont
  {Jahn}}, \bibinfo {author} {\bibfnamefont {M.}~\bibnamefont {Munsch}},
  \bibinfo {author} {\bibfnamefont {L.}~\bibnamefont {B{\'{e}}guin}}, \bibinfo
  {author} {\bibfnamefont {A.~V.}\ \bibnamefont {Kuhlmann}}, \bibinfo {author}
  {\bibfnamefont {M.}~\bibnamefont {Renggli}}, \bibinfo {author} {\bibfnamefont
  {Y.}~\bibnamefont {Huo}}, \bibinfo {author} {\bibfnamefont {F.}~\bibnamefont
  {Ding}}, \bibinfo {author} {\bibfnamefont {R.}~\bibnamefont {Trotta}},
  \bibinfo {author} {\bibfnamefont {M.}~\bibnamefont {Reindl}}, \bibinfo
  {author} {\bibfnamefont {O.~G.}\ \bibnamefont {Schmidt}}, \bibinfo {author}
  {\bibfnamefont {A.}~\bibnamefont {Rastelli}}, \bibinfo {author}
  {\bibfnamefont {P.}~\bibnamefont {Treutlein}}, \ and\ \bibinfo {author}
  {\bibfnamefont {R.~J.}\ \bibnamefont {Warburton}},\ }\bibfield  {title}
  {\enquote {\bibinfo {title} {An artificial rb atom in a semiconductor with
  lifetime-limited linewidth},}\ }\href
  {https://link.aps.org/doi/10.1103/PhysRevB.92.245439} {\bibfield  {journal}
  {\bibinfo  {journal} {Physical Review B}\ }\textbf {\bibinfo {volume} {92}},\
  \bibinfo {pages} {245439} (\bibinfo {year} {2015})}\BibitemShut {NoStop}%
\bibitem [{\citenamefont {Santori}\ \emph {et~al.}(2001)\citenamefont
  {Santori}, \citenamefont {Pelton}, \citenamefont {Solomon}, \citenamefont
  {Dale},\ and\ \citenamefont {Yamamoto}}]{Santori2001}%
  \BibitemOpen
  \bibfield  {author} {\bibinfo {author} {\bibfnamefont {C.}~\bibnamefont
  {Santori}}, \bibinfo {author} {\bibfnamefont {M.}~\bibnamefont {Pelton}},
  \bibinfo {author} {\bibfnamefont {G.}~\bibnamefont {Solomon}}, \bibinfo
  {author} {\bibfnamefont {Y.}~\bibnamefont {Dale}}, \ and\ \bibinfo {author}
  {\bibfnamefont {Y.}~\bibnamefont {Yamamoto}},\ }\bibfield  {title} {\enquote
  {\bibinfo {title} {Triggered single photons from a quantum dot},}\ }\href
  {\doibase 10.1103/physrevlett.86.1502} {\bibfield  {journal} {\bibinfo
  {journal} {Physical Review Letters}\ }\textbf {\bibinfo {volume} {86}},\
  \bibinfo {pages} {1502} (\bibinfo {year} {2001})}\BibitemShut {NoStop}%
\bibitem [{\citenamefont {Zhai}\ \emph {et~al.}(2020)\citenamefont {Zhai},
  \citenamefont {Löbl}, \citenamefont {Nguyen}, \citenamefont {Ritzmann},
  \citenamefont {Javadi}, \citenamefont {Spinnler}, \citenamefont {Wieck},
  \citenamefont {Ludwig},\ and\ \citenamefont {Warburton}}]{Zhai2020}%
  \BibitemOpen
  \bibfield  {author} {\bibinfo {author} {\bibfnamefont {L.}~\bibnamefont
  {Zhai}}, \bibinfo {author} {\bibfnamefont {M.~C.}\ \bibnamefont {Löbl}},
  \bibinfo {author} {\bibfnamefont {G.~N.}\ \bibnamefont {Nguyen}}, \bibinfo
  {author} {\bibfnamefont {J.}~\bibnamefont {Ritzmann}}, \bibinfo {author}
  {\bibfnamefont {A.}~\bibnamefont {Javadi}}, \bibinfo {author} {\bibfnamefont
  {C.}~\bibnamefont {Spinnler}}, \bibinfo {author} {\bibfnamefont {A.~D.}\
  \bibnamefont {Wieck}}, \bibinfo {author} {\bibfnamefont {A.}~\bibnamefont
  {Ludwig}}, \ and\ \bibinfo {author} {\bibfnamefont {R.~J.}\ \bibnamefont
  {Warburton}},\ }\bibfield  {title} {\enquote {\bibinfo {title} {Low-noise
  {GaAs} quantum dots for quantum photonics},}\ }\href@noop {} {\bibfield
  {journal} {\bibinfo  {journal} {Nature Communications}\ }\textbf {\bibinfo
  {volume} {11}} (\bibinfo {year} {2020})}\BibitemShut {NoStop}%
\bibitem [{\citenamefont {Nauerth}\ \emph {et~al.}(2013)\citenamefont
  {Nauerth}, \citenamefont {Moll}, \citenamefont {Rau}, \citenamefont {Fuchs},
  \citenamefont {Horwath}, \citenamefont {Frick},\ and\ \citenamefont
  {Weinfurter}}]{Nauerth2013}%
  \BibitemOpen
  \bibfield  {author} {\bibinfo {author} {\bibfnamefont {S.}~\bibnamefont
  {Nauerth}}, \bibinfo {author} {\bibfnamefont {F.}~\bibnamefont {Moll}},
  \bibinfo {author} {\bibfnamefont {M.}~\bibnamefont {Rau}}, \bibinfo {author}
  {\bibfnamefont {C.}~\bibnamefont {Fuchs}}, \bibinfo {author} {\bibfnamefont
  {J.}~\bibnamefont {Horwath}}, \bibinfo {author} {\bibfnamefont
  {S.}~\bibnamefont {Frick}}, \ and\ \bibinfo {author} {\bibfnamefont
  {H.}~\bibnamefont {Weinfurter}},\ }\bibfield  {title} {\enquote {\bibinfo
  {title} {Air-to-ground quantum communication},}\ }\href {\doibase
  10.1038/nphoton.2013.46} {\bibfield  {journal} {\bibinfo  {journal} {Nature
  Photonics}\ }\textbf {\bibinfo {volume} {7}},\ \bibinfo {pages} {382--386}
  (\bibinfo {year} {2013})}\BibitemShut {NoStop}%
\bibitem [{\citenamefont {Liao}\ \emph {et~al.}(2017)\citenamefont {Liao},
  \citenamefont {Cai}, \citenamefont {Liu}, \citenamefont {Zhang},
  \citenamefont {Li}, \citenamefont {Ren}, \citenamefont {Yin}, \citenamefont
  {Shen}, \citenamefont {Cao}, \citenamefont {Li}, \citenamefont {Li},
  \citenamefont {Chen}, \citenamefont {Sun}, \citenamefont {Jia}, \citenamefont
  {Wu}, \citenamefont {Jiang}, \citenamefont {Wang}, \citenamefont {Huang},
  \citenamefont {Wang}, \citenamefont {Zhou}, \citenamefont {Deng},
  \citenamefont {Xi}, \citenamefont {Ma}, \citenamefont {Hu}, \citenamefont
  {Zhang}, \citenamefont {Chen}, \citenamefont {Liu}, \citenamefont {Wang},
  \citenamefont {Zhu}, \citenamefont {Lu}, \citenamefont {Shu}, \citenamefont
  {Peng}, \citenamefont {Wang},\ and\ \citenamefont {Pan}}]{Liao2017}%
  \BibitemOpen
  \bibfield  {author} {\bibinfo {author} {\bibfnamefont {S.-K.}\ \bibnamefont
  {Liao}}, \bibinfo {author} {\bibfnamefont {W.-Q.}\ \bibnamefont {Cai}},
  \bibinfo {author} {\bibfnamefont {W.-Y.}\ \bibnamefont {Liu}}, \bibinfo
  {author} {\bibfnamefont {L.}~\bibnamefont {Zhang}}, \bibinfo {author}
  {\bibfnamefont {Y.}~\bibnamefont {Li}}, \bibinfo {author} {\bibfnamefont
  {J.-G.}\ \bibnamefont {Ren}}, \bibinfo {author} {\bibfnamefont
  {J.}~\bibnamefont {Yin}}, \bibinfo {author} {\bibfnamefont {Q.}~\bibnamefont
  {Shen}}, \bibinfo {author} {\bibfnamefont {Y.}~\bibnamefont {Cao}}, \bibinfo
  {author} {\bibfnamefont {Z.-P.}\ \bibnamefont {Li}}, \bibinfo {author}
  {\bibfnamefont {F.-Z.}\ \bibnamefont {Li}}, \bibinfo {author} {\bibfnamefont
  {X.-W.}\ \bibnamefont {Chen}}, \bibinfo {author} {\bibfnamefont {L.-H.}\
  \bibnamefont {Sun}}, \bibinfo {author} {\bibfnamefont {J.-J.}\ \bibnamefont
  {Jia}}, \bibinfo {author} {\bibfnamefont {J.-C.}\ \bibnamefont {Wu}},
  \bibinfo {author} {\bibfnamefont {X.-J.}\ \bibnamefont {Jiang}}, \bibinfo
  {author} {\bibfnamefont {J.-F.}\ \bibnamefont {Wang}}, \bibinfo {author}
  {\bibfnamefont {Y.-M.}\ \bibnamefont {Huang}}, \bibinfo {author}
  {\bibfnamefont {Q.}~\bibnamefont {Wang}}, \bibinfo {author} {\bibfnamefont
  {Y.-L.}\ \bibnamefont {Zhou}}, \bibinfo {author} {\bibfnamefont
  {L.}~\bibnamefont {Deng}}, \bibinfo {author} {\bibfnamefont {T.}~\bibnamefont
  {Xi}}, \bibinfo {author} {\bibfnamefont {L.}~\bibnamefont {Ma}}, \bibinfo
  {author} {\bibfnamefont {T.}~\bibnamefont {Hu}}, \bibinfo {author}
  {\bibfnamefont {Q.}~\bibnamefont {Zhang}}, \bibinfo {author} {\bibfnamefont
  {Y.-A.}\ \bibnamefont {Chen}}, \bibinfo {author} {\bibfnamefont {N.-L.}\
  \bibnamefont {Liu}}, \bibinfo {author} {\bibfnamefont {X.-B.}\ \bibnamefont
  {Wang}}, \bibinfo {author} {\bibfnamefont {Z.-C.}\ \bibnamefont {Zhu}},
  \bibinfo {author} {\bibfnamefont {C.-Y.}\ \bibnamefont {Lu}}, \bibinfo
  {author} {\bibfnamefont {R.}~\bibnamefont {Shu}}, \bibinfo {author}
  {\bibfnamefont {C.-Z.}\ \bibnamefont {Peng}}, \bibinfo {author}
  {\bibfnamefont {J.-Y.}\ \bibnamefont {Wang}}, \ and\ \bibinfo {author}
  {\bibfnamefont {J.-W.}\ \bibnamefont {Pan}},\ }\bibfield  {title} {\enquote
  {\bibinfo {title} {Satellite-to-ground quantum key distribution},}\ }\href
  {\doibase 10.1038/nature23655} {\bibfield  {journal} {\bibinfo  {journal}
  {Nature}\ }\textbf {\bibinfo {volume} {549}},\ \bibinfo {pages} {43--47}
  (\bibinfo {year} {2017})}\BibitemShut {NoStop}%
\bibitem [{\citenamefont {Yin}\ \emph {et~al.}(2017)\citenamefont {Yin},
  \citenamefont {Cao}, \citenamefont {Li}, \citenamefont {Ren}, \citenamefont
  {Liao}, \citenamefont {Zhang}, \citenamefont {Cai}, \citenamefont {Liu},
  \citenamefont {Li}, \citenamefont {Dai}, \citenamefont {Li}, \citenamefont
  {Huang}, \citenamefont {Deng}, \citenamefont {Li}, \citenamefont {Zhang},
  \citenamefont {Liu}, \citenamefont {Chen}, \citenamefont {Lu}, \citenamefont
  {Shu}, \citenamefont {Peng}, \citenamefont {Wang},\ and\ \citenamefont
  {Pan}}]{Yin2017}%
  \BibitemOpen
  \bibfield  {author} {\bibinfo {author} {\bibfnamefont {J.}~\bibnamefont
  {Yin}}, \bibinfo {author} {\bibfnamefont {Y.}~\bibnamefont {Cao}}, \bibinfo
  {author} {\bibfnamefont {Y.-H.}\ \bibnamefont {Li}}, \bibinfo {author}
  {\bibfnamefont {J.-G.}\ \bibnamefont {Ren}}, \bibinfo {author} {\bibfnamefont
  {S.-K.}\ \bibnamefont {Liao}}, \bibinfo {author} {\bibfnamefont
  {L.}~\bibnamefont {Zhang}}, \bibinfo {author} {\bibfnamefont {W.-Q.}\
  \bibnamefont {Cai}}, \bibinfo {author} {\bibfnamefont {W.-Y.}\ \bibnamefont
  {Liu}}, \bibinfo {author} {\bibfnamefont {B.}~\bibnamefont {Li}}, \bibinfo
  {author} {\bibfnamefont {H.}~\bibnamefont {Dai}}, \bibinfo {author}
  {\bibfnamefont {M.}~\bibnamefont {Li}}, \bibinfo {author} {\bibfnamefont
  {Y.-M.}\ \bibnamefont {Huang}}, \bibinfo {author} {\bibfnamefont
  {L.}~\bibnamefont {Deng}}, \bibinfo {author} {\bibfnamefont {L.}~\bibnamefont
  {Li}}, \bibinfo {author} {\bibfnamefont {Q.}~\bibnamefont {Zhang}}, \bibinfo
  {author} {\bibfnamefont {N.-L.}\ \bibnamefont {Liu}}, \bibinfo {author}
  {\bibfnamefont {Y.-A.}\ \bibnamefont {Chen}}, \bibinfo {author}
  {\bibfnamefont {C.-Y.}\ \bibnamefont {Lu}}, \bibinfo {author} {\bibfnamefont
  {R.}~\bibnamefont {Shu}}, \bibinfo {author} {\bibfnamefont {C.-Z.}\
  \bibnamefont {Peng}}, \bibinfo {author} {\bibfnamefont {J.-Y.}\ \bibnamefont
  {Wang}}, \ and\ \bibinfo {author} {\bibfnamefont {J.-W.}\ \bibnamefont
  {Pan}},\ }\bibfield  {title} {\enquote {\bibinfo {title} {Satellite-to-ground
  entanglement-based quantum key distribution},}\ }\href
  {https://link.aps.org/doi/10.1103/PhysRevLett.119.200501} {\bibfield
  {journal} {\bibinfo  {journal} {Physical Review Letters}\ }\textbf {\bibinfo
  {volume} {119}} (\bibinfo {year} {2017})}\BibitemShut {NoStop}%
\bibitem [{\citenamefont {Yin}\ \emph {et~al.}(2020)\citenamefont {Yin},
  \citenamefont {Li}, \citenamefont {Liao}, \citenamefont {Yang}, \citenamefont
  {Cao}, \citenamefont {Zhang}, \citenamefont {Ren}, \citenamefont {Cai},
  \citenamefont {Liu}, \citenamefont {Li}, \citenamefont {Shu}, \citenamefont
  {Huang}, \citenamefont {Deng}, \citenamefont {Li}, \citenamefont {Zhang},
  \citenamefont {Liu}, \citenamefont {Chen}, \citenamefont {Lu}, \citenamefont
  {Wang}, \citenamefont {Xu}, \citenamefont {Wang}, \citenamefont {Peng},
  \citenamefont {Ekert},\ and\ \citenamefont {Pan}}]{Yin2020}%
  \BibitemOpen
  \bibfield  {author} {\bibinfo {author} {\bibfnamefont {J.}~\bibnamefont
  {Yin}}, \bibinfo {author} {\bibfnamefont {Y.-H.}\ \bibnamefont {Li}},
  \bibinfo {author} {\bibfnamefont {S.-K.}\ \bibnamefont {Liao}}, \bibinfo
  {author} {\bibfnamefont {M.}~\bibnamefont {Yang}}, \bibinfo {author}
  {\bibfnamefont {Y.}~\bibnamefont {Cao}}, \bibinfo {author} {\bibfnamefont
  {L.}~\bibnamefont {Zhang}}, \bibinfo {author} {\bibfnamefont {J.-G.}\
  \bibnamefont {Ren}}, \bibinfo {author} {\bibfnamefont {W.-Q.}\ \bibnamefont
  {Cai}}, \bibinfo {author} {\bibfnamefont {W.-Y.}\ \bibnamefont {Liu}},
  \bibinfo {author} {\bibfnamefont {S.-L.}\ \bibnamefont {Li}}, \bibinfo
  {author} {\bibfnamefont {R.}~\bibnamefont {Shu}}, \bibinfo {author}
  {\bibfnamefont {Y.-M.}\ \bibnamefont {Huang}}, \bibinfo {author}
  {\bibfnamefont {L.}~\bibnamefont {Deng}}, \bibinfo {author} {\bibfnamefont
  {L.}~\bibnamefont {Li}}, \bibinfo {author} {\bibfnamefont {Q.}~\bibnamefont
  {Zhang}}, \bibinfo {author} {\bibfnamefont {N.-L.}\ \bibnamefont {Liu}},
  \bibinfo {author} {\bibfnamefont {Y.-A.}\ \bibnamefont {Chen}}, \bibinfo
  {author} {\bibfnamefont {C.-Y.}\ \bibnamefont {Lu}}, \bibinfo {author}
  {\bibfnamefont {X.-B.}\ \bibnamefont {Wang}}, \bibinfo {author}
  {\bibfnamefont {F.}~\bibnamefont {Xu}}, \bibinfo {author} {\bibfnamefont
  {J.-Y.}\ \bibnamefont {Wang}}, \bibinfo {author} {\bibfnamefont {C.-Z.}\
  \bibnamefont {Peng}}, \bibinfo {author} {\bibfnamefont {A.~K.}\ \bibnamefont
  {Ekert}}, \ and\ \bibinfo {author} {\bibfnamefont {J.-W.}\ \bibnamefont
  {Pan}},\ }\bibfield  {title} {\enquote {\bibinfo {title} {Entanglement-based
  secure quantum cryptography over 1,120 kilometres},}\ }\href {\doibase
  10.1038/s41586-020-2401-y} {\bibfield  {journal} {\bibinfo  {journal}
  {Nature}\ }\textbf {\bibinfo {volume} {582}},\ \bibinfo {pages} {501--505}
  (\bibinfo {year} {2020})}\BibitemShut {NoStop}%
\bibitem [{\citenamefont {Cai}\ and\ \citenamefont {Scarani}(2009)}]{Cai_2009}%
  \BibitemOpen
  \bibfield  {author} {\bibinfo {author} {\bibfnamefont {R.~Y.~Q.}\
  \bibnamefont {Cai}}\ and\ \bibinfo {author} {\bibfnamefont {V.}~\bibnamefont
  {Scarani}},\ }\bibfield  {title} {\enquote {\bibinfo {title} {Finite-key
  analysis for practical implementations of quantum key distribution},}\ }\href
  {\doibase 10.1088/1367-2630/11/4/045024} {\bibfield  {journal} {\bibinfo
  {journal} {New Journal of Physics}\ }\textbf {\bibinfo {volume} {11}},\
  \bibinfo {pages} {045024} (\bibinfo {year} {2009})}\BibitemShut {NoStop}%
\bibitem [{\citenamefont {Rickert}\ \emph {et~al.}(2019)\citenamefont
  {Rickert}, \citenamefont {Kupko}, \citenamefont {Rodt}, \citenamefont
  {Reitzenstein},\ and\ \citenamefont {Heindel}}]{Rickert2019}%
  \BibitemOpen
  \bibfield  {author} {\bibinfo {author} {\bibfnamefont {L.}~\bibnamefont
  {Rickert}}, \bibinfo {author} {\bibfnamefont {T.}~\bibnamefont {Kupko}},
  \bibinfo {author} {\bibfnamefont {S.}~\bibnamefont {Rodt}}, \bibinfo {author}
  {\bibfnamefont {S.}~\bibnamefont {Reitzenstein}}, \ and\ \bibinfo {author}
  {\bibfnamefont {T.}~\bibnamefont {Heindel}},\ }\bibfield  {title} {\enquote
  {\bibinfo {title} {Optimized designs for telecom-wavelength quantum light
  sources based on hybrid circular bragg gratings},}\ }\href {\doibase
  10.1364/oe.27.036824} {\bibfield  {journal} {\bibinfo  {journal} {Optics
  Express}\ }\textbf {\bibinfo {volume} {27}},\ \bibinfo {pages} {36824}
  (\bibinfo {year} {2019})}\BibitemShut {NoStop}%
\bibitem [{\citenamefont {Dada}\ \emph {et~al.}(2016)\citenamefont {Dada},
  \citenamefont {Santana}, \citenamefont {Malein}, \citenamefont
  {Koutroumanis}, \citenamefont {Ma}, \citenamefont {Zajac}, \citenamefont
  {Lim}, \citenamefont {Song},\ and\ \citenamefont {Gerardot}}]{Dada2016}%
  \BibitemOpen
  \bibfield  {author} {\bibinfo {author} {\bibfnamefont {A.~C.}\ \bibnamefont
  {Dada}}, \bibinfo {author} {\bibfnamefont {T.~S.}\ \bibnamefont {Santana}},
  \bibinfo {author} {\bibfnamefont {R.~N.~E.}\ \bibnamefont {Malein}}, \bibinfo
  {author} {\bibfnamefont {A.}~\bibnamefont {Koutroumanis}}, \bibinfo {author}
  {\bibfnamefont {Y.}~\bibnamefont {Ma}}, \bibinfo {author} {\bibfnamefont
  {J.~M.}\ \bibnamefont {Zajac}}, \bibinfo {author} {\bibfnamefont {J.~Y.}\
  \bibnamefont {Lim}}, \bibinfo {author} {\bibfnamefont {J.~D.}\ \bibnamefont
  {Song}}, \ and\ \bibinfo {author} {\bibfnamefont {B.~D.}\ \bibnamefont
  {Gerardot}},\ }\bibfield  {title} {\enquote {\bibinfo {title}
  {Indistinguishable single photons with flexible electronic triggering},}\
  }\href {\doibase 10.1364/OPTICA.3.000493} {\bibfield  {journal} {\bibinfo
  {journal} {Optica}\ }\textbf {\bibinfo {volume} {3}},\ \bibinfo {pages}
  {493--498} (\bibinfo {year} {2016})}\BibitemShut {NoStop}%
\bibitem [{\citenamefont {Munnelly}\ \emph {et~al.}(2017)\citenamefont
  {Munnelly}, \citenamefont {Heindel}, \citenamefont {Thoma}, \citenamefont
  {Kamp}, \citenamefont {Höfling}, \citenamefont {Schneider},\ and\
  \citenamefont {Reitzenstein}}]{Munnelly2017}%
  \BibitemOpen
  \bibfield  {author} {\bibinfo {author} {\bibfnamefont {P.}~\bibnamefont
  {Munnelly}}, \bibinfo {author} {\bibfnamefont {T.}~\bibnamefont {Heindel}},
  \bibinfo {author} {\bibfnamefont {A.}~\bibnamefont {Thoma}}, \bibinfo
  {author} {\bibfnamefont {M.}~\bibnamefont {Kamp}}, \bibinfo {author}
  {\bibfnamefont {S.}~\bibnamefont {Höfling}}, \bibinfo {author}
  {\bibfnamefont {C.}~\bibnamefont {Schneider}}, \ and\ \bibinfo {author}
  {\bibfnamefont {S.}~\bibnamefont {Reitzenstein}},\ }\bibfield  {title}
  {\enquote {\bibinfo {title} {Electrically tunable single-photon source
  triggered by a monolithically integrated quantum dot microlaser},}\ }\href
  {\doibase 10.1021/acsphotonics.7b00119} {\bibfield  {journal} {\bibinfo
  {journal} {{ACS} Photonics}\ }\textbf {\bibinfo {volume} {4}},\ \bibinfo
  {pages} {790--794} (\bibinfo {year} {2017})}\BibitemShut {NoStop}%
\bibitem [{\citenamefont {Lee}\ \emph {et~al.}(2017)\citenamefont {Lee},
  \citenamefont {Murray}, \citenamefont {Bennett}, \citenamefont {Ellis},
  \citenamefont {Dangel}, \citenamefont {Farrer}, \citenamefont {Spencer},
  \citenamefont {Ritchie},\ and\ \citenamefont {Shields}}]{Lee2017}%
  \BibitemOpen
  \bibfield  {author} {\bibinfo {author} {\bibfnamefont {J.~P.}\ \bibnamefont
  {Lee}}, \bibinfo {author} {\bibfnamefont {E.}~\bibnamefont {Murray}},
  \bibinfo {author} {\bibfnamefont {A.~J.}\ \bibnamefont {Bennett}}, \bibinfo
  {author} {\bibfnamefont {D.~J.~P.}\ \bibnamefont {Ellis}}, \bibinfo {author}
  {\bibfnamefont {C.}~\bibnamefont {Dangel}}, \bibinfo {author} {\bibfnamefont
  {I.}~\bibnamefont {Farrer}}, \bibinfo {author} {\bibfnamefont
  {P.}~\bibnamefont {Spencer}}, \bibinfo {author} {\bibfnamefont {D.~A.}\
  \bibnamefont {Ritchie}}, \ and\ \bibinfo {author} {\bibfnamefont {A.~J.}\
  \bibnamefont {Shields}},\ }\bibfield  {title} {\enquote {\bibinfo {title}
  {Electrically driven and electrically tunable quantum light sources},}\
  }\href {https://doi.org/10.1063/1.4976197} {\bibfield  {journal} {\bibinfo
  {journal} {Applied Physics Letters}\ }\textbf {\bibinfo {volume} {110}},\
  \bibinfo {pages} {071102} (\bibinfo {year} {2017})}\BibitemShut {NoStop}%
\bibitem [{\citenamefont {Xiang}\ \emph {et~al.}(2020)\citenamefont {Xiang},
  \citenamefont {Huwer}, \citenamefont {Skiba-Szymanska}, \citenamefont
  {Stevenson}, \citenamefont {Ellis}, \citenamefont {Farrer}, \citenamefont
  {Ward}, \citenamefont {Ritchie},\ and\ \citenamefont {Shields}}]{Xiang2020}%
  \BibitemOpen
  \bibfield  {author} {\bibinfo {author} {\bibfnamefont {Z.-H.}\ \bibnamefont
  {Xiang}}, \bibinfo {author} {\bibfnamefont {J.}~\bibnamefont {Huwer}},
  \bibinfo {author} {\bibfnamefont {J.}~\bibnamefont {Skiba-Szymanska}},
  \bibinfo {author} {\bibfnamefont {R.~M.}\ \bibnamefont {Stevenson}}, \bibinfo
  {author} {\bibfnamefont {D.~J.~P.}\ \bibnamefont {Ellis}}, \bibinfo {author}
  {\bibfnamefont {I.}~\bibnamefont {Farrer}}, \bibinfo {author} {\bibfnamefont
  {M.~B.}\ \bibnamefont {Ward}}, \bibinfo {author} {\bibfnamefont {D.~A.}\
  \bibnamefont {Ritchie}}, \ and\ \bibinfo {author} {\bibfnamefont {A.~J.}\
  \bibnamefont {Shields}},\ }\bibfield  {title} {\enquote {\bibinfo {title} {A
  tuneable telecom wavelength entangled light emitting diode deployed in an
  installed fibre network},}\ }\href
  {https://doi.org/10.1038/s42005-020-0390-7} {\bibfield  {journal} {\bibinfo
  {journal} {Communications Physics}\ }\textbf {\bibinfo {volume} {3}},\
  \bibinfo {pages} {121} (\bibinfo {year} {2020})}\BibitemShut {NoStop}%
\bibitem [{\citenamefont {Thoma}\ \emph {et~al.}(2016)\citenamefont {Thoma},
  \citenamefont {Schnauber}, \citenamefont {Gschrey}, \citenamefont {Seifried},
  \citenamefont {Wolters}, \citenamefont {Schulze}, \citenamefont
  {Strittmatter}, \citenamefont {Rodt}, \citenamefont {Carmele}, \citenamefont
  {Knorr}, \citenamefont {Heindel},\ and\ \citenamefont
  {Reitzenstein}}]{Thoma2016}%
  \BibitemOpen
  \bibfield  {author} {\bibinfo {author} {\bibfnamefont {A.}~\bibnamefont
  {Thoma}}, \bibinfo {author} {\bibfnamefont {P.}~\bibnamefont {Schnauber}},
  \bibinfo {author} {\bibfnamefont {M.}~\bibnamefont {Gschrey}}, \bibinfo
  {author} {\bibfnamefont {M.}~\bibnamefont {Seifried}}, \bibinfo {author}
  {\bibfnamefont {J.}~\bibnamefont {Wolters}}, \bibinfo {author} {\bibfnamefont
  {J.-H.}\ \bibnamefont {Schulze}}, \bibinfo {author} {\bibfnamefont
  {A.}~\bibnamefont {Strittmatter}}, \bibinfo {author} {\bibfnamefont
  {S.}~\bibnamefont {Rodt}}, \bibinfo {author} {\bibfnamefont {A.}~\bibnamefont
  {Carmele}}, \bibinfo {author} {\bibfnamefont {A.}~\bibnamefont {Knorr}},
  \bibinfo {author} {\bibfnamefont {T.}~\bibnamefont {Heindel}}, \ and\
  \bibinfo {author} {\bibfnamefont {S.}~\bibnamefont {Reitzenstein}},\
  }\bibfield  {title} {\enquote {\bibinfo {title} {Exploring dephasing of a
  solid-state quantum emitter via time- and temperature-dependent
  hong-ou-mandel experiments},}\ }\href
  {https://link.aps.org/doi/10.1103/PhysRevLett.116.033601} {\bibfield
  {journal} {\bibinfo  {journal} {Physical Review Letters}\ }\textbf {\bibinfo
  {volume} {116}},\ \bibinfo {pages} {033601} (\bibinfo {year}
  {2016})}\BibitemShut {NoStop}%
\bibitem [{\citenamefont {Braunstein}\ and\ \citenamefont
  {Pirandola}(2012)}]{Braunstein2012}%
  \BibitemOpen
  \bibfield  {author} {\bibinfo {author} {\bibfnamefont {S.~L.}\ \bibnamefont
  {Braunstein}}\ and\ \bibinfo {author} {\bibfnamefont {S.}~\bibnamefont
  {Pirandola}},\ }\bibfield  {title} {\enquote {\bibinfo {title}
  {Side-channel-free quantum key distribution},}\ }\href {\doibase
  10.1103/PhysRevLett.108.130502} {\bibfield  {journal} {\bibinfo  {journal}
  {Physical Review Letters}\ }\textbf {\bibinfo {volume} {108}},\ \bibinfo
  {pages} {130502} (\bibinfo {year} {2012})}\BibitemShut {NoStop}%
\bibitem [{\citenamefont {Lo}, \citenamefont {Curty},\ and\ \citenamefont
  {Qi}(2012)}]{Lo2012}%
  \BibitemOpen
  \bibfield  {author} {\bibinfo {author} {\bibfnamefont {H.-K.}\ \bibnamefont
  {Lo}}, \bibinfo {author} {\bibfnamefont {M.}~\bibnamefont {Curty}}, \ and\
  \bibinfo {author} {\bibfnamefont {B.}~\bibnamefont {Qi}},\ }\bibfield
  {title} {\enquote {\bibinfo {title} {Measurement-device-independent quantum
  key distribution},}\ }\href {\doibase 10.1103/PhysRevLett.108.130503}
  {\bibfield  {journal} {\bibinfo  {journal} {Physical Review Letters}\
  }\textbf {\bibinfo {volume} {108}},\ \bibinfo {pages} {130503} (\bibinfo
  {year} {2012})}\BibitemShut {NoStop}%
\end{thebibliography}


%


\section*{Supplemental Material}\label{SM}

\setcounter{figure}{0}
\renewcommand{\thefigure}{S\arabic{figure}}

\onecolumngrid
\subsection{Blinking behavior}
\begin{figure*}[h]
	\includegraphics{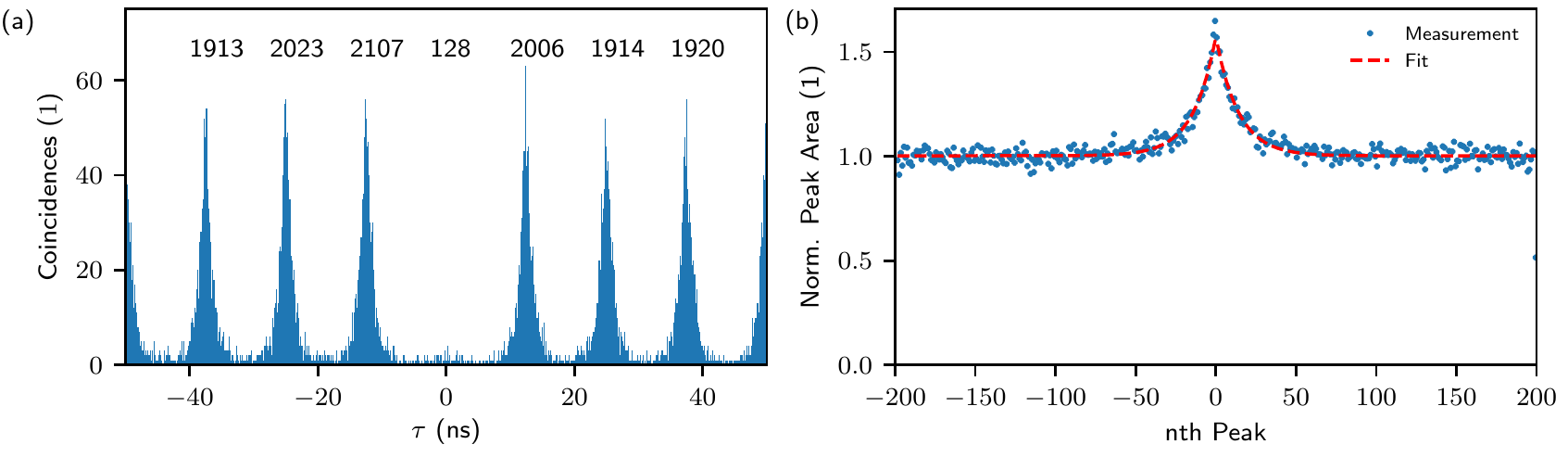}
	\caption{Second-order autocorrelation $g^{(2)}(\tau)$ histogram (\SI{50}{\pico\second} bin width) under pulsed resonant p-shell excitation at \SI{1247.9}{\nano\meter}. (a)~Histogram from Fig.~2~(b) in the main text with the number of integrated coincidences shown for each peak. (b)~Normalized number of integrated coincidences of the peaks in a wider temporal window revealing blinking behavior of the quantum dot. The time constants $\tau_{\mathrm{on}} = \SI{482.3(25)}{\nano\second}$ and $\tau_{\mathrm{off}} = \SI{275.1(10)}{\nano\second}$ are extracted from the model-fit following Ref.\cite{Santori2001}.}
	\label{fig:blinking}
\end{figure*}

\newpage
\subsection{Measurement data for QBER and signal fraction under temporal filtering}
\begin{figure*}[h]
	\includegraphics{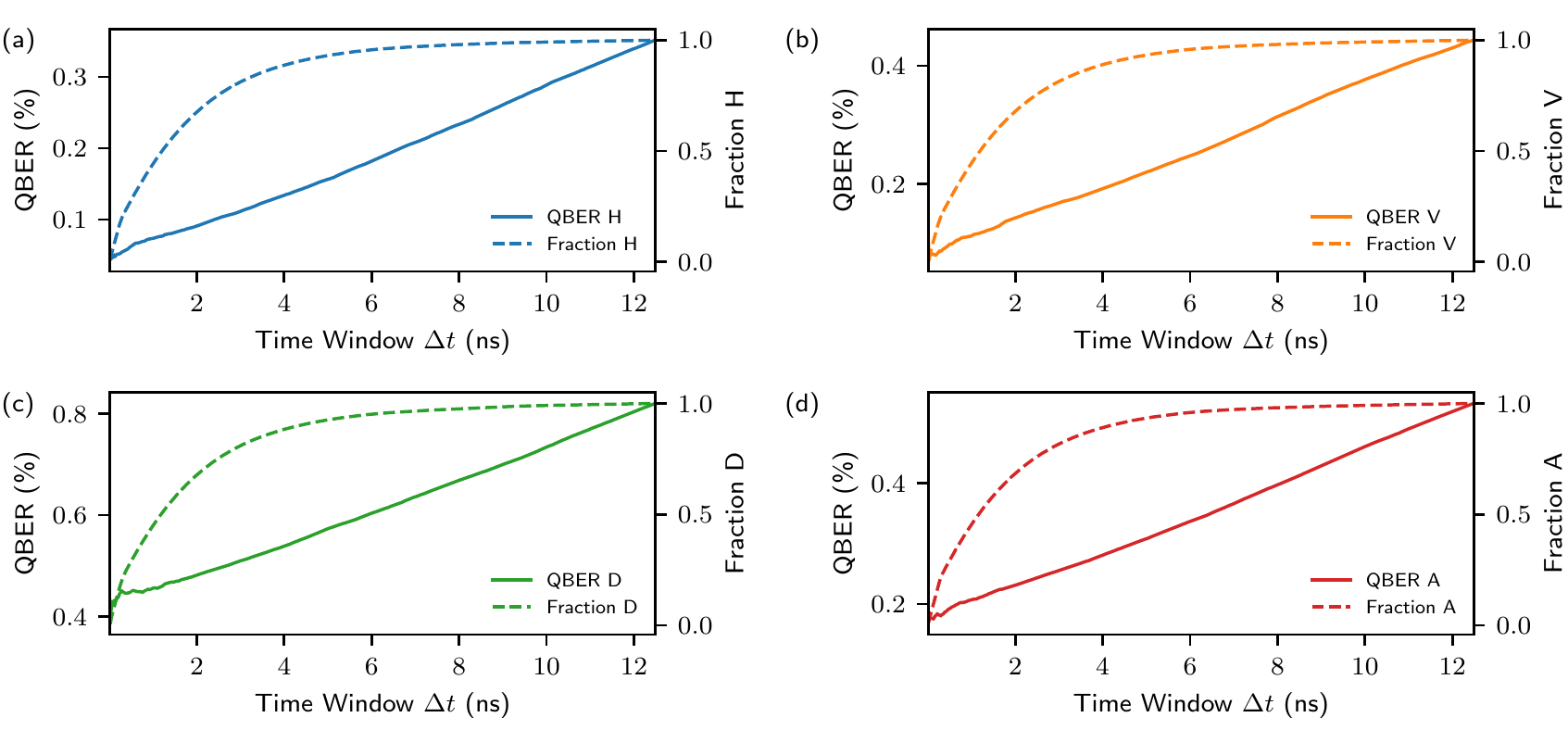}
	\caption{The effect of temporal filtering on the QBER and Fraction of the overall signal for each polarization. From the photon arrival time distributions of Fig.~2\,(a) in the main text the QBERs for all four possible input polarizations are calculated. See Ref.\cite{Kupko2020} for detailed explanations.}
	\label{fig:QBER_Frac}
\end{figure*}

\newpage

\subsection{Parameter Optimization}
\begin{figure*}[h]
	\includegraphics{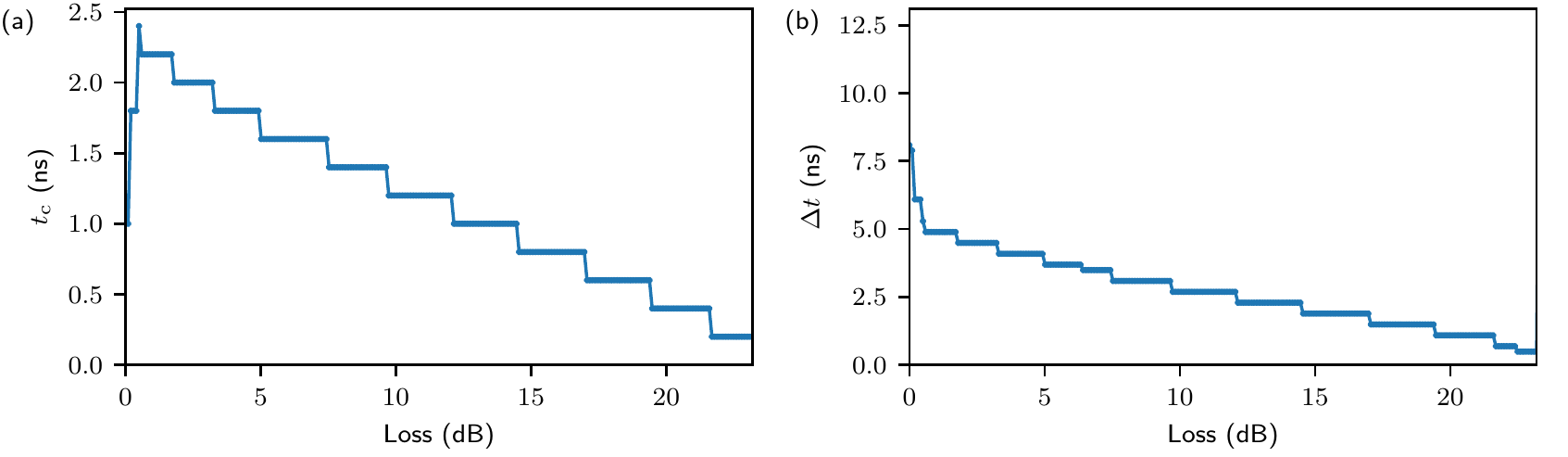}
	\caption{Parameter sets for (a) center $t_{\mathrm{c}}$ and (b) width $\Delta_t$ of the 2D temporal filters optimized for different losses underlying the asymptotic key rate in Fig.~4\,(a) of the main text (data displayed by points, lines are a guide to the eye). The value of $t_{\mathrm{c}}$ reduces with decreasing $\Delta_t$, i.e. the optimal signal to noise ratio can be achieved around the maximum of the photon arrival time distribution. The step-like behavior results from the finite statistics of the data and the finite step size for optimization.}
	\label{fig:parameter_opt}
\end{figure*}

\end{document}